\documentclass[sigconf]{acmart}
\def\BibTeX{{\rm B\kern-.05em{\sc i\kern-.025em b}\kern-.08emT\kern-.1667em\lower.7ex\hbox{E}\kern-.125emX}}
    \pdfoutput=1
\usepackage{nicefrac}
\usepackage{siunitx}
\usepackage{array,framed}
\usepackage{booktabs}
\usepackage{soul}
\usepackage[inline]{enumitem}
\usepackage{amsmath}
\usepackage{amsfonts}
\usepackage{balance}
\usepackage{soul}
\usepackage{textcomp}
\usepackage{xcolor}
\usepackage{graphicx}
\usepackage{float}
\usepackage{caption}
\usepackage{subcaption}
\usepackage{array}

\usepackage{textcomp}
\usepackage{setspace}
\usepackage{latexsym,fancyhdr,url}
\usepackage{enumerate}
\usepackage{algorithm2e}
\usepackage{algpseudocode}
\usepackage{graphics}
\usepackage{xparse} % argument parsing -- \edist
\usepackage{xspace}
\usepackage{multirow}
\usepackage{csvsimple}
\usepackage{balance}
\usepackage{
  tikz,
  pgfplots,
  pgfplotstable
}

\usetikzlibrary{
  shapes.geometric,
  arrows,
  external,
  pgfplots.groupplots,
  matrix
}

\pgfplotsset{compat=1.9}
\usepackage{mathtools,}

\DeclareMathAlphabet{\mathcal}{OMS}{cmsy}{m}{n}
\DeclareGraphicsExtensions{%
    .png,.PNG,%
    .pdf,.PDF,%
    .jpg,.mps,.jpeg,.jbig2,.jb2,.JPG,.JPEG,.JBIG2,.JB2}

\usepackage{xparse}
\newcommand{\bnm}{\begin{newmath}}
\newcommand{\enm}{\end{newmath}}

\newcommand{\bea}{\begin{eqnarray*}}%
\newcommand{\eea}{\end{eqnarray*}}%

\newcommand{\bne}{\begin{newequation}}
\newcommand{\ene}{\end{newequation}}

\newcommand{\bal}{\begin{newalign}}
\newcommand{\eal}{\end{newalign}}

\newenvironment{newalign}{\begin{align}%
\setlength{\abovedisplayskip}{4pt}%
\setlength{\belowdisplayskip}{4pt}%
\setlength{\abovedisplayshortskip}{6pt}%
\setlength{\belowdisplayshortskip}{6pt} }{\end{align}}

\newenvironment{newmath}{\begin{displaymath}%
\setlength{\abovedisplayskip}{4pt}%
\setlength{\belowdisplayskip}{4pt}%
\setlength{\abovedisplayshortskip}{6pt}%
\setlength{\belowdisplayshortskip}{6pt} }{\end{displaymath}}

\newenvironment{newequation}{\begin{equation}%
\setlength{\abovedisplayskip}{4pt}%
\setlength{\belowdisplayskip}{4pt}%
\setlength{\abovedisplayshortskip}{6pt}%
\setlength{\belowdisplayshortskip}{6pt} }{\end{equation}}

\newcounter{ctr}

\newcounter{mytable}
\def\mytable{\begin{centering}\refstepcounter{mytable}}
\def\endmytable{\end{centering}}

\newcounter{myfig}
\def\myfig{\begin{centering}\refstepcounter{myfig}}
\def\endmyfig{\end{centering}}

\newlength{\saveparindent}
\newlength{\saveparskip}
\newcommand{\E}{{\rm I\kern-.3em E}}

\renewcommand{\eqref}[1]{\mbox{Equation~(\ref{#1})}}

\def \part {part}

\DeclareMathOperator*{\argmin}{argmin}

\renewcommand{\paragraph}[1]{\vspace*{6pt}\noindent\textbf{#1}\;}

\def \blackslug{\hbox{\hskip 1pt \vrule width 4pt height 8pt
    depth 1.5pt \hskip 1pt}}
\def \qed{\quad\blackslug\lower 8.5pt\null\par}
\newcounter{mynote}[section]

\newcommand\ignore[1]{}

\newcounter{rcnote}[section]

\newcounter{mrnote}[section]

\newcounter{fknote}[section]

\newcounter{anote}[section]

\DeclareMathSymbol{\mlq}{\mathord}{operators}{``}
\DeclareMathSymbol{\mrq}{\mathord}{operators}{`'}

\newcommand{\rhf}[2]{R_{f, \gamma}}

\DeclareDocumentCommand{\edist}{o o}{
  \ensuremath{
    \IfNoValueTF{#1}{{d}}{{\sf d}(#1,#2)}
  }
}

\newcommand{\olrk}[1]{\ifx\nursymbol#1\else\!\!\mskip4.5mu plus 0.5mu\left(\mskip0.5mu plus0.5mu #1\mskip1.5mu plus0.5mu \right)\fi}

\NewDocumentCommand{\indseq}{ O{1} O{r} }{{#1}\ldots {#2}}

\begin{document}
%\fontfamily{lmr}\selectfont
% \def\thetitle{A Practical Way to Generate Strong Keys from Noisy Data}
\fancyhead{}
\def\thetitle{MDTD: A Multi-Domain Trojan Detector for\\ Deep Neural Networks}
\title{\thetitle}

% \author{Arezoo Rajabi}
% \author{Surudhi Asokraj}
% \author{Fengqing Jiang}
% \author{Luyao Niu}
% \affiliation{%
%   \institution{University of Washington}
%   \streetaddress{University of Washington}
%   \city{Seattle}
%   \country{USA}}
% \email{rajabia@uw.edu}
% \email{surudh22@uw.edu}
% \email{fqjiang@uw.edu}
% \email{luyaoniu@uw.edu}

\author{Arezoo Rajabi}
\affiliation{%
  \institution{University of Washington}
  \streetaddress{University of Washington}
  \city{Seattle}
  \country{USA}}
\email{rajabia@uw.edu}

\author{Surudhi Asokraj}
\affiliation{%
  \institution{University of Washington}
  \streetaddress{University of Washington}
  \city{Seattle}
  \country{USA}}
\email{surudh22@uw.edu}

\author{Fengqing Jiang}
\affiliation{%
  \institution{University of Washington}
  \streetaddress{University of Washington}
  \city{Seattle}
  \country{USA}}
\email{fqjiang@uw.edu}

\author{Luyao Niu}
\affiliation{%
  \institution{University of Washington}
  \streetaddress{University of Washington}
  \city{Seattle}
  \country{USA}}
\email{luyaoniu@uw.edu}

\author{Bhaskar Ramasubramanian}
\affiliation{%
  \institution{Western Washington University}
  \streetaddress{Western Washington University}
  \city{Bellingham}
  \country{USA}}
\email{ramasub@wwu.edu}

% \author{Jim Ritcey, Radha Poovendran}
% \affiliation{%
%   \institution{University of Washington}
%   \streetaddress{University of Washington}
%   \city{Seattle}
%   \country{USA}}
% \email{(jar7, rp3)@uw.edu}

\author{Jim Ritcey}
\affiliation{%
  \institution{University of Washington}
  \streetaddress{University of Washington}
  \city{Seattle}
  \country{USA}}
\email{jar7@uw.edu}

\author{Radha Poovendran}
\affiliation{%
  \institution{University of Washington}
  \streetaddress{University of Washington}
  \city{Seattle}
  \country{USA}}
\email{rp3@uw.edu}
% \author{Rahul Chatterjee}
% \affiliation{\small{UW--Madison}}

\date{}

\begin{abstract}
Machine learning models that use deep neural networks (DNNs) are %have been shown to be 
vulnerable to backdoor attacks. 
An adversary carrying out a backdoor attack embeds a predefined perturbation called a trigger into a small subset of input samples and trains the DNN  
%The DNN can then be trained 
such that the presence of the trigger in the input results in an adversary-desired output class. %, while %. 
Such adversarial retraining however needs to ensure that %the adversarial training ensures that 
outputs for inputs without the trigger remain unaffected and provide high classification accuracy on clean samples. 
%Backdoor attacks can have severe implications in safety-critical applications where an attacker can adversely influence behavior of the DNN. 
Existing defenses against backdoor attacks are %known to be 
computationally expensive, and their success has been demonstrated primarily on image-based inputs. 
The increasing popularity of deploying pretrained DNNs to reduce costs of re/training large models makes defense mechanisms that aim to detect `suspicious' input samples preferable. 

% Moreover, with the increasing deployment of pretrained DNN models 
% Considering the popularity of pretrained models to reduce costs of re/training large models, defences that aim to detect `suspicious' input samples will be more practical. 
%avoid training cost of large models, (ii) the high accuracy of these models on benign inputs, and (iii) the computational cost of re/training large models, the defences tend to detect suspicious inputs are more interesting in practice. 

%The performance of  existing defenses for detecting Trojan inputs  on other domain than images is less understood or computationally ineffective.
% \st{The performance of such defenses on other inputs is less understood.}

In this paper, we propose \emph{MDTD}, a Multi-Domain Trojan Detector for DNNs, which detects inputs containing a Trojan trigger at testing time. 
MDTD does not require knowledge of trigger-embedding strategy of the attacker and can be applied to a pretrained DNN model with image, audio, or graph-based inputs. 
%computationally inexpensive, does not require knowledge of attacker strategy, 
%, graph, and text-based inputs. 
%MDTD is the first known Trojan detection mechanism for graph-based inputs. 
MDTD leverages an insight that input samples containing a Trojan trigger are located relatively farther away from a decision boundary than clean samples. 
MDTD estimates the distance to a decision boundary using adversarial learning methods and
% and then computes the smallest magnitude of noise required for the model to misclassify a sample. 
uses this distance to infer whether a test-time input sample is Trojaned or not. 

%
%MDTD learns a threshold for the distance to the decision boundary, and uses this threshold to flag a sample as possibly Trojaned. 
We evaluate MDTD against state-of-the-art Trojan detection methods across \emph{five} widely used image-based datasets- CIFAR100, CIFAR10, GTSRB, SVHN, and Flowers102, \emph{four} graph-based datasets- AIDS, WinMal, Toxicant, and COLLAB, and the SpeechCommand audio dataset. Our results show that MDTD effectively identifies samples that contain different types of Trojan triggers. 
We further evaluate MDTD against 
adaptive attacks where an adversary trains a robust DNN to increase (decrease) distance of benign (Trojan) inputs from a decision boundary. 
%an adaptive attack where an adversary trains a robust DNN model to increase the distance of benign inputs from a decision boundary. 
Although such training by the adversary reduces the detection rate of MDTD, %for larger magnitudes of adversarial noise perturbations, 
this is accomplished at the expense of reducing classification accuracy 
or adversary success rate, 
%to below $50\%$, 
thus rendering the resulting model unfit for use. 
%Such an attack significantly reduces the detection rate of MDTD, but this comes at the cost of classification accuracy on benign inputs, which may cause a user to discard the model.
%This attack can significantly mitigate the MDTD detection rate at the cost of losing the model's performance on benign inputs which causes users do not have any intention to use/deploy these models.   

%, \emph{four} graph-based and \emph{two} text-based datasets. 
%We also evaluate the performance of \emph{MDTD} on \emph{four} graph-based. 
% Our results show that MDTD effectively identifies samples that contain different types of Trojan triggers. We also show that an adversary who trains robust DNN models using a combination of clean and Trojaned samples does not cause significant deterioration in MDTD performance without also significantly reducing classification accuracy of the DNN model. %and attack success rate. 
\end{abstract}

\begin{CCSXML}
<ccs2012>
<concept>
<concept_id>10002978</concept_id>
<concept_desc>Security and privacy</concept_desc>
<concept_significance>500</concept_significance>
</concept>
</ccs2012>
\end{CCSXML}

\ccsdesc[500]{Security and privacy}

\maketitle

%\input{abstract}

% \begin{abstract}
%     this is the abstract. check one two
% \end{abstract}

\keywords{
MDTD, Trojan detection, backdoor attack
}

% Section I
\section{Introduction}

Advances in cost-effective storage and computing have resulted in the widespread use of deep neural networks (DNNs) to solve complex tasks. Examples of such tasks include image classification~\cite{yu2018artificial}, text generation~\cite{zhu2018texygen}, and safety-critical applications such as 
autonomous driving~\cite{grigorescu2020survey}. 
% \st{The success of DNN models in reasoning about unseen inputs relies on the models being trained on large and diverse sets of data. }
However, training such large models requires significant computational resources, which may not be available to most DNN users. 
%and most of DNN users do not own powerfull computational resources.
Two approaches that have been proposed to overcome the challenge of significant computational resources required to train such models are 
(i) using publicly shared pre-trained DNN models~\cite{kurita2020weight} and (ii) training large models on online machine learning platforms~\cite{AWS, BigML, Caffe}. However, when the end-user of a DNN is different from the entity that trained the model (e.g., in online ML platforms), 
%(e.g., in online ML platforms~\cite{AWS, BigML, Caffe}) 
it is possible for an adversary to launch adversarial attacks such as a backdoor attack~\cite{backddorsurvey} and share the defective model for use by the public. %a defective train a model 
%Increasing the size of the training dataset by `outsourcing', albeit from not fully trusted sources, have had limited success~\cite{badnet}. 
%To increase the size of the training dataset, methods to `outsource' the training set, albeit from not fully trusted sources, were shown to have limited success~\cite{badnet}. 
% \st{However, DNNs operate in adversarial environments, which make them susceptible to different types of attacks}~\cite{goodfellow2014FGS,icml2021,backddorsurvey}. 
% \st{Attacks on DNN models performing classification tasks can be carried out at test-time or during training. }

\begin{figure}
    \centering
    \begin{tabular}{c c c c c c}
         \includegraphics[scale=1.1, trim={0.2cm 0cm 0.0cm 0}]{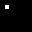} &
         \includegraphics[scale=1.1, trim={0.2cm 0cm 0.0cm 0}]{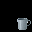}&
         \includegraphics[scale=1.1, trim={0.2cm 0cm 0.0cm 0}]{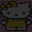}& 
         \includegraphics[scale=1.1, trim={0.2cm 0cm 0.0cm 0}]{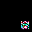}&
         \includegraphics[scale=1.1, trim={0.2cm 0cm 0.0cm 0}]{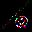}&
         \includegraphics[scale=1.1, trim={0.2cm 0cm 0.0cm 0}]{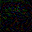}\\

         \includegraphics[scale=1.1, trim={0.2cm 0cm 0.0cm 0}]{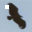} &
         \includegraphics[scale=1.1, trim={0.2cm 0cm 0.0cm 0}]{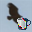}&
         \includegraphics[scale=1.1, trim={0.2cm 0cm 0.0cm 0}]{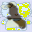}& 
         \includegraphics[scale=1.1, trim={0.2cm 0cm 0.0cm 0}]{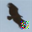}&
         \includegraphics[scale=1.1, trim={0.2cm 0cm 0.0cm 0}]{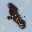}&
         \includegraphics[scale=1.1, trim={0.2cm 0cm 0.0cm 0}]{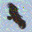}\\

        {\scriptsize BadNets} & {\scriptsize Nature} & {\scriptsize Blend }&  {\scriptsize SQ } & {\scriptsize WM} &{ \scriptsize L2 inv}  \\
          
    \end{tabular}
    \caption{(Top Row) Different types of Trojan triggers that we examine for %image-based 
    inputs from the CIFAR10 dataset that consists of $10$ classes. (Bottom Row) Image of a bird (\emph{Class 2}) embedded with the trigger. After %As a result of retraining, 
    the DNN identifies the bird embedded with a trigger as a frog (\emph{Class 6}).
    %The top row shows the different types of Trojan triggers that we examine for image-based inputs from the CIFAR10 dataset that consists of $10$ classes. The bottom row shows the image of a bird (\emph{Class 2}) embedded with the trigger. As a result of retraining, the DNN identifies the bird embedded with a trigger as a frog (\emph{Class 6}).
    } 
    %Different Trojan triggers and Trojan samples. First row shows the trigger and the second row shows the image of bird (class 2) perturbed by trigger which will be classified to class frog (class 6). }
    \label{fig:TrojanSamples}
\end{figure}
% \st{An adversarial learning attack}~\cite{szegedy2013intriguing, carlini2017adversarial} \st{is a test-time attack where an adversary generates input samples at test-time that are perceptually similar to the actual input. 
% These perceptually similar samples were misclassified by well-trained DNNs}~\cite{szegedy2013intriguing, carlini2017adversarial}. 

%In this paper, we focus on backdoor attacks~\cite{backddorsurvey}, in which 
An adversary carrying out a backdoor attack inserts a predefined perturbation called a \emph{trigger} into a small set of input samples~\cite{backddorsurvey}. 
The DNN models can then be trained so that the presence of the trigger in an input will result in an output label that is different from the correct output label (an existing class or a completely new class not known to the user~\cite{backddorsurvey}). Adversarial training also ensures that output labels corresponding to clean inputs (inputs without the trigger) remain unaffected. 
A DNN model that misclassifies inputs that contain a trigger is termed \emph{Trojaned}. 
Backdoor attacks are highly effective when only classification output labels of models are available to the users. 
In \cite{badnet}, DNN models used in autonomous driving applications were shown to incorrectly identify a STOP sign with a small sticker on it (the trigger) as a `speed limit' sign. 
%
%For example, DNN models used for traffic sign detection~\cite{badnet} have been shown to produce an incorrect output label when the traffic sign has a trigger- for e.g., a STOP sign with a small sticker on it is identified (incorrectly) as a `speed limit' sign. 
Backdoor attacks with different types of triggers embedded in inputs from the CIFAR10 dataset are illustrated in Fig.~\ref{fig:TrojanSamples}.

% \st{Developing defenses against adversarial attacks on DNNs have been examined from multiple perspectives. 
% Defenses against test-time attacks were proposed in}~\cite{buckman2018thermometer, dhillon2018stochastic, xie2018mitigating, yan2018deep}, \st{which use gradient-based methods}~\cite{papernot2018towards} \st{to enable detection of adversarial examples. }
Defenses against backdoor attacks involve (i) pruning or retraining the DNN (e.g., \emph{Fine-Pruning}~\cite{liu2018fine}), (ii) developing Trojan model detectors (e.g., \emph{Neural Cleanse}~\cite{wang2019neuralcleans}) to detect an embedded backdoor, or (iii) detecting Trojan samples at inference~\cite{gao2019strip} or training time~\cite{tran2018samplefilter1}.
Performance of these methods were examined %in \cite{backddorsurvey} 
and reported to be computationally expensive in \cite{backddorsurvey} (Sec. VIII). These methods also assume that the user of the model has adequate resources to (re)train the DNN model~\cite{liu2018fine, tran2018samplefilter1} or identify and reconstruct an embedded trigger~\cite{wang2019neuralcleans}. 
These challenges underpin a need to develop %computationally inexpensive 
mechanisms to effectively detect input samples that have been embedded with a Trojan trigger and discard such inputs before they can be provided to a DNN model. 

In this paper, we propose \emph{MDTD}, a mechanism to detect Trojan samples that uses a distance-based reasoning to overcome the challenges listed above.
MDTD is a computationally inexpensive inference-time detection mechanism that can be applied to pre-trained DNN models. MDTD does not aim to inspect and remove an embedded backdoor from the given DNN model; rather, it determines if an input sample to the given DNN contains a trigger with high probability, and discards such inputs. 
%it discards inputs that might contain a Trojan trigger. 
The effectiveness of MDTD is underscored by the fact that it is agnostic to the specific trigger-embedding strategy employed by the adversary.

MDTD uses the insight that samples containing a Trojan trigger will typically be located %relatively 
farther away from the decision boundary compared to a clean sample. Consequently, a larger magnitude of noise will need to be added to a Trojaned sample to move it across the decision boundary so that it will be misclassified by the DNN. 
Since it makes use of the distance metric from a decision boundary, MDTD is applicable to different input data modalities.
%Therefore, to quantifiably distinguish between clean and Trojaned input samples, MDTD makes use of the distances of given samples to a decision boundary.

%We provide an illustration of the MDTD insight 
We illustrate the insight and motivation behind %developing 
MDTD 
using t-SNE visualization techniques~\cite{van2008visualizing} to demonstrate that embedding of a Trojan trigger can be qualitatively examined through the lens of feature values at intermediate DNN layers. To quantitatively verify this insight, we compare distances of clean and Trojan samples to a decision boundary using the notion of a 
certified radius~\cite{cohen2019certified}. %of Trojan and clean samples.
However, computing the certified radius %is known to 
can incur large costs~\cite{li2020CertifiedComputationalCost}. 
To overcome this challenge, 
%To overcome large costs associated with computing the certified radius~\cite{li2020CertifiedComputationalCost}, 
MDTD 
%Then we describe how MDTD 
estimates the distance of Trojan and clean samples from a decision boundary using adversarial learning \cite{goodfellow2014FGS} in a computationally efficient manner. MDTD then determines a threshold on the computed distance using only a small number of clean samples without making any assumption about the adversary's trigger-embedding strategy.
Our contributions are:

\begin{itemize}

%    \item We make a case for the design of MDTD using t-SNE visualizations~\cite{van2008visualizing} to show that intermediate layers of DNNs generate different feature values for clean and Trojan samples. 
    %, we use t-SNE visualizations~\cite{van2008visualizing} to show that intermediate layers of DNNs generate different feature values for clean and Trojan samples. 
    %We use t-SNE visualizations~\cite{van2008visualizing} to show that intermediate layers of DNNs generate different feature values for clean and Trojan samples. 
 %   We show that Trojan samples exhibit greater robustness to noise than clean samples, which enables distinguishing between them. 
    %To motivate our insight, we first indicate intermediate layers of DNNs generate different feature values for Trojan samples and these samples have more robustness to noise compared to clean samples. Therefore, they are distinguishable from clean samples.
    
    \item We propose \emph{MDTD}, a practical %unified 
    framework to detect Trojaned inputs in image, graph, and audio-based input domains in a computationally inexpensive manner. 
    %multiple input domains such as image, graph, and text. 
%    \emph{MDTD} makes no assumptions on attacker strategy, and can effectively distinguish between clean and Trojaned inputs using only a small number of clean samples.
    %a computationally inexpensive and novel input Trojan detection defense called \emph{MDTD}, which is applicable for different input domains e.g., image, graph and text. MDTD does not have any assumption for attack strategy and can effectively distinguish clean samples from Trojans using a set of  few clean samples.
    
    \item We demonstrate the effectiveness of \emph{MDTD} through comprehensive evaluations and comparisons with state-of-the-art (SOTA) Trojan input detection methods for different types of Trojan triggers across \emph{five} image-based input datasets: \textbf{CIFAR100, CIFAR10, GTSRB, SVHN, and Flowers102}. 

   % \emph{five} image-based datasets- CIFAR100, CIFAR10, GTSRB, SVHN, and Flowers102- and \emph{four} graph-based datasets- AIDS, WinMal, Toxicant, and COLLAB. 
%    and \emph{two} text-based input datasets. 
    
    \item We examine the performance of \emph{MDTD} on \emph{four} graph-based input domains: \textbf{AIDS, WinMal, Toxicant, and COLLAB} and \emph{one} audio dataset \textbf{SpeechCommand}. %\emph{MDTD} is the first known Trojan detection mechanism for graph and audio-based datasets. 

    \item We evaluate MDTD against adaptive Trojan trigger-embedding attacks where the adversary has complete knowledge about the details of MDTD-based detection and aims to construct new Trojan trigger embeddings to bypass detection. We empirically show that although such retraining of the DNN reduces the detection rate of MDTD, it simultaneously lowers classification accuracy of the DNN on clean samples below $50\%$, thus making the DNN unfit for use.

\end{itemize}

Sec.~\ref{sec:preliminary} provides background on DNNs and backdoor attacks. Sec.~\ref{sec:threatmodel} describes our threat model, and Sec.~\ref{UserAbility} specifies assumptions on user capability. 
We motivate and describe the design of \emph{MDTD} in Sec.~\ref{sec:motivation}. %describe its working in Sec.~\ref{sec:ourapproach}. 
We evaluate \emph{MDTD} in Sec.~\ref{sec:evauations} and discuss MDTD in Sec.~\ref{sec:discussion}. 
Sec.~\ref{sec:RelatedWork} presents related work and Sec. \ref{sec:Conclusion} concludes the paper.

% basic introduction
% \begin{figure}[ht]
%     \centering
%     \begin{tabular}{c c c  }%c c c c c
%         \includegraphics{images/samples/Trojan_sample_source.png}&
%           \includegraphics{images/samples/Trojan_sample_EuroSAT_3_53.287506103515625.png}&
%           \includegraphics{images/samples/Trojan_sample.png}\\ 
%           \shortstack{(a) Clean\\Sample} &\shortstack{ (b) Natural\\ Trigger}&\shortstack{(c) Embedded \\ Trigger}
%     \end{tabular}
%     \caption{Clean sample and Trojan samples with natural and embedded triggers. Natural trigger is trained on a clean model, while embedded trigger (red square on top left corner of the image) is a predefined trigger by attacker in Trojan model.}
%     \label{fig:TrojanSamples}
% \end{figure}
\section{Preliminaries} \label{sec:preliminary}

% The vulnerability of deep neural networks (DNNs) to backdoor attacks is well investigated in the literature~\cite{backddorsurvey}. 

This section gives an overview of deep neural networks (DNNs),  
graph neural networks (GNNs), and %, and text models, 
long-short-term-memory (LSTM) %audio classification 
models.  
We describe how an adversary can carry out a backdoor attack on these models, and specify metrics to evaluate effectiveness of the attack and our proposed defense \emph{MDTD}.

\subsection{DNNs and Backdoor Attacks}
Deep neural networks (DNNs) are complex machine learning models (ML) developed for tasks with high-dimensional input spaces (e.g., image classification, text generation) \cite{goodfellow2016deep}. 
These models take an input $x$, compose %the input 
$x$ through %several 
layers ($f(x):=l_1\circ l_2\circ\cdots \circ l_k(x)$), and return output $y$. 
For e.g., in image classification, $x\in [0,1]^{(W\times H)}$ is an image, and the DNN returns an output $y\in \{1,\cdots, C\}$ where $W \times H$ is the resolution of $x$ and $C$ is the number of classes. 

DNN models for classification tasks are known to be vulnerable to backdoor attacks~\cite{backddorsurvey}. %which we describe below. 
%\subsection{Backdoor Attack}
An adversary carrying out a backdoor attack can retrain a DNN to return a different output label when a trigger is embedded into the input sample, %by an adversary, 
 while recognizing clean samples %(without the trigger) 
 correctly. 
% recognizing normal (unperturbed and clean) inputs correctly, while returning a different output when a trigger is embedded into the input sample by an adversary. 
%We categorize the backdoor attack into two categories of (i) Adversarial embedded backdoor, and (ii) Natural backdoor attacks~\cite{tao2022modelorthogonalization}. 
An adversary can carry out a backdoor attack by embedding triggers into a subset of samples in the training data~\cite{badnet,chen2017targetedbackdoor} or manipulating weights in layers of the DNN~\cite{li2019invisible} to induce erroneous behavior at test time for inputs that contain the trigger. 
%An adversary carrying out a backdoor attack manipulates weights (also c in one or more layers of the DNN hyperparameters of the DNN model (e.g., by corrupting training data~\cite{badnet,chen2017targetedbackdoor}) to induce erroneous behavior at test time for inputs that contain the trigger. 
For e.g., in image classification using the CIFAR10 dataset (Fig.~\ref{fig:TrojanSamples}), the adversary manipulates the model $f$ to return a desired label corresponding to \emph{frog} (\emph{Class 6}) that is different from the true label of \emph{bird} (\emph{Class 2}) for inputs that contain a predefined trigger.

\subsection{GNNs and Backdoor Attacks}

Graph neural networks (GNNs) are a class of deep learning models designed to make inferences on graph-based data~\cite{hamilton2020}. 
The input to a GNN is a graph $\mathcal{G} = (V,E)$ where $V$ is the set of individual nodes, and $E$ is the set of edges between pairs of nodes. 
Each node $v \in V$ has an associated set of $d$ features, denoted $x_v \in \mathbb{R}^d$. 
We let $X \in \mathbb{R}^{|V| \times d}$ be the feature representation matrix associated with graph $\mathcal{G}$. 
In this paper, we focus on a recently proposed backdoor attack on GNNs that use a message passing paradigm~\cite{hamilton2017}, and graph classification tasks where the goal is to predict the class that an input graph belongs to. 
Under the message passing paradigm, at each iteration $t$, $x_v$ is updated as follows: $x^{(t)}_v = \mathcal{U}(x^{(t-1)}_v, \mathcal{A}({x^{(t-1)}_u, \forall u \in \mathcal{N}(v)}))$, where $\mathcal{N}(v)$ is the set of neighbours of $v$, $\mathcal{A}$ is an aggregate function that takes the feature representations of each node from $v$'s neighbors as input, and \ $\mathcal{U}$ is an update function that takes $x_v$ at iteration $(t-1)$ and the output of the aggregate function $\mathcal{A}$ as inputs. 
After $T$ iterations, individual node representations are pooled to generate a graph representation $x_\mathcal{G} = f(\mathcal{G},X)$. 
The graph classification task can then be expressed as $h: f(\cdot,\cdot) \rightarrow \{1,2,\dots,C\}$. 

An attacker carrying out a backdoor attack on GNNs uses a subgraph (a subset of vertices and edges associated to vertices) of $\mathcal{G}$ as the trigger. 
We adopt the Graph Trojan Attack (GTA) proposed in~\cite{xi2021} to generate Trojaned GNN models. 
GTA uses trigger-embedded graphs to update GNN parameters. 
The updated GNN model is passed to a trigger generation network which generates the trigger for the next iteration of the message passing procedure. 
The trigger generation network consists of a topology generator that updates the subgraph, %nodes and edges, 
and a feature generator that updates features associated to nodes in the Trojaned subgraph. 
The goal of the adversary is to ensure that the Trojaned GNN model returns a desired label $y^d$ that is different from the true label for graph inputs that are embedded with the predefined `triggered' subgraph.

\subsection{Audio Models and Backdoor Attacks}
Long-Short-Term-Memory (LSTM) models %are a widely used class of ML models that 
enable processing and reasoning about sequences of information, e.g., in audio and speech processing~\cite{speechDNN}. 
We use an LSTM combined with a DNN model to train an audio classifier. 
The model takes an audio input $x^{1\times H}$, and returns an output $y\in \{1,\cdots, C\}$, where $1 \times H$ is the resolution of $x$ and $C$ is the number of classes. 
An audio input is comprised of a set of frequencies, which could be different for each input. 

For a sample audio $x$, its Trojaned version $x_T$ can be generated using two different backdoor attacks: (i) \underline{Modified}, where a small part of the audio was replaced by an arbitrary (but fixed) audio noise pattern, and (ii) \underline{Blend}, where a randomly generated (but fixed) audio noise trigger was mixed into a part of the audio. 
Specifically, at time-interval $[i,i+w]$, $x_T [i:i+w] = (1-\alpha) \times x[i:i+w]+\alpha\times Trigger$, where $\alpha = 1$ for Modified, and $\alpha \in [0.05,0.2]$ for Blend. 

\subsection{Metrics}\label{subsec:metrics}
We describe metrics used to evaluate %the effectiveness of 
backdoor attacks and Trojan sample detection methods. 
We assume that defense mechanisms return a \emph{positive} label if they identify a sample as Trojan. 
In the literature~\cite{james2013introduction}, such positive identification of the Trojan is called a true positive rate. SImilarly, when the defense mechanism incorrectly labels a clean sample as Trojan, it is called a false positive or false alarm. We now define suitable metrics below: 

\noindent{\it \underline{True Positive Rate (TPR)}} is the fraction of Trojan samples that received a positive label. 

\noindent{\it \underline{False Positive Rate (FPR)}} is the fraction of clean samples that incorrectly received a positive label (raising a \emph{false alarm}). 

\noindent{\it \underline{$F_1$-score}}: An effective Trojan detection method has a high detection accuracy (TPR) and low false alarm (FPR). Defining $TNR=1-FPR$, the $F_1$-score combines $TPR$ and $TNR$ as: 
\begin{equation}
    F_1= \frac{2*TPR* TNR}{TPR + TNR}. \label{eqn:F1score}
\end{equation}
%where $TNR=1-FPR$. 
The $F_1$-score is widely used to compare Trojan detection methods, and a higher $F_1$-score indicates a better detector~\cite{chen2018detecting}. 

\noindent{\it \underline{Attack Success Rate (ASR)}} is the fraction of Trojan samples that result in the DNN model returning the attacker's desired output~\cite{backddorsurvey}.

% \section{User and Threat Models}\label{sec:threatmodel}
% In this section, we introduce the threat model and describe our assumptions on the capabilities of the attacker and user (defender).

\section{Threat Model} \label{sec:threatmodel}
In this section, we introduce the threat model we consider, and describe our assumptions on capabilities of the attacker.% and specify metrics used to evaluate attack performance. 
%
%\noindent{\bf Threat Model:}
%\noindent{\bf Model:}
%\subsection{Adversary Model}
%Collecting large amounts of training data and obtaining 
% Since computational resources required to train large DNN models can be unavailable for a user with limited capabilities,
% deploying pre-trained DNN models and using cloud-based ML platforms and services are becoming increasingly popular. % to overcome these barriers. 
% However, the trustworthiness of publicly available models, and limited computational resources can be leveraged by an adversary to launch a backdoor attack. 

\noindent 
\emph{\textbf{Adversary Assumptions}}: 
We assume that the adversary has access to a pretrained model and adequate data is available such that a subset of these samples can be used by the adversary to embed a Trojan trigger into them. The adversary is also assumed to have sufficient computational resources required to (re)train the DNN using both, the Trojan trigger-embedded and clean samples.% as well as the clean samples. 

\noindent 
\emph{\textbf{Adversary Goals and Actions}}: The adversary's aim is to use Trojan-embedded as well as clean samples to train a Trojaned DNN model such that: (i) for Trojan trigger-embedded inputs, the output of the DNN is an adversary-desired target class, which may be different from the true class, and (ii) for clean inputs, the classification accuracy is as close to the accuracy of the un-Trojaned DNN.

\noindent
\emph{\textbf{Performance Metrics}}: 
Performance metrics defined in Sec. \ref{subsec:metrics}, including the true/ false positive rates, and the attack success rate (ASR) can be computed by the adversary after it trains the Trojaned DNN model. We use these metrics to characterize and empirically evaluate the effectiveness of an attack. 
The objective of the adversary is typically to ensure that the value of ASR is high, while the $F_1$-score in our context, which quantifies the defender's ability to detect a sample containing a Trojan trigger is as small as possible. \\

The adversary is assumed to perform trigger embedding in a stealthy manner such that its specific attack strategy and parameters, including the nature and location of the Trojan triggers embedded in the sample, as well as information about data samples that have been embedded with Trojan triggers is not revealed. 

\begin{figure*}
\centering
\begin{tabular}{c c c c c c c }
   
    \rotatebox{90}{$\:\:\:\:\:\:\:\:$CIFAR100}&
    \includegraphics[scale=0.11, trim={4.5cm 2.5cm 4.5cm 4.5cm}]{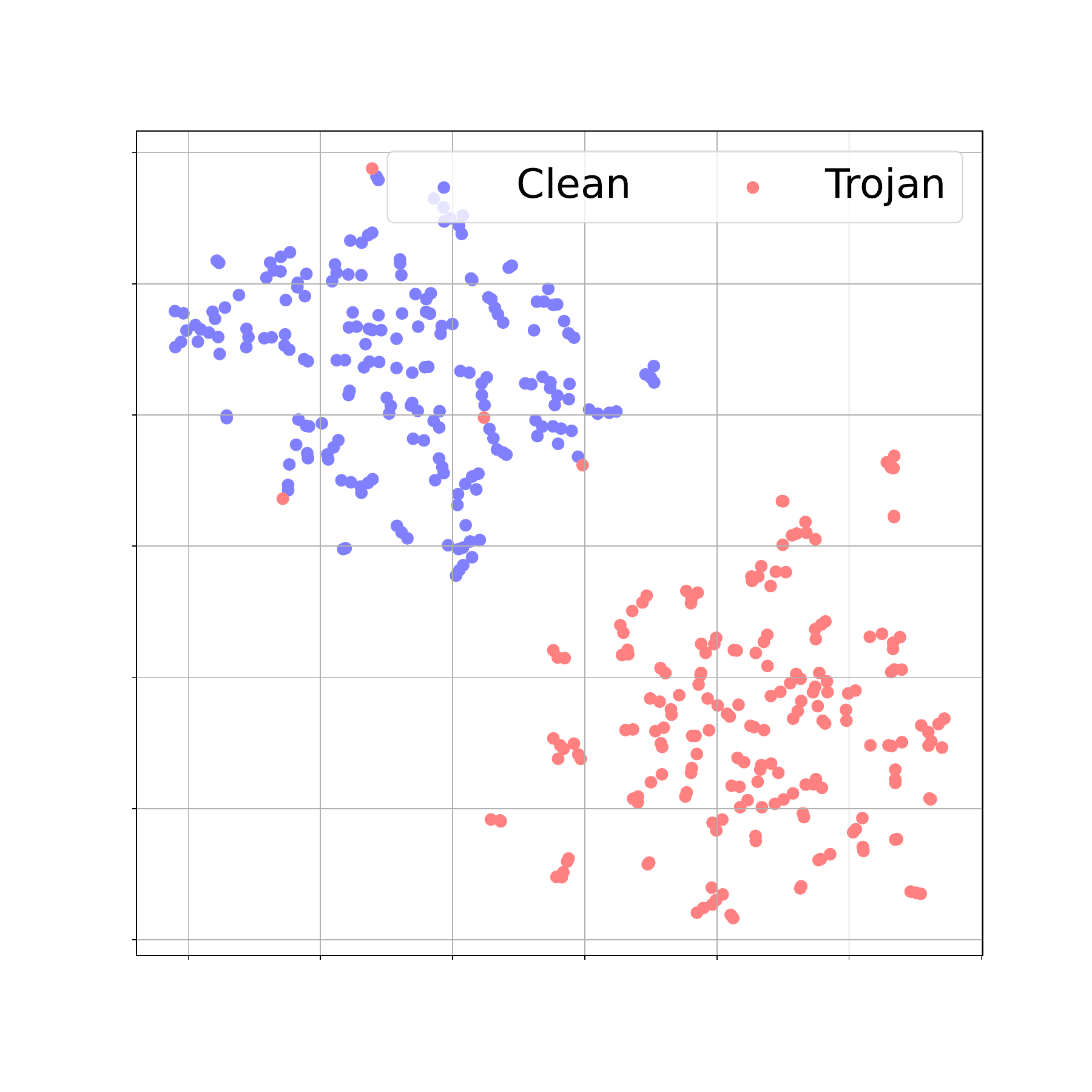}&
    \includegraphics[scale=0.11, trim={4.5cm 2.5cm 4.5cm 4.5cm}]{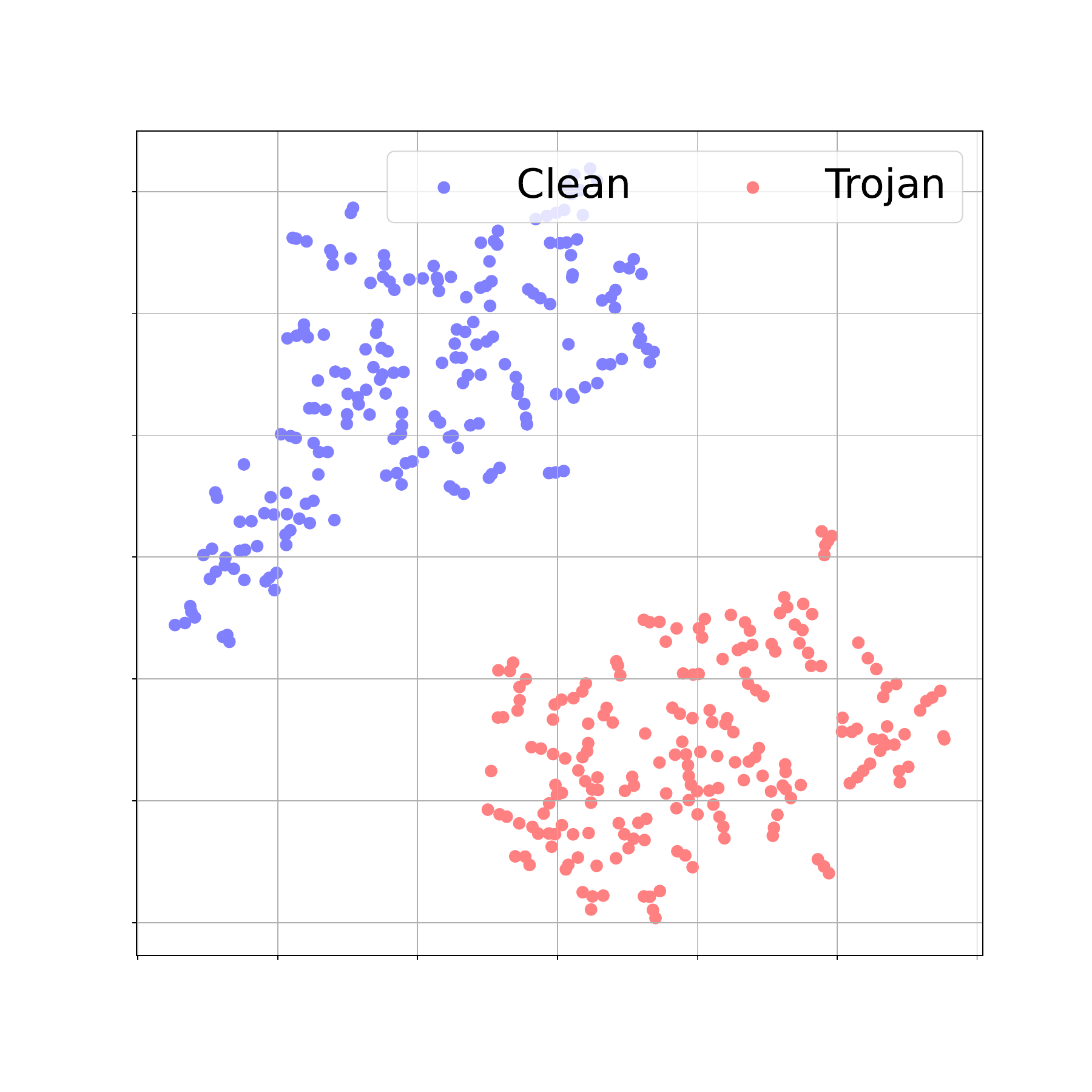}&
     \includegraphics[scale=0.11, trim={4.5cm 2.5cm 4.5cm 4.5cm}]{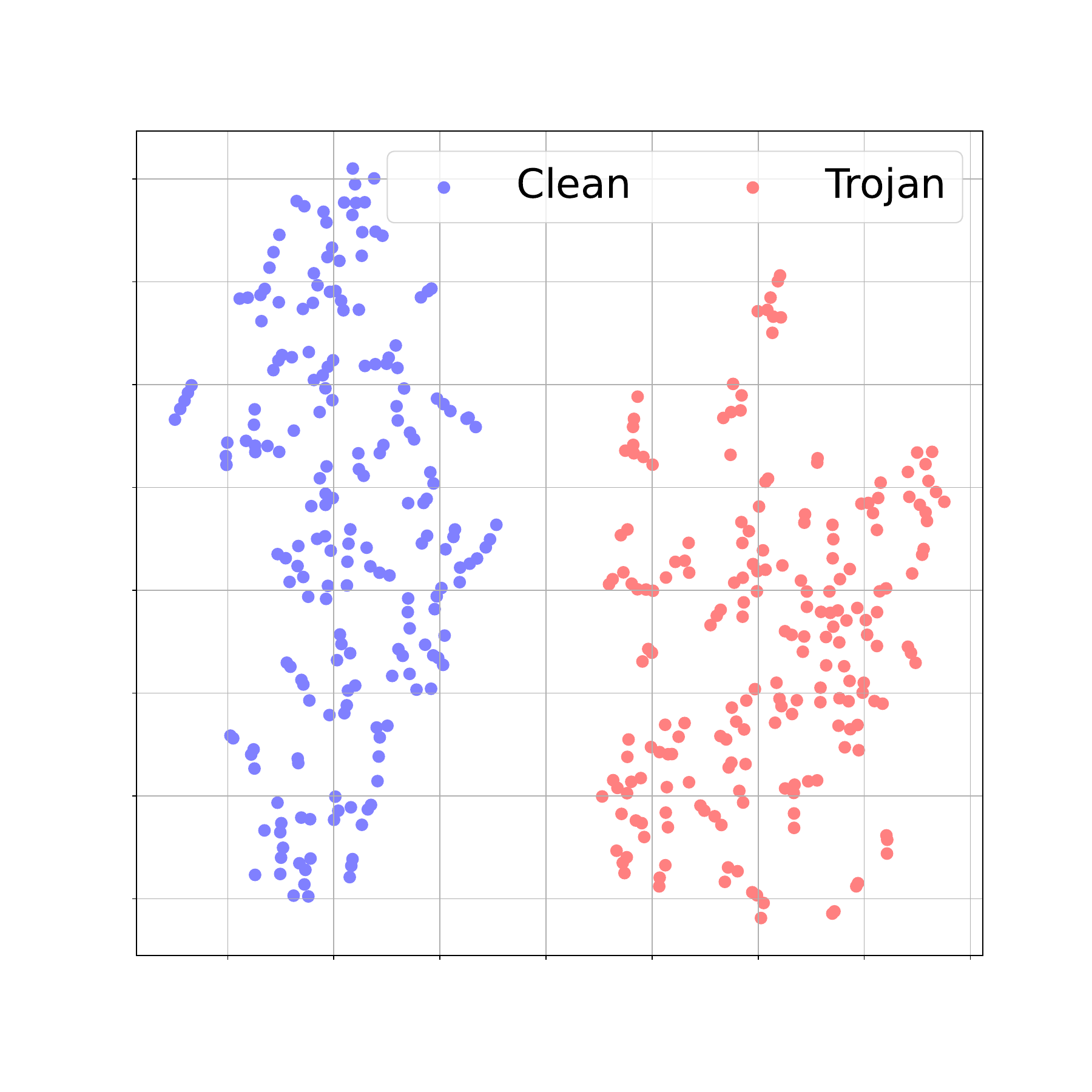}&
    \includegraphics[scale=0.11, trim={4.5cm 2.5cm 4.5cm 4.5cm}]{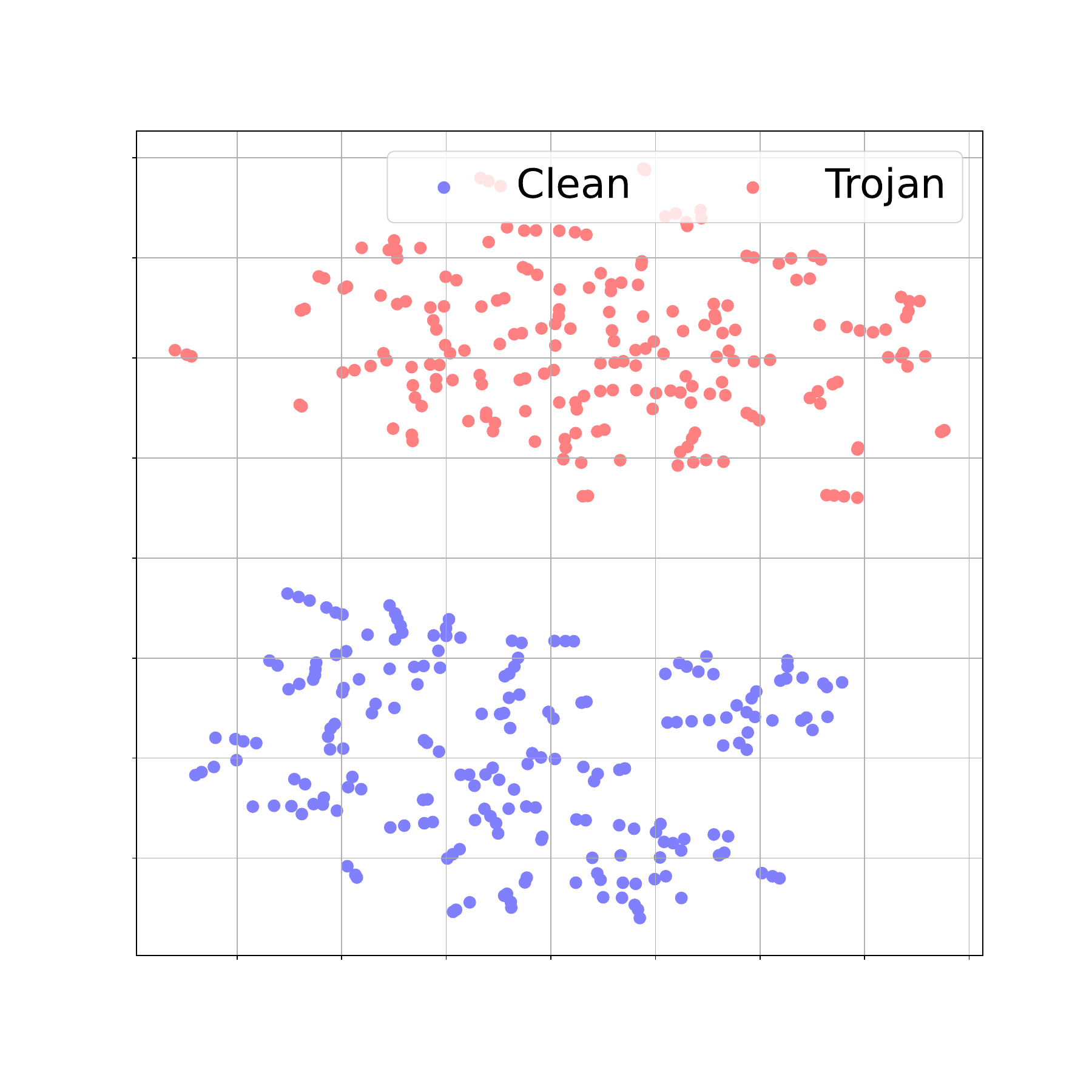}&
    \includegraphics[scale=0.11, trim={4.5cm 2.5cm 4.5cm 4.5cm}]{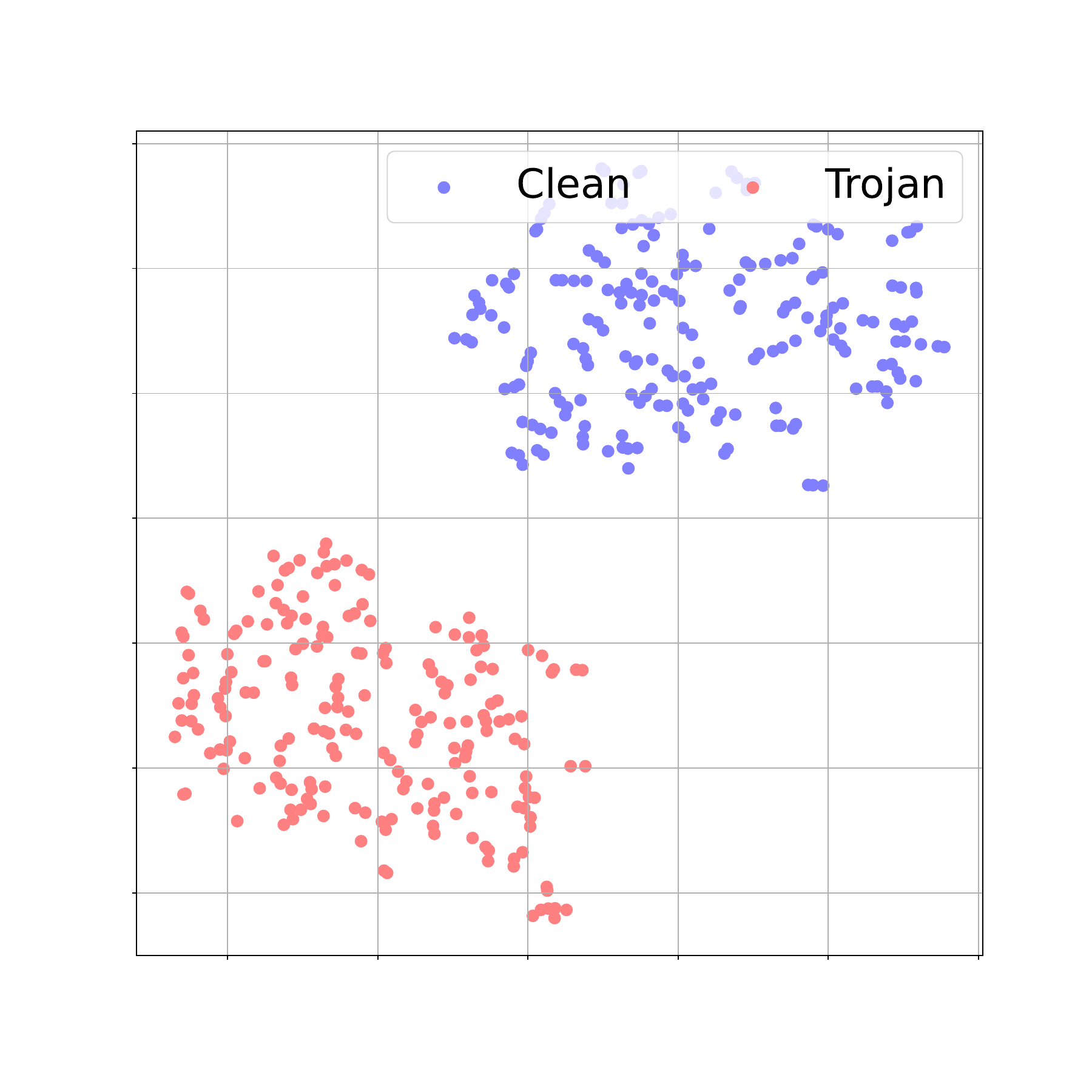}&
    \includegraphics[scale=0.11, trim={4.5cm 2.5cm 4.5cm 4.5cm}]{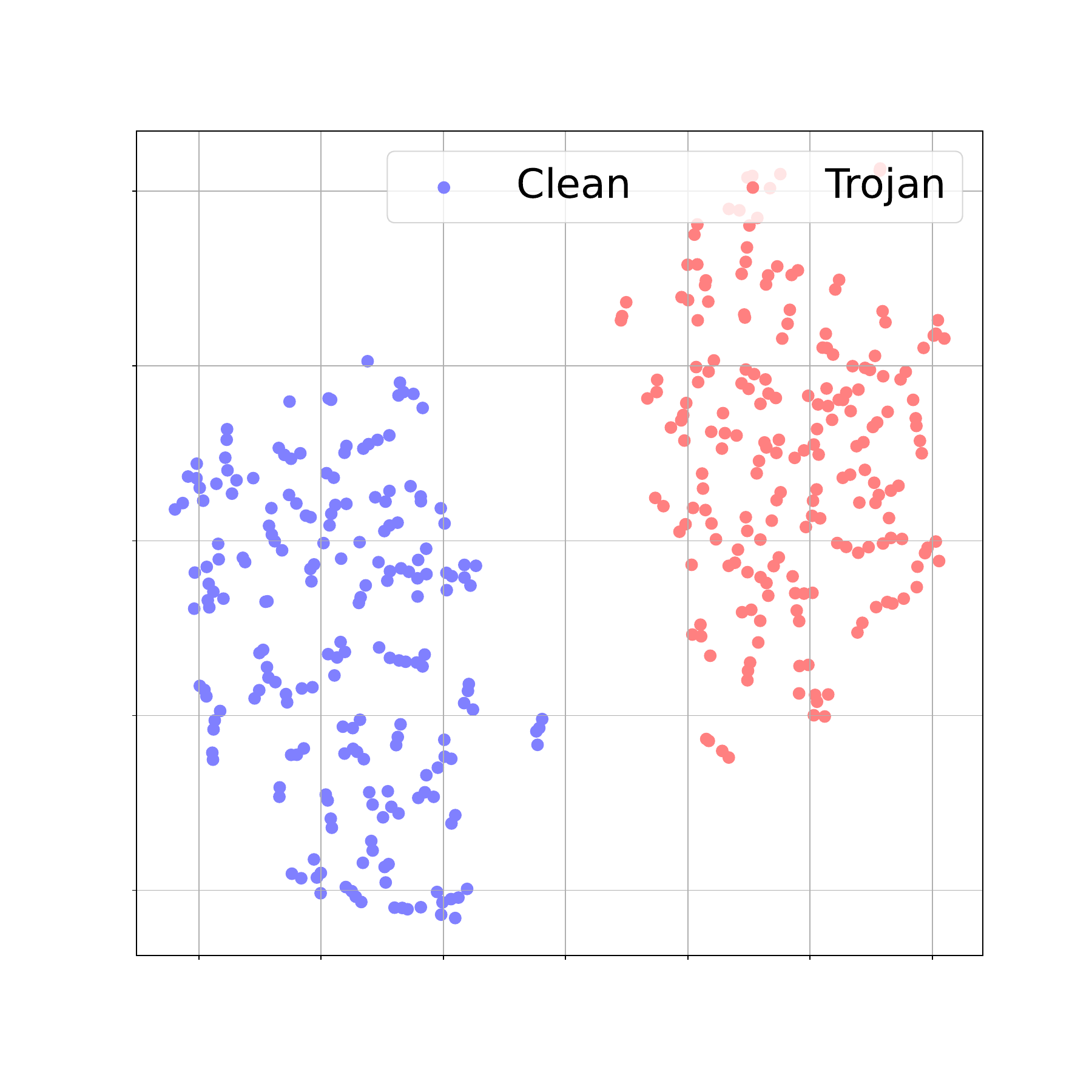}\\

    \rotatebox{90}{$\:\:\:\:\:\:\:\:\:$CIFAR10}&
     \includegraphics[scale=0.11, trim={4.5cm 2.5cm 4.5cm 4.5cm}]{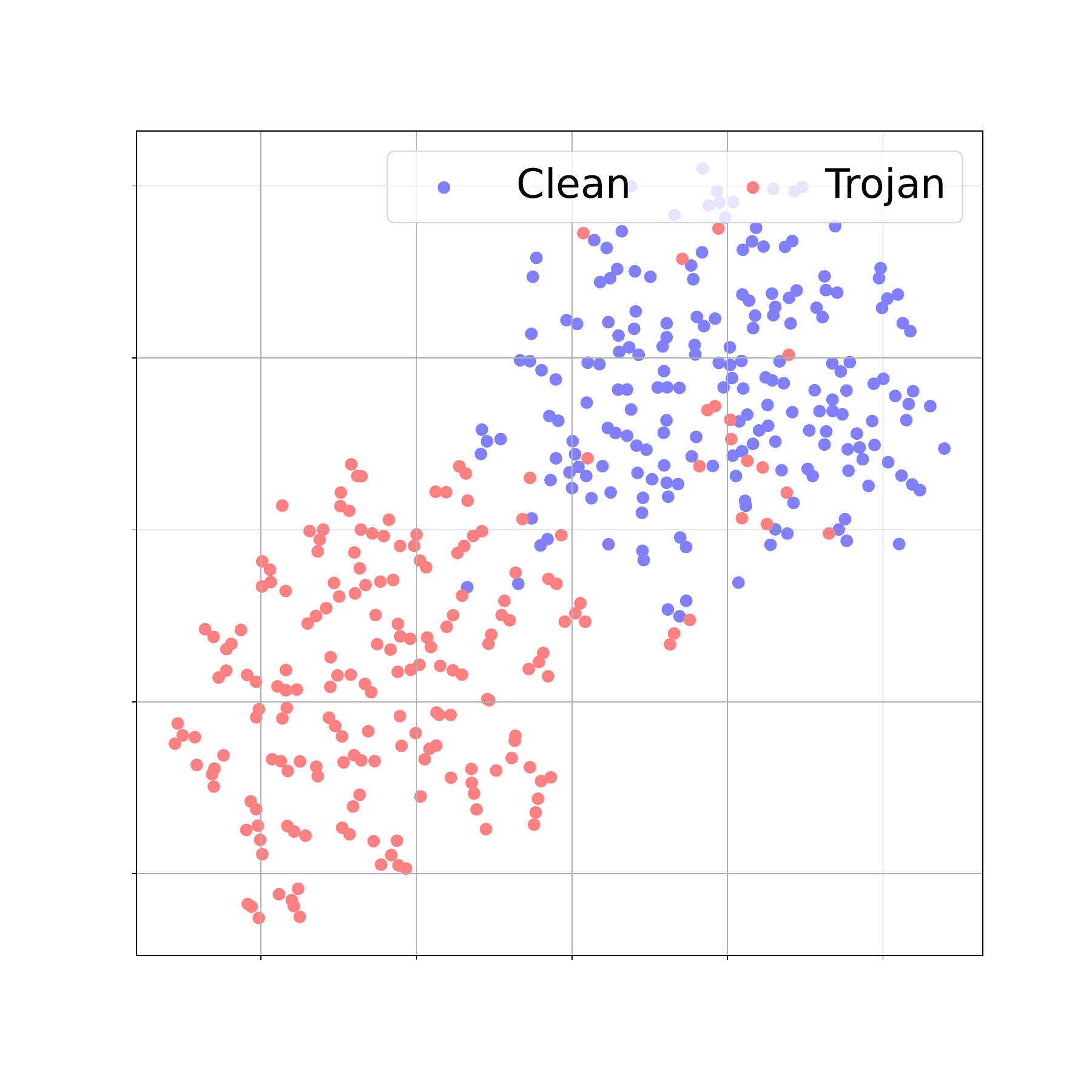}&
      \includegraphics[scale=0.11, trim={4.5cm 2.5cm 4.5cm 4.5cm}]{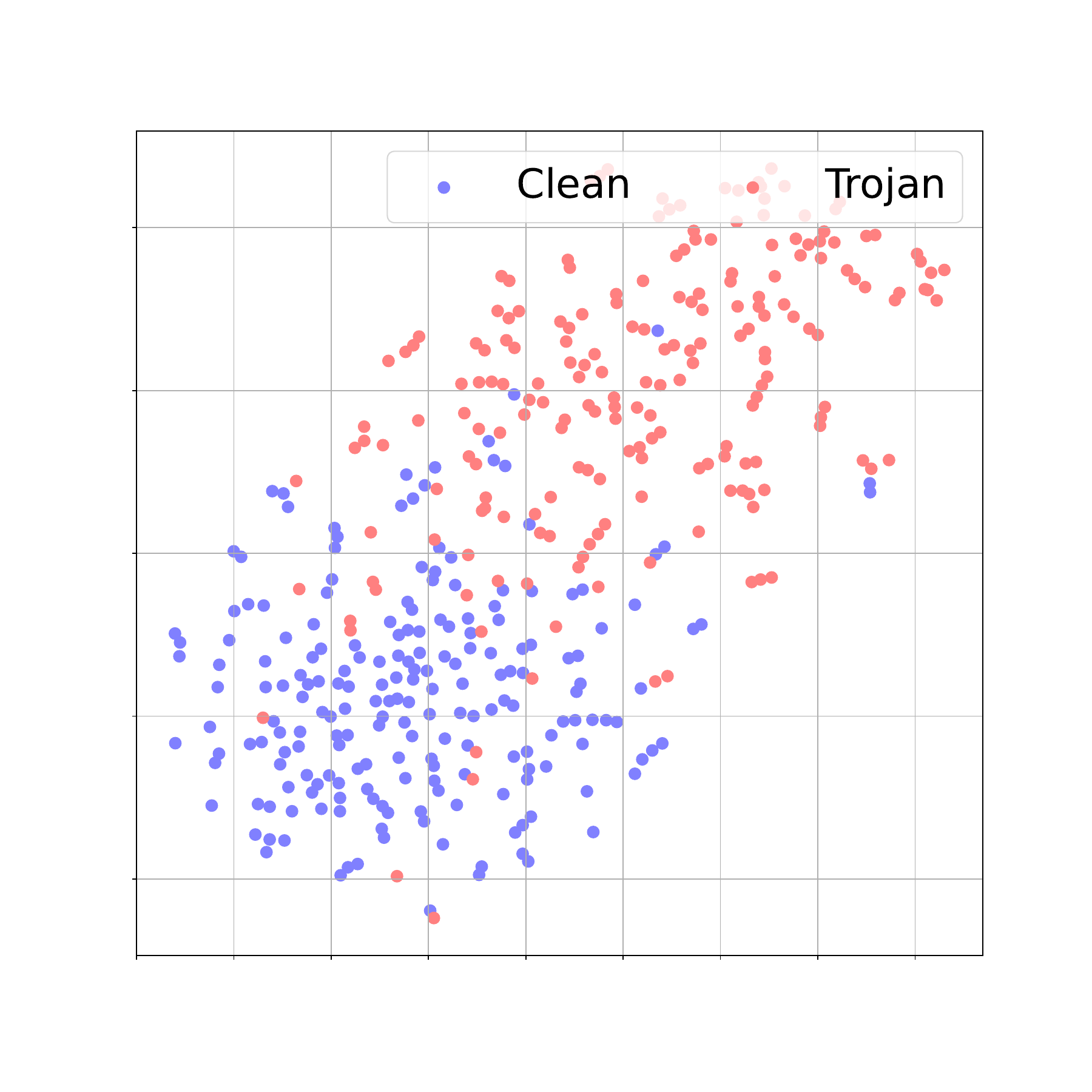}&
     \includegraphics[scale=0.11, trim={4.5cm 2.5cm 4.5cm 4.5cm}]{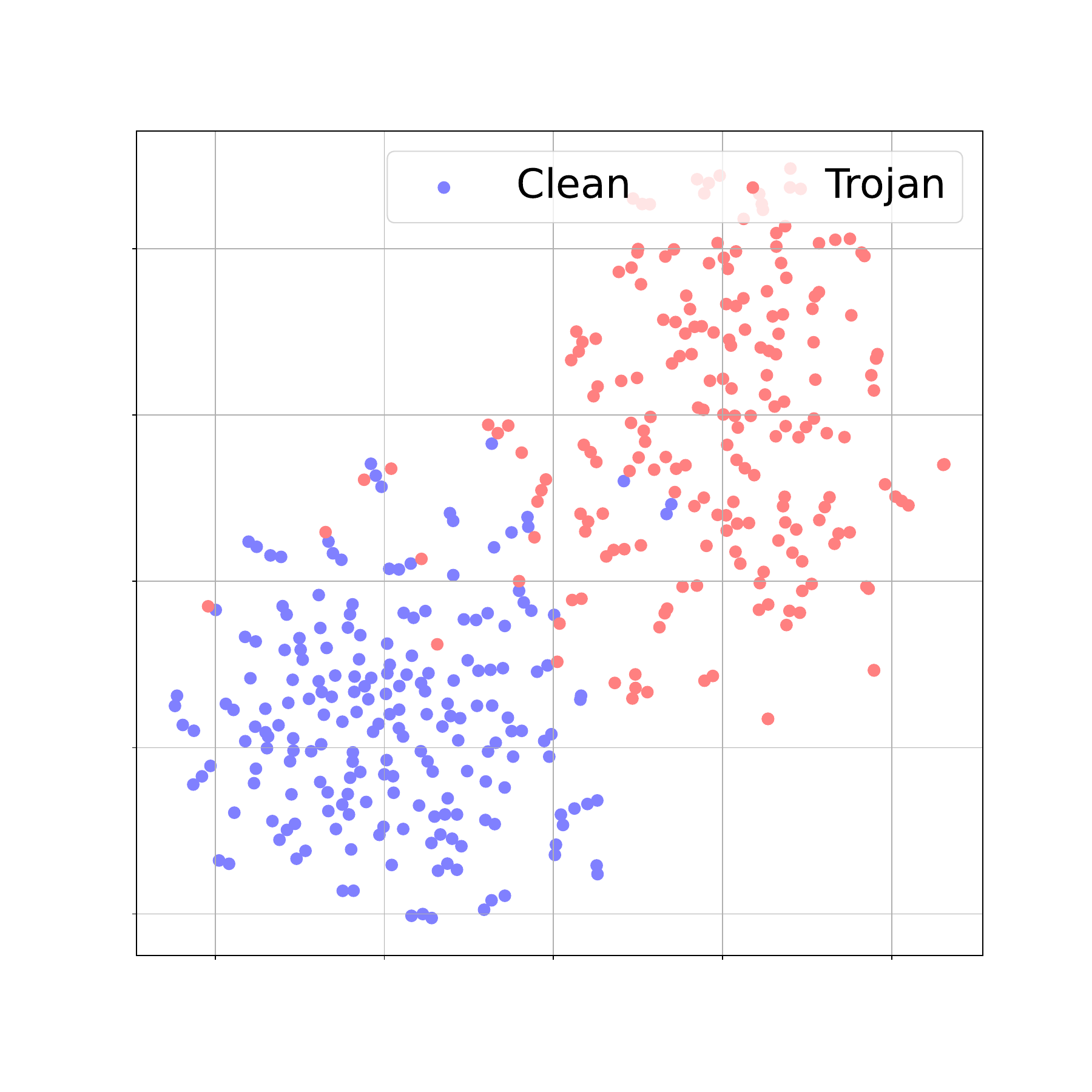}&
    \includegraphics[scale=0.11, trim={4.5cm 2.5cm 4.5cm 4.5cm}]{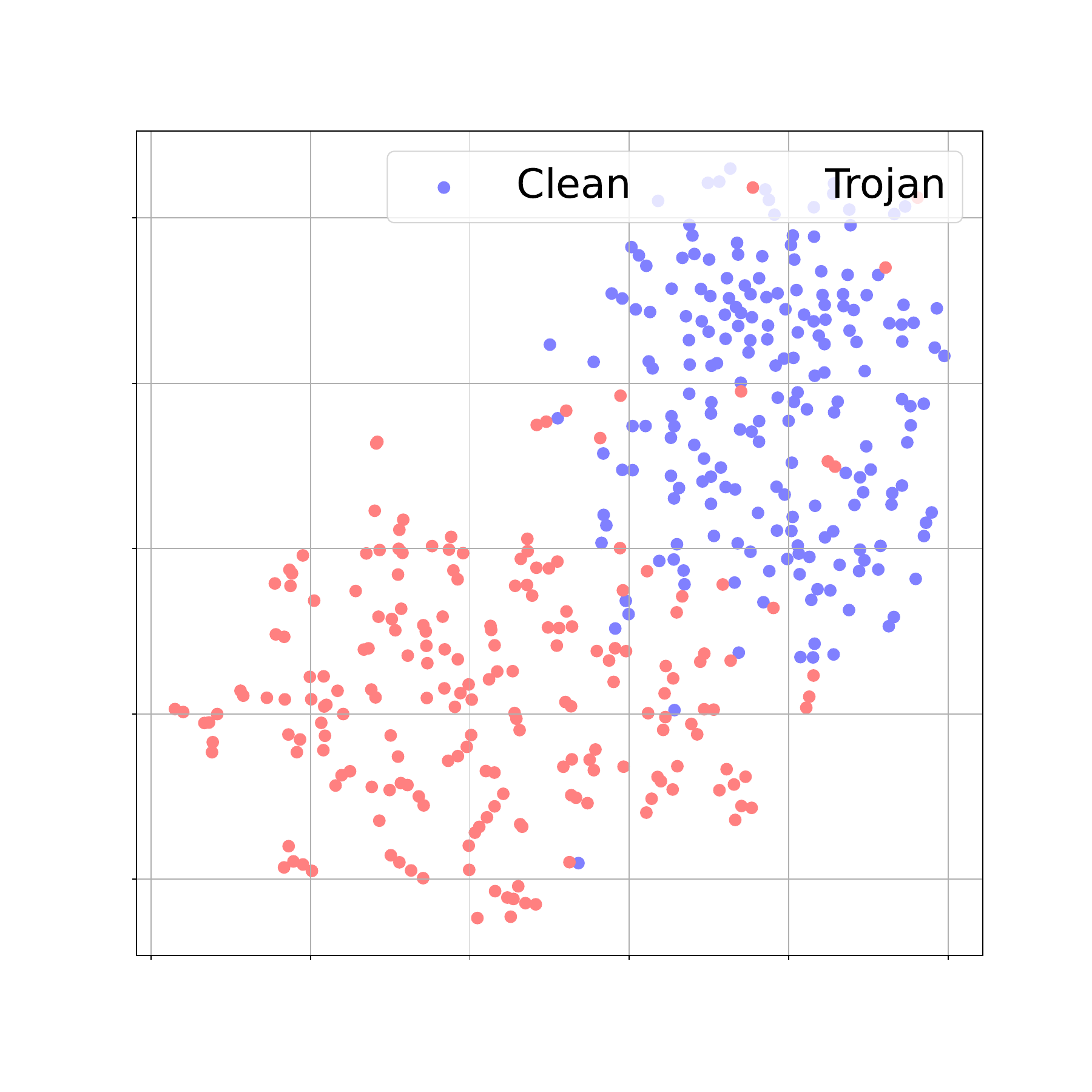}&
    \includegraphics[scale=0.11, trim={4.5cm 2.5cm 4.5cm 4.5cm}]{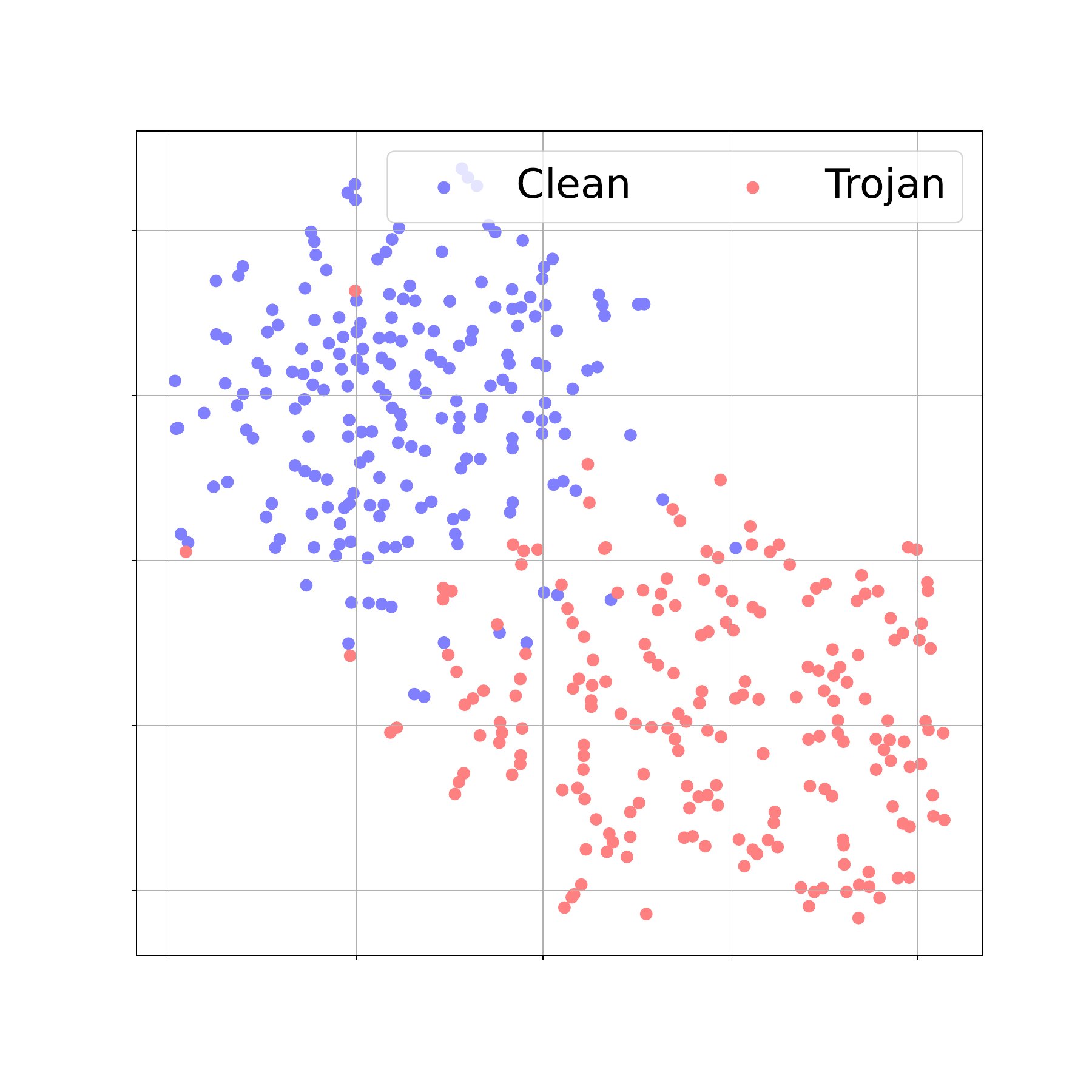}&
    \includegraphics[scale=0.11, trim={4.5cm 2.5cm 4.5cm 4.5cm}]{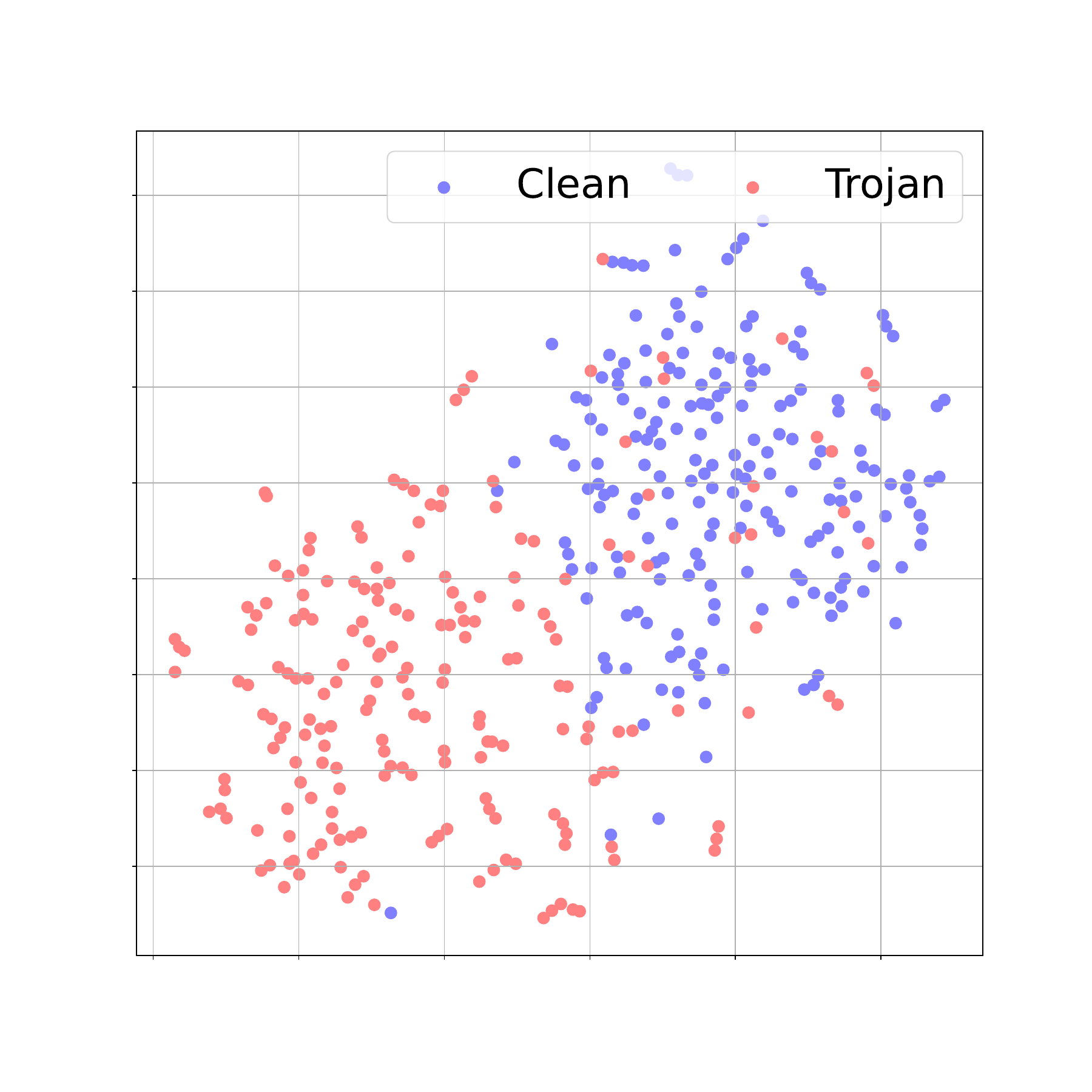}\\

    \rotatebox{90}{$\:\:\:\:\:\:\:\:\:\:\:$GTSRB}&
     \includegraphics[scale=0.11, trim={4.5cm 2.5cm 4.5cm 4.5cm}]{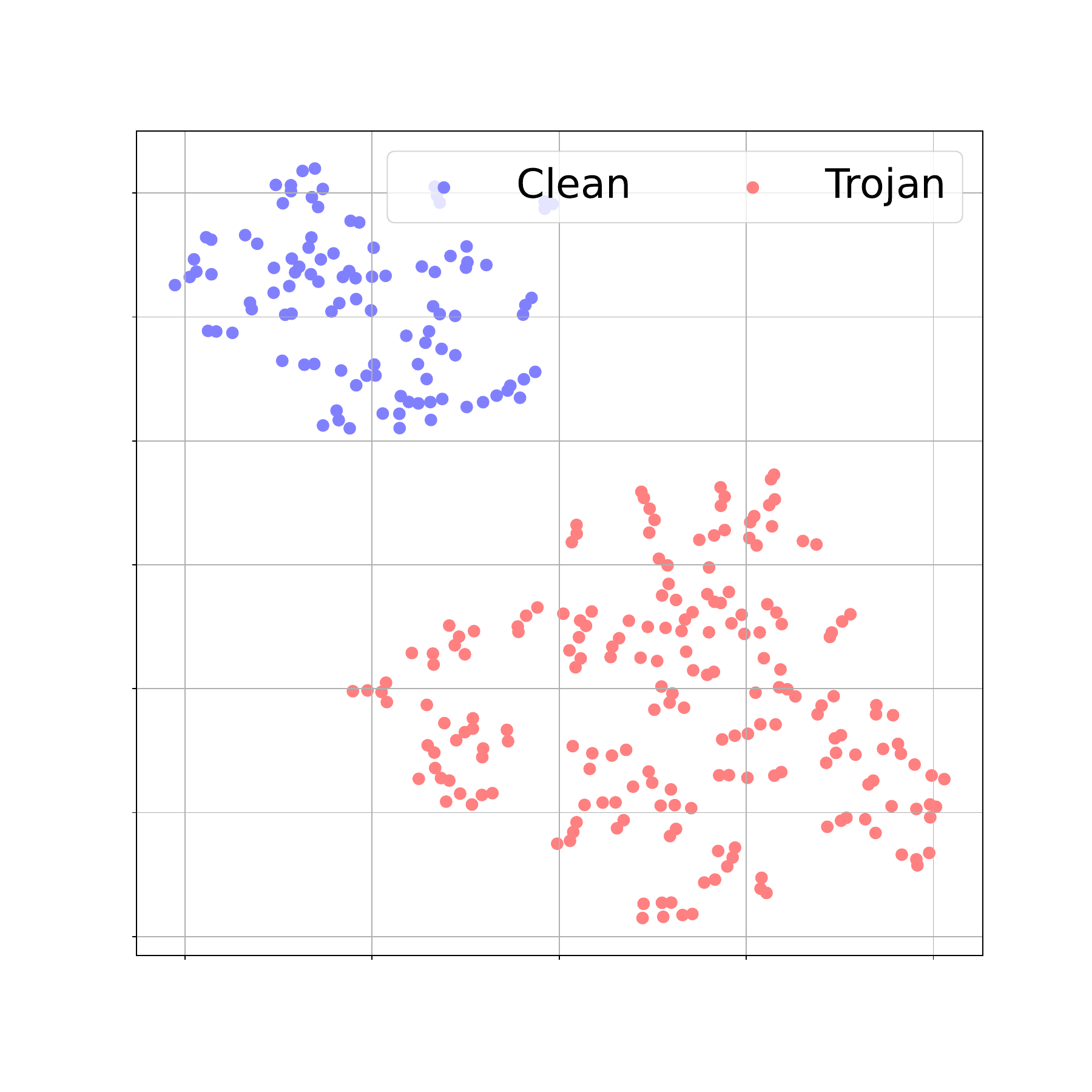}&
     \includegraphics[scale=0.11, trim={4.5cm 2.5cm 4.5cm 4.5cm}]{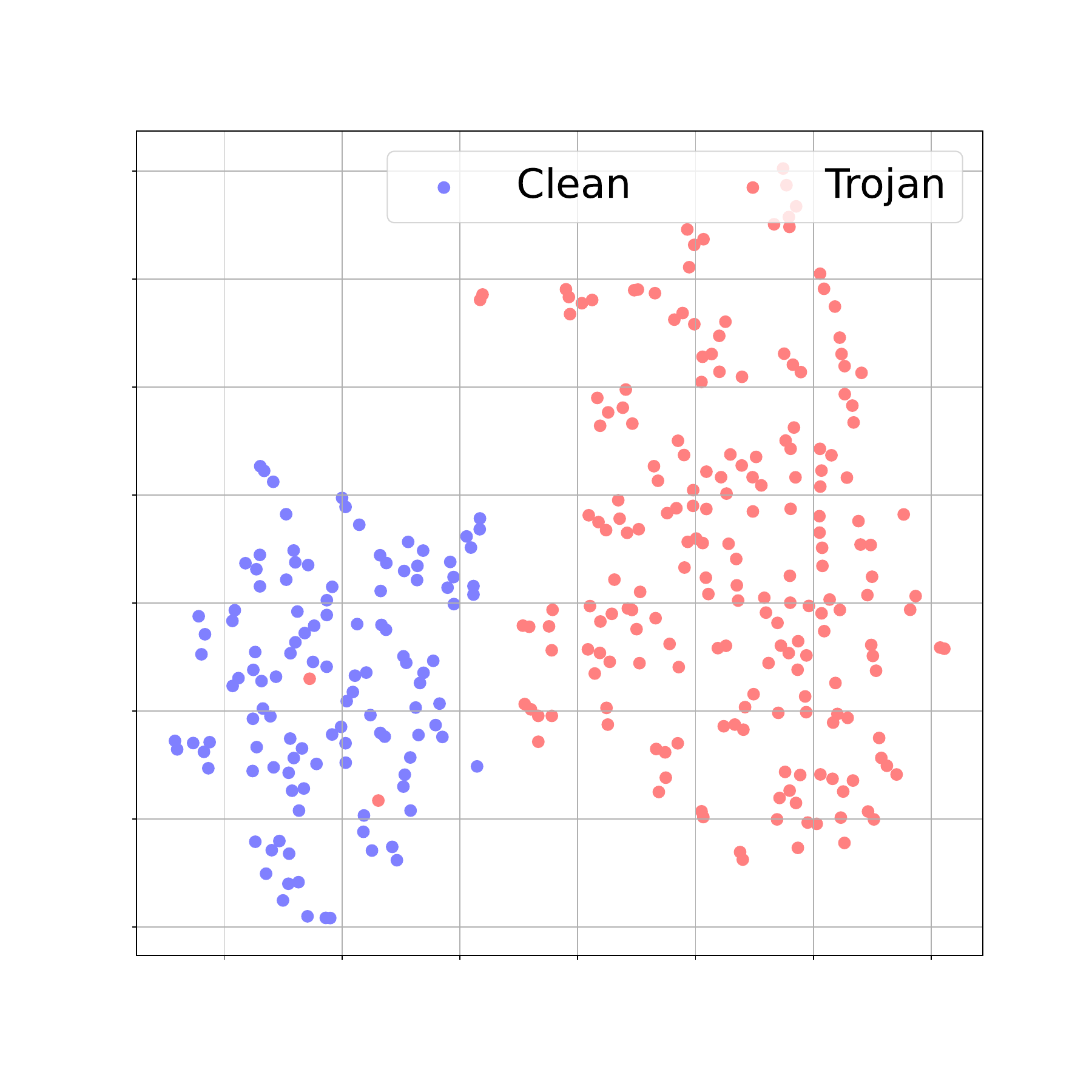}&
     \includegraphics[scale=0.11, trim={4.5cm 2.5cm 4.5cm 4.5cm}]{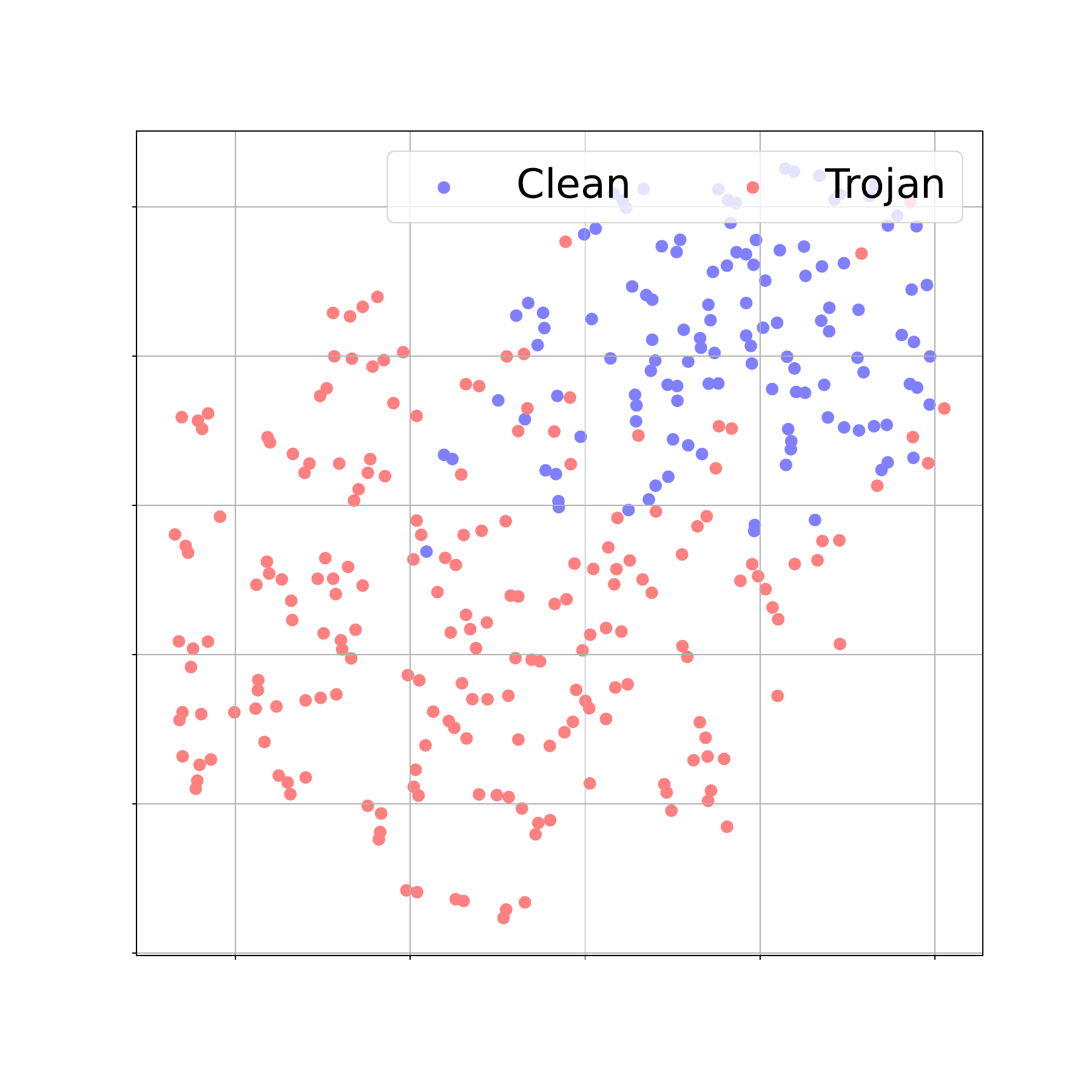}&
    \includegraphics[scale=0.11, trim={4.5cm 2.5cm 4.5cm 4.5cm}]{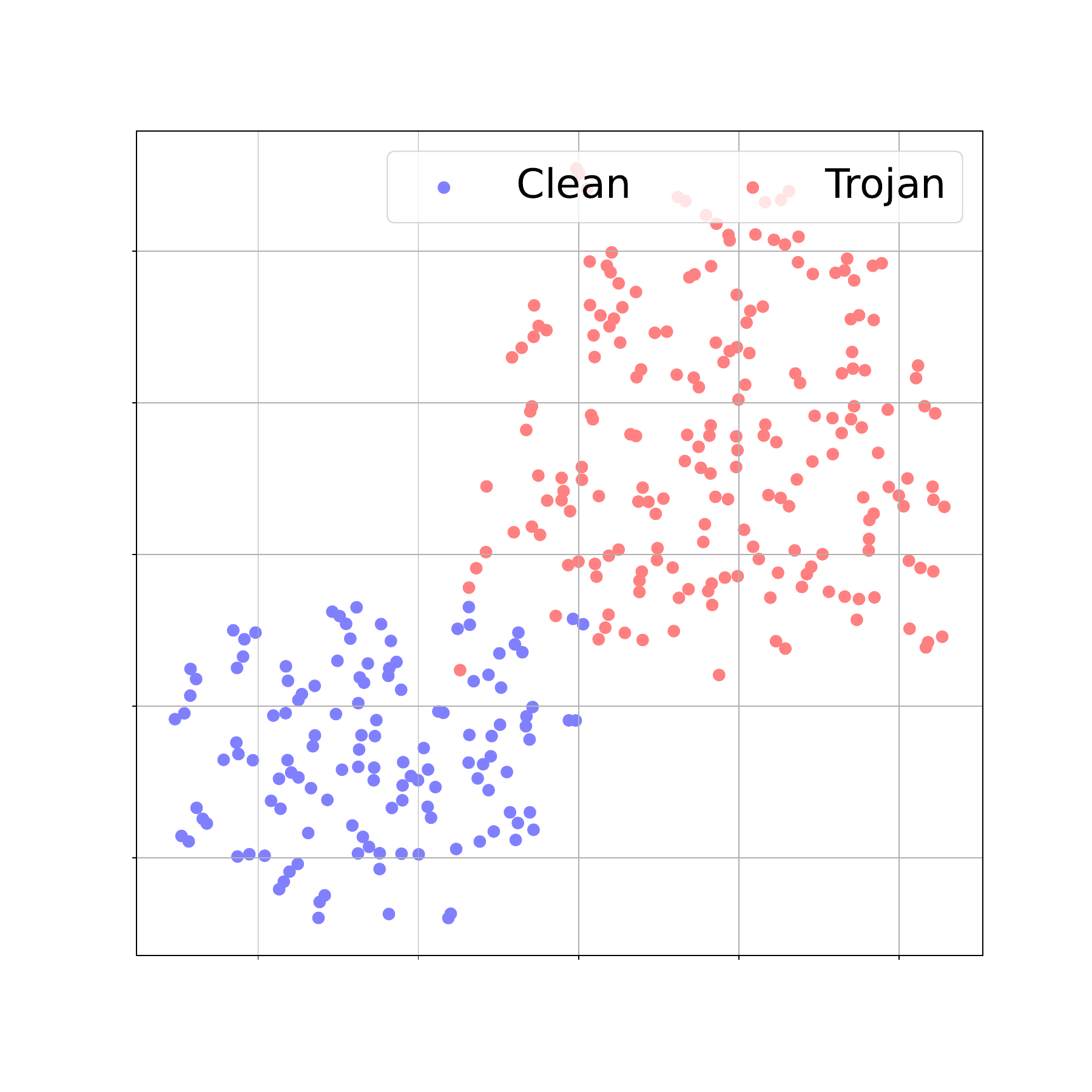}&
    \includegraphics[scale=0.11, trim={4.5cm 2.5cm 4.5cm 4.5cm}]{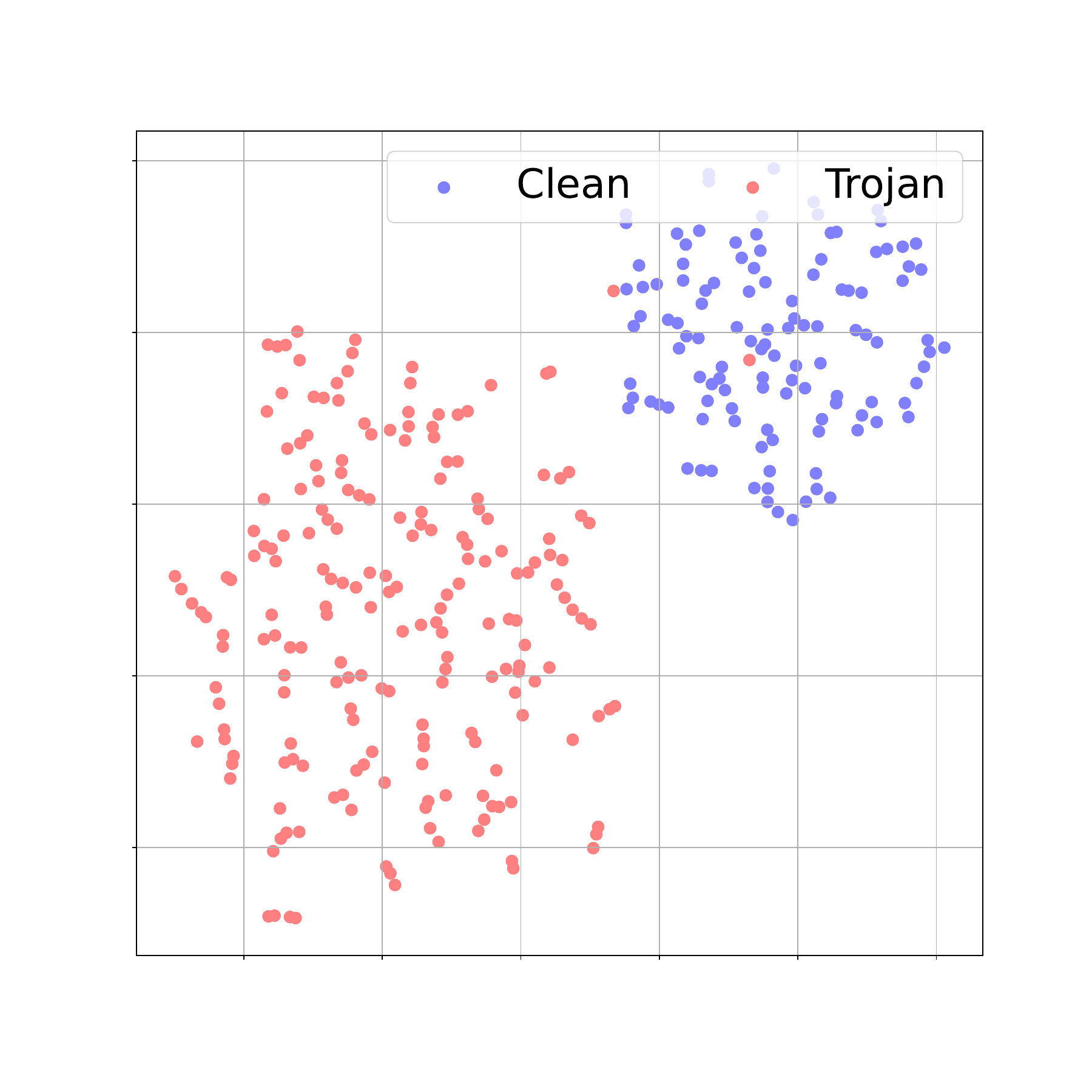}&
    \includegraphics[scale=0.11, trim={4.5cm 2.5cm 4.5cm 4.5cm}]{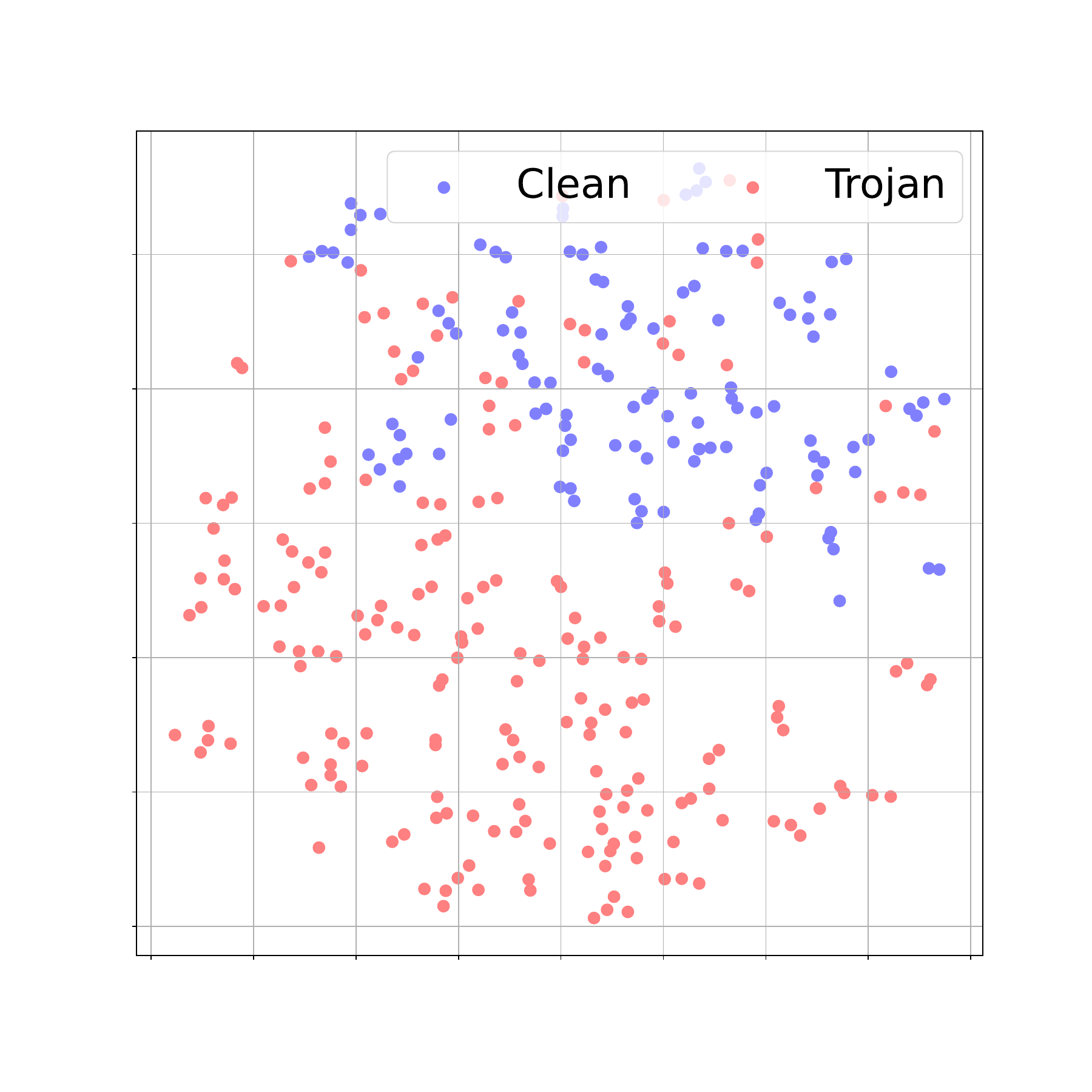}\\
     \rotatebox{90}{$\:\:\:\:\:\:\:$ SVHN}&
     \includegraphics[scale=0.11, trim={4.5cm 2.5cm 4.5cm 4.5cm}]{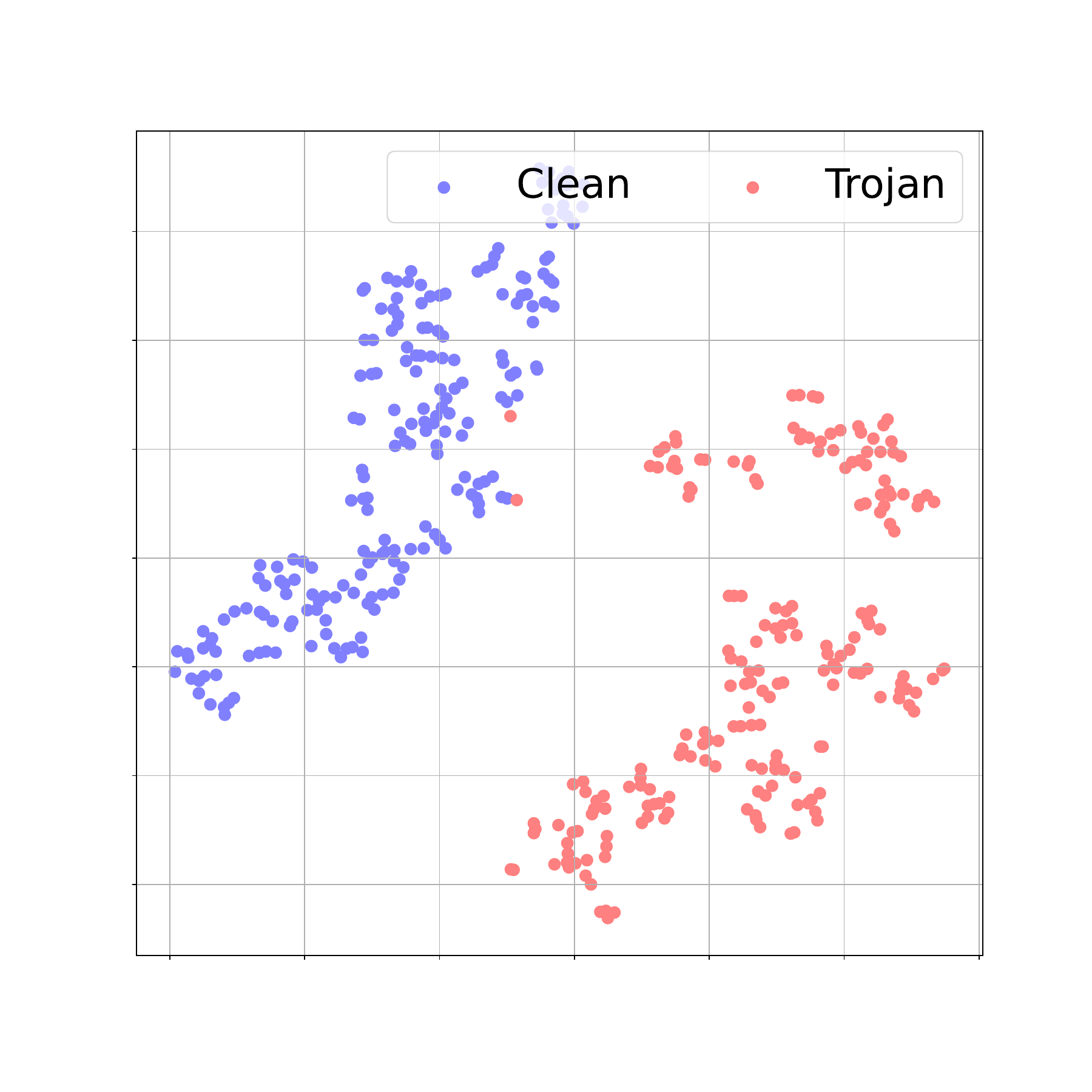}&
    \includegraphics[scale=0.11, trim={4.5cm 2.5cm 4.5cm 4.5cm}]{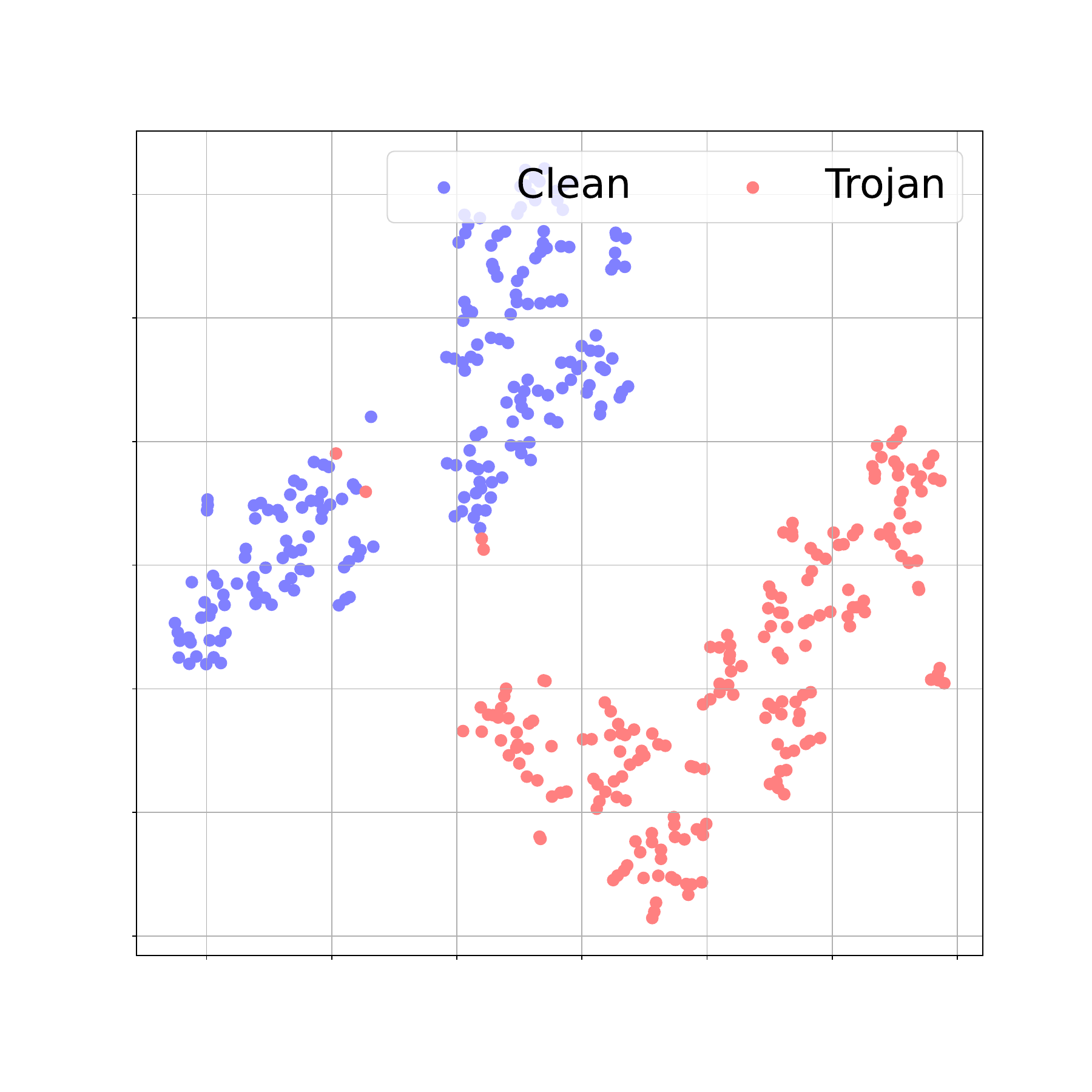}&
     \includegraphics[scale=0.11, trim={4.5cm 2.5cm 4.5cm 4.5cm}]{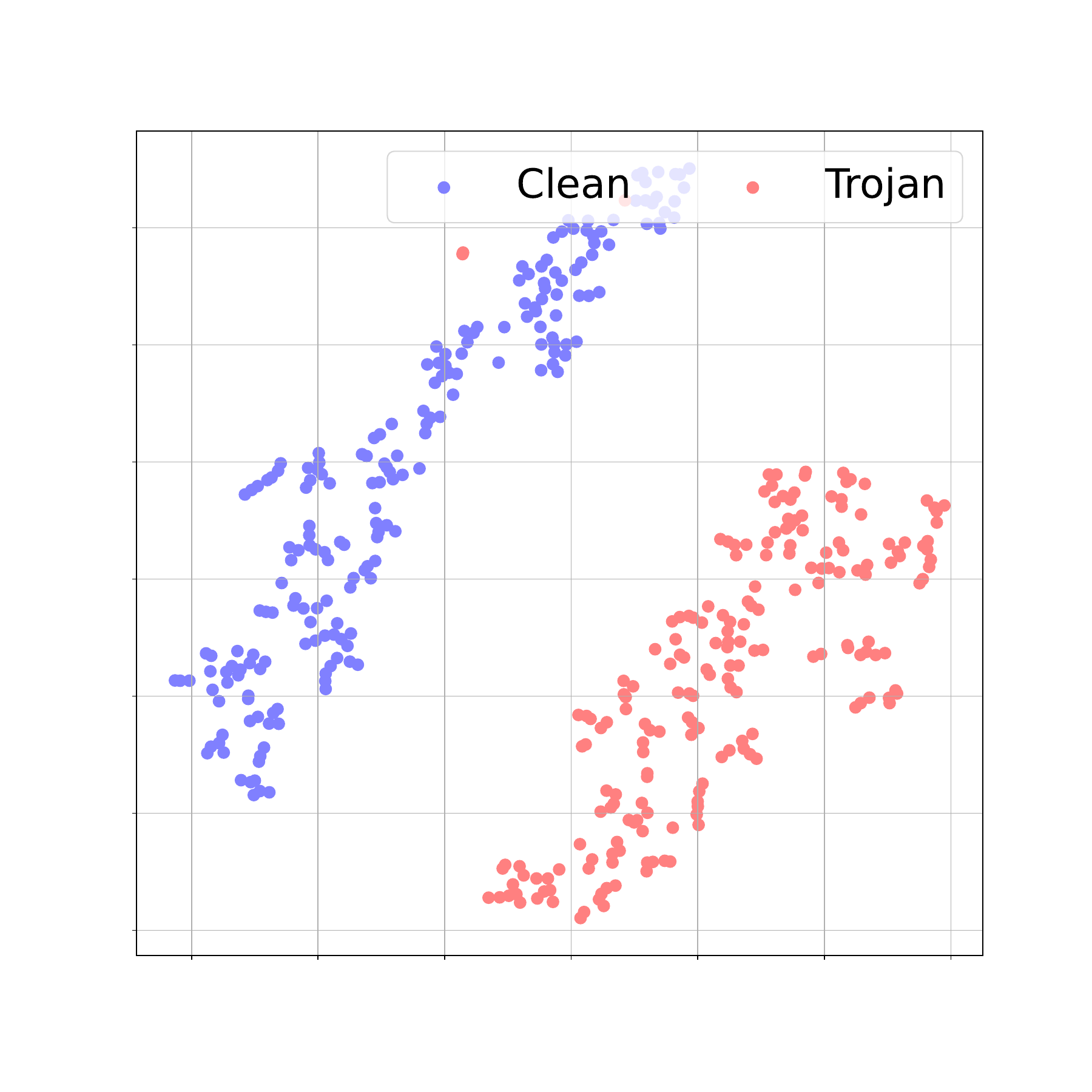}&
    \includegraphics[scale=0.11, trim={4.5cm 2.5cm 4.5cm 4.5cm}]{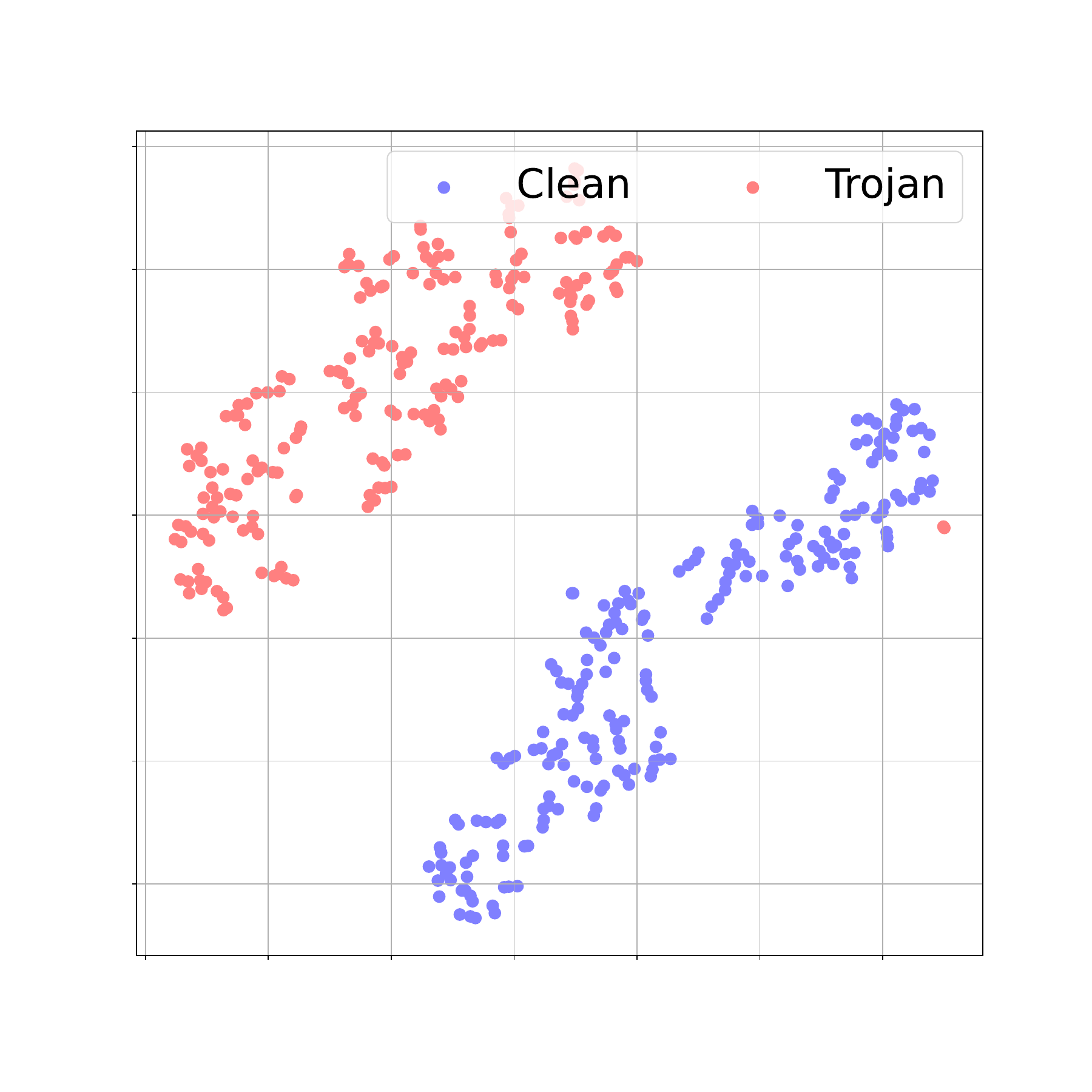}&
    \includegraphics[scale=0.11, trim={4.5cm 2.5cm 4.5cm 4.5cm}]{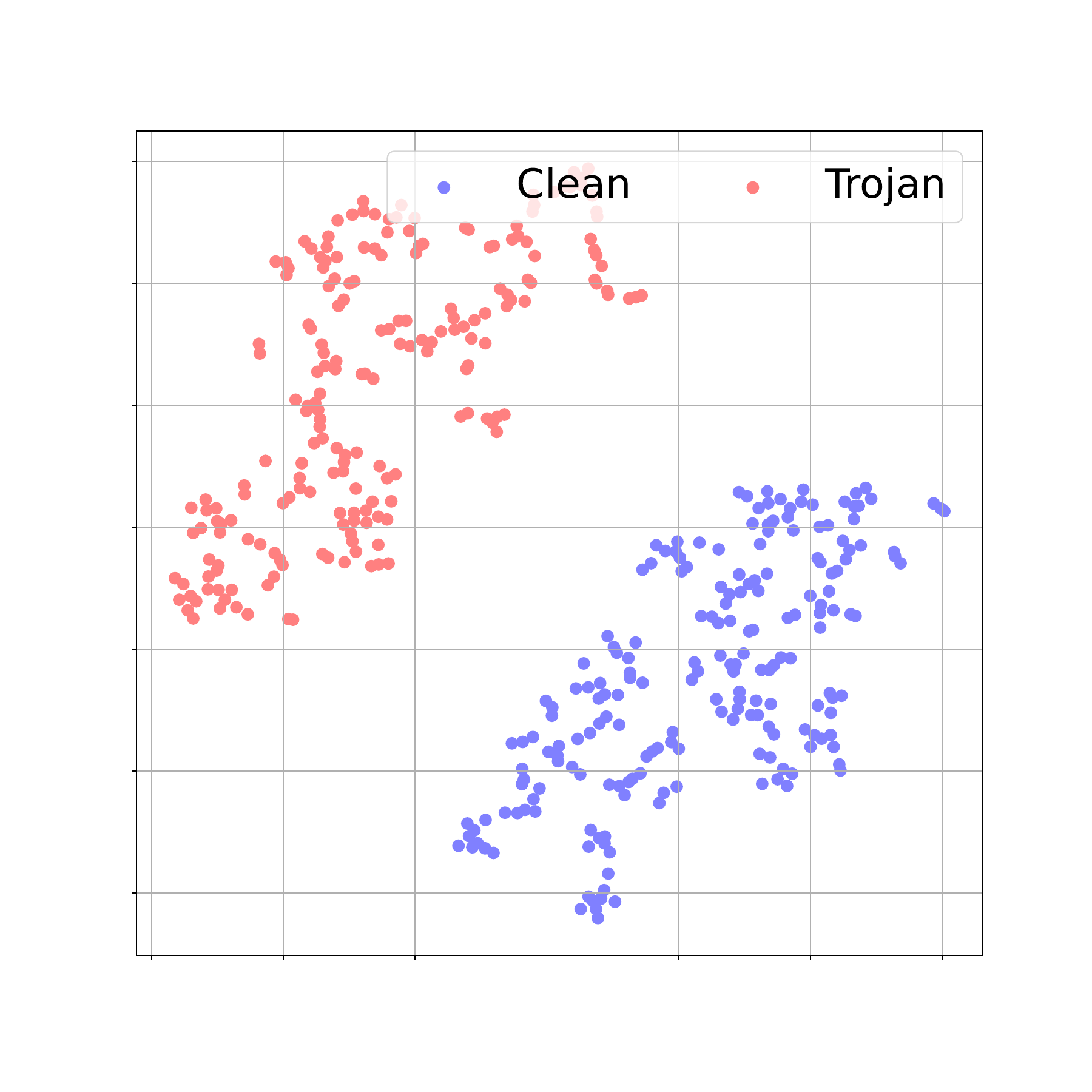}&
    \includegraphics[scale=0.11, trim={4.5cm 2.5cm 4.5cm 4.5cm}]
    {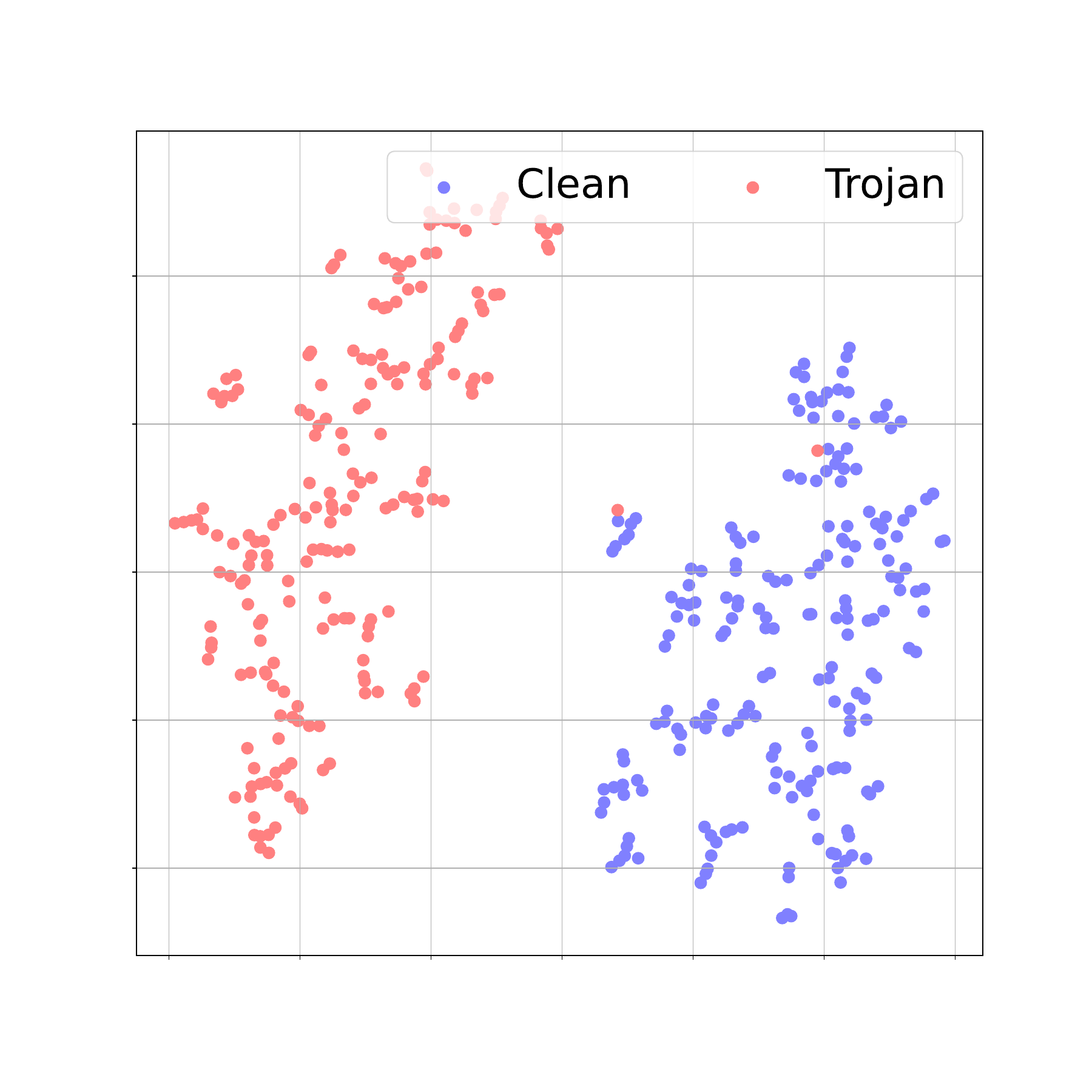}\\
    
    \rotatebox{90}{$\:\:\:\:\:\:\:$ Flower102}&
     \includegraphics[scale=0.11, trim={4.5cm 2.5cm 4.5cm 4.5cm}]{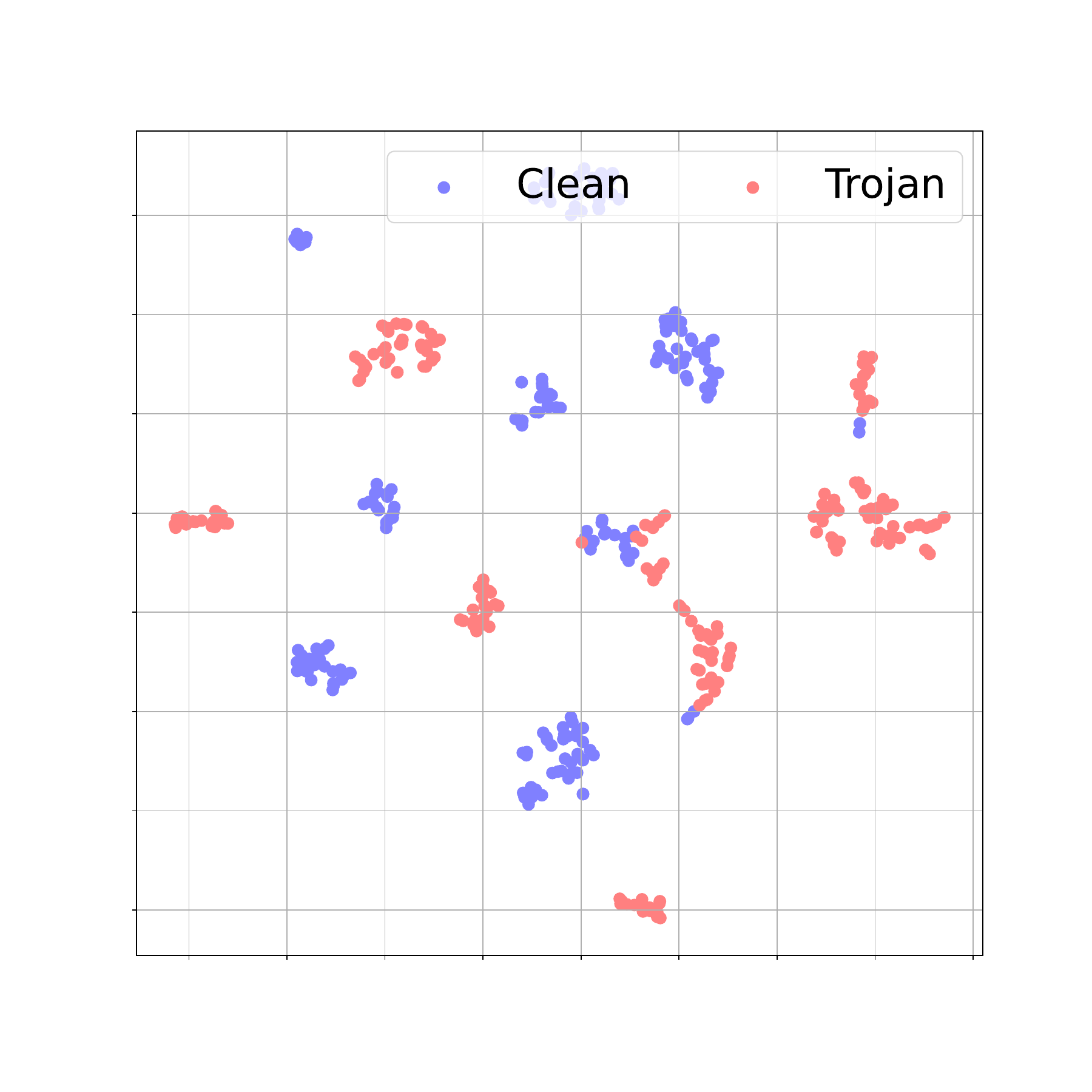}&
     \includegraphics[scale=0.11, trim={4.5cm 2.5cm 4.5cm 4.5cm}]{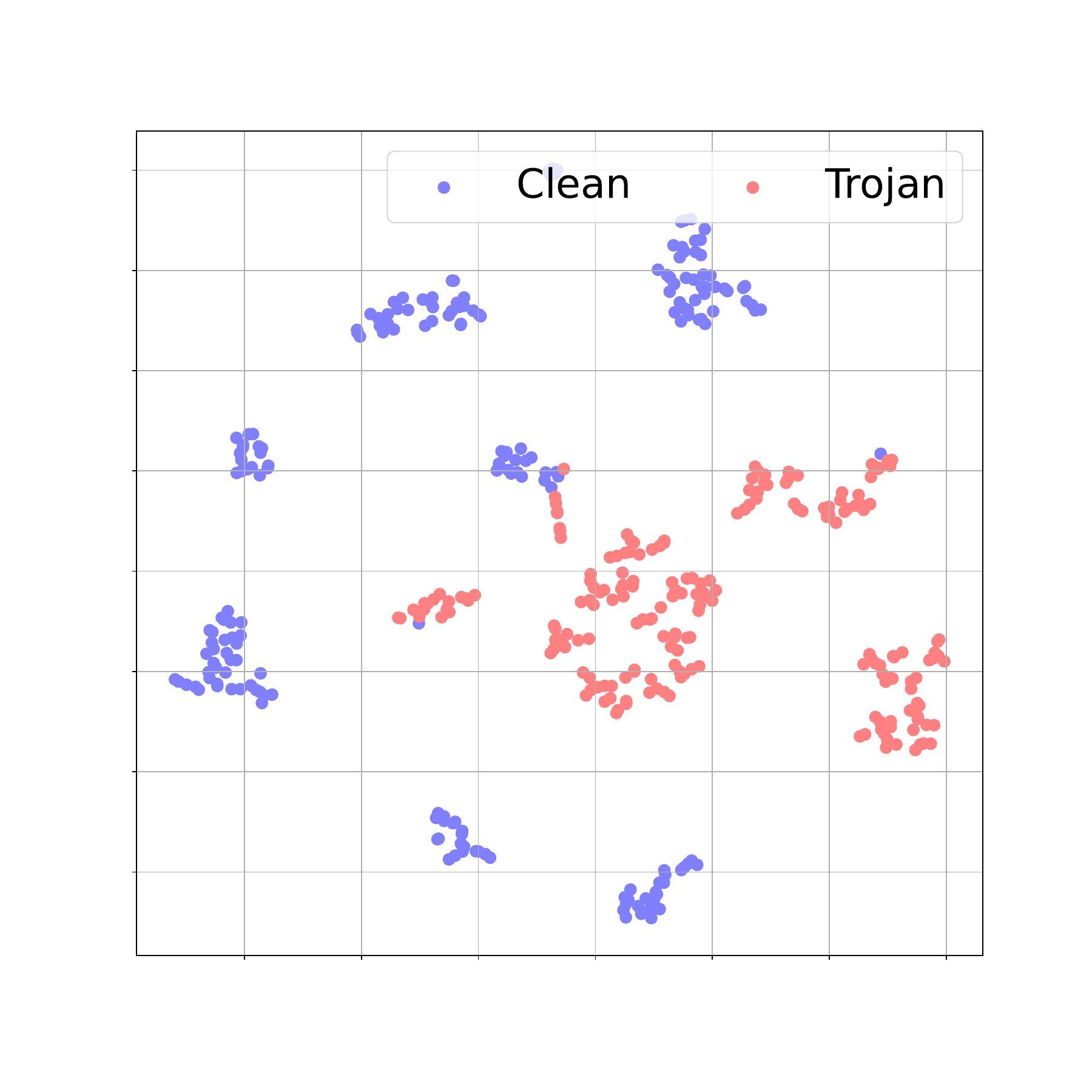}&
     \includegraphics[scale=0.11, trim={4.5cm 2.5cm 4.5cm 4.5cm}]{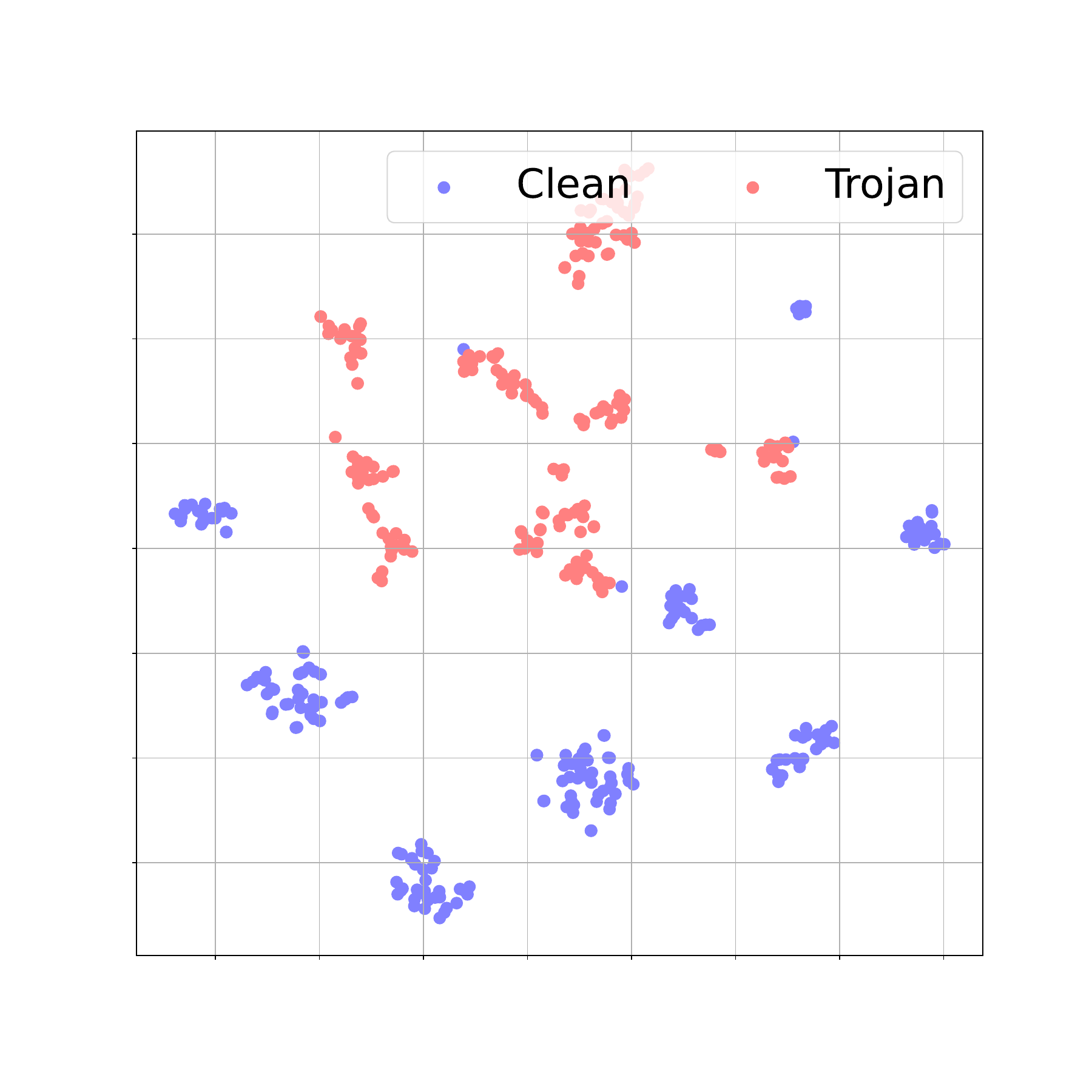}&
    \includegraphics[scale=0.11, trim={4.5cm 2.5cm 4.5cm 4.5cm}]{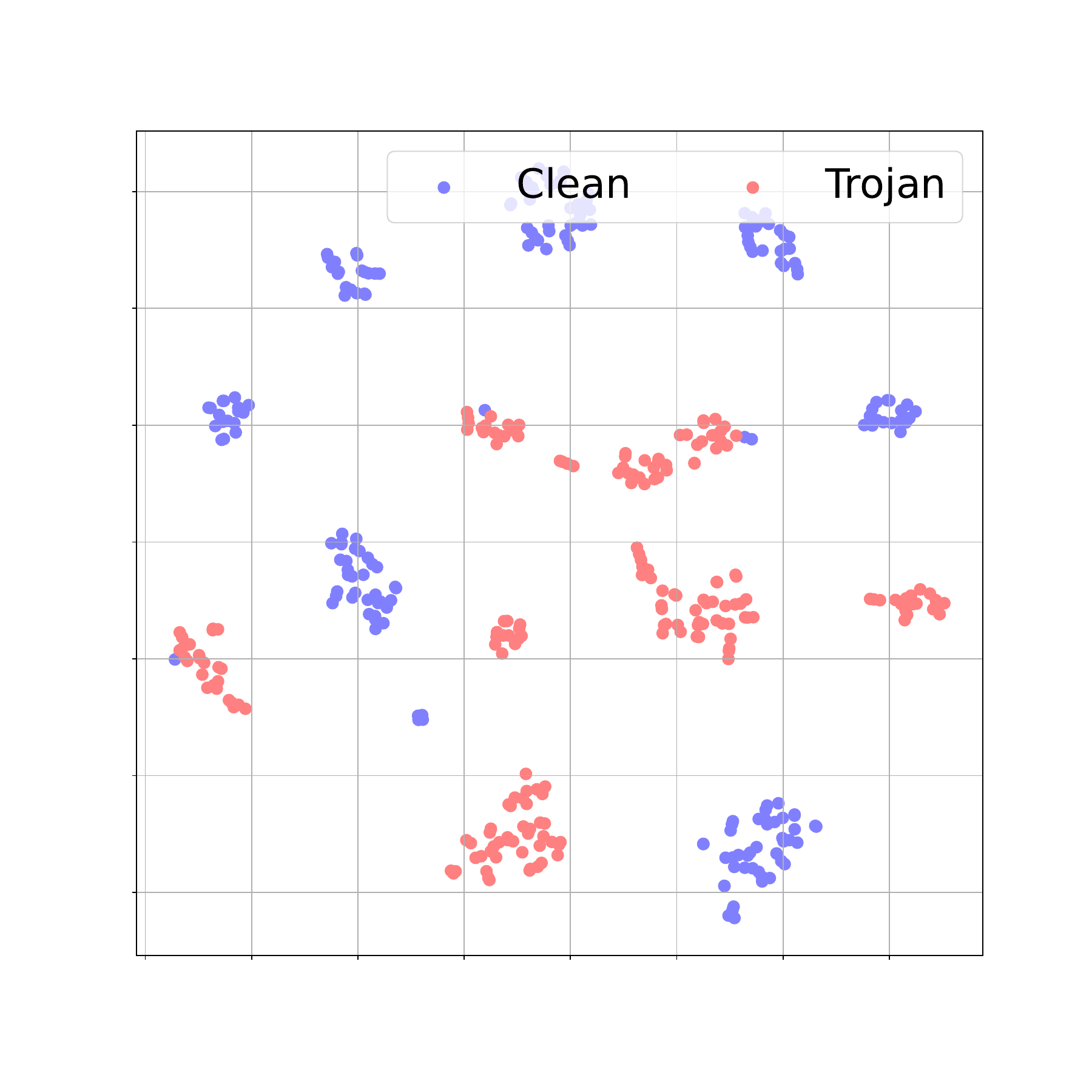}&
    \includegraphics[scale=0.11, trim={4.5cm 2.5cm 4.5cm 4.5cm}]{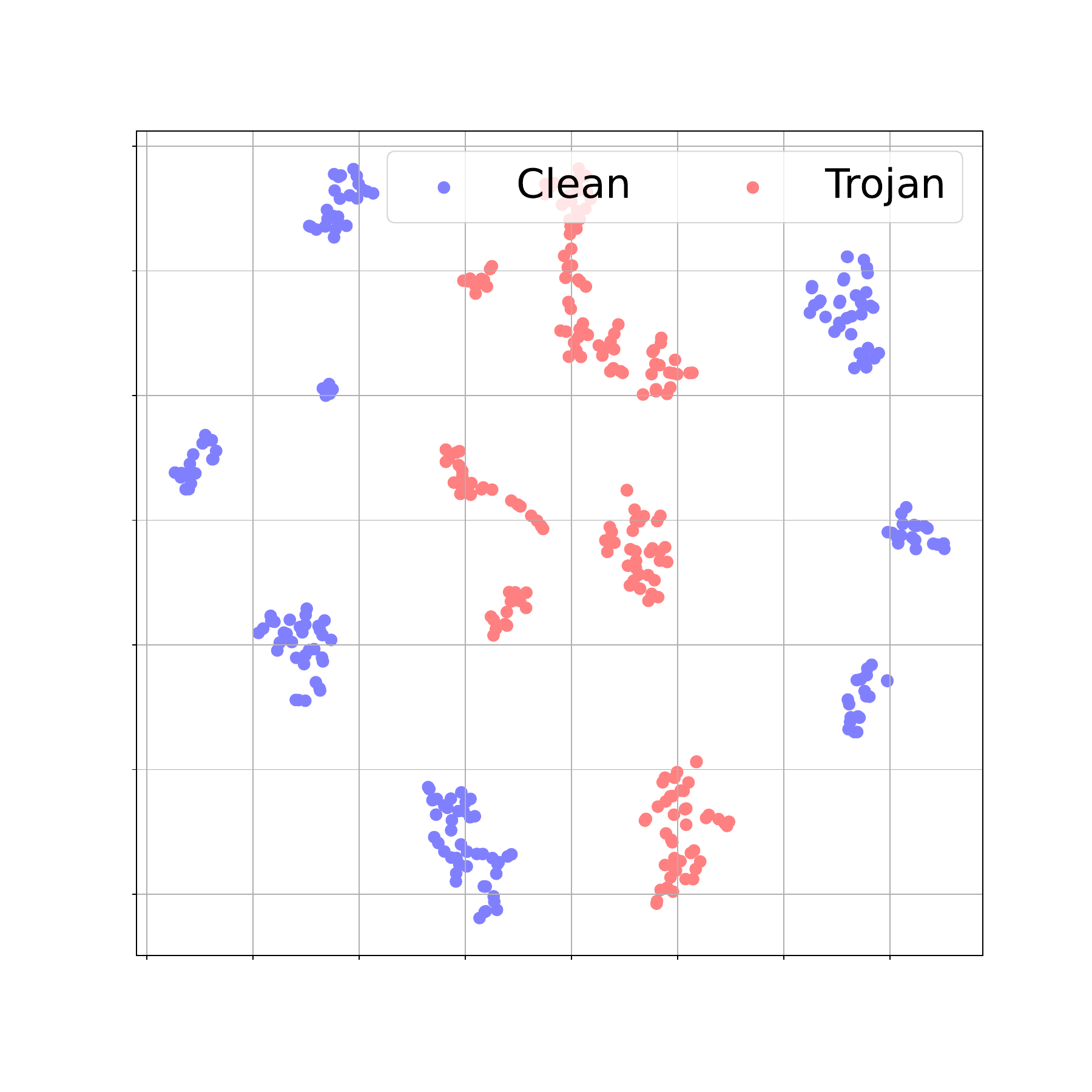}&
    \includegraphics[scale=0.11, trim={4.5cm 2.5cm 4.5cm 4.5cm}]
    {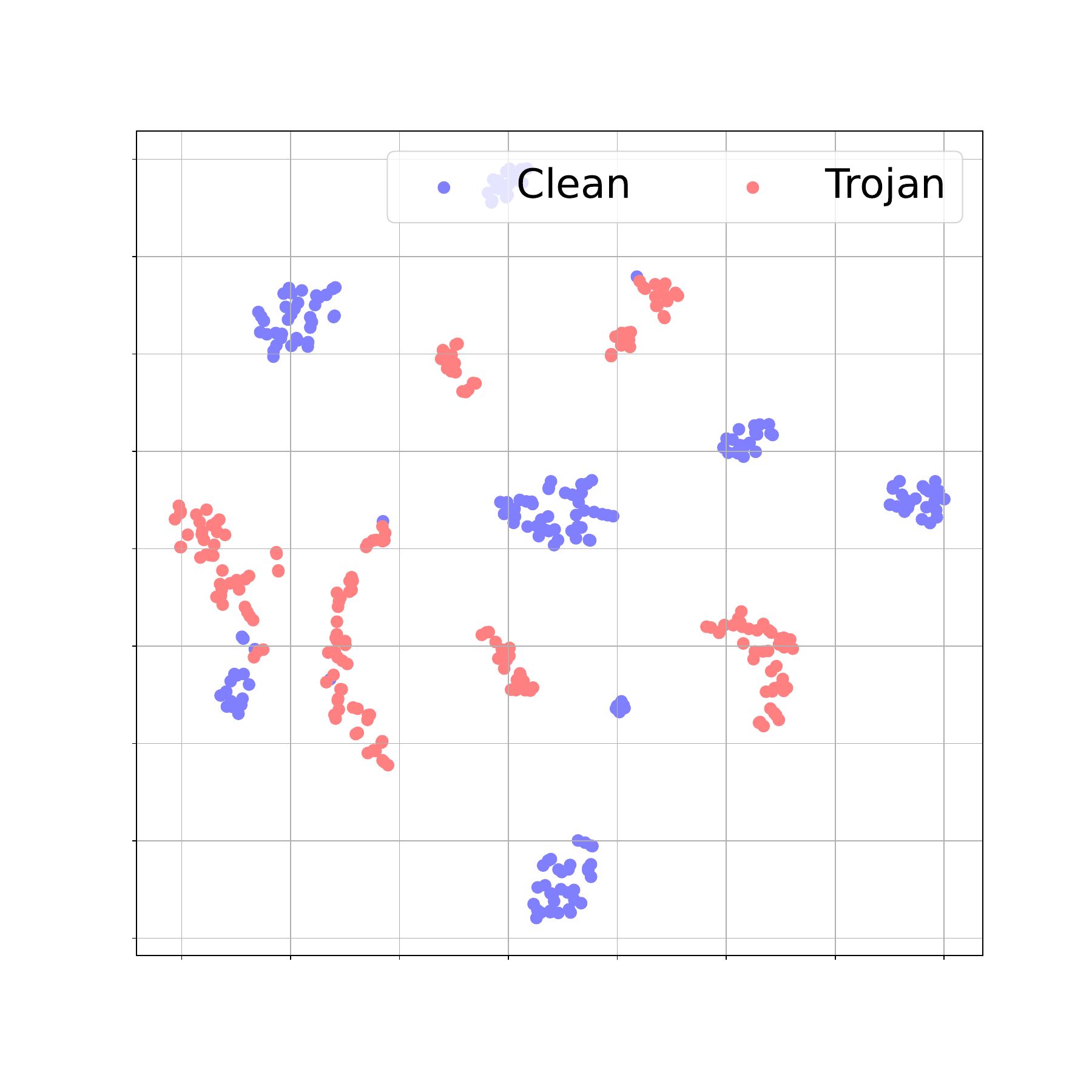}\\

    % \includegraphics[scale=0.09, trim={4.5cm 2.5cm 4.5cm 4.5cm}]{images/tsne/svhn_natural_svhn_trojan_tsne.pdf}\\
    
    %  \rotatebox{90}{$\:\:\:\:\:\:$EuroSAT}&
    % \includegraphics[scale=0.09, trim={4.5cm 2.5cm 4.5cm 4.5cm}]{images/tsne/EuroSAT_blend_trojan_tsne.pdf}&
    % \includegraphics[scale=0.09, trim={4.5cm 2.5cm 4.5cm 4.5cm}]{images/tsne/EuroSAT_l2_inv_trojan_tsne.pdf}&
    % \includegraphics[scale=0.09, trim={4.5cm 2.5cm 4.5cm 4.5cm}]{images/tsne/EuroSAT_nature_trojan_tsne.pdf}&
    % \includegraphics[scale=0.09, trim={4.5cm 2.5cm 4.5cm 4.5cm}]{images/tsne/EuroSAT_trojan_sq_trojan_tsne.pdf}&
    % \includegraphics[scale=0.09, trim={4.5cm 2.5cm 4.5cm 4.5cm}]{images/tsne/EuroSAT_trojan_wm_trojan_tsne.pdf}&
    % \includegraphics[scale=0.09, trim={4.5cm 2.5cm 4.5cm 4.5cm}]{images/tsne/EuroSAT_badnets_trojan_tsne.pdf}&
    % \includegraphics[scale=0.09, trim={4.5cm 2.5cm 4.5cm 4.5cm}]{images/tsne/EuroSAT_natural_EuroSAT_trojan_tsne.pdf}\\
    & BadNets &  Blend &  Nature & Trojan SQ & Watermark & L2 inv 
\end{tabular}
% {\scriptsize Badnets} & {\scriptsize Nature} & {\scriptsize Blend }&  {\scriptsize SQ } & {\scriptsize WM} &{\scriptsize L2 inv} & {\scriptsize Natural} 
\caption{This figure shows t-SNE visualizations for outputs of the penultimate layer of the DNN for $200$ clean samples (\textcolor{blue}{blue dots}) from the target class (\emph{Class 6}) and $200$ Trojaned samples (\textcolor{red}{red dots}) misclassified to the target class. We examine the CIFAR10, CIFAR100, GTSRB, SVHN, and Flower102 datasets, and \textbf{six Trojan triggers} (Badnets, Blend, Nature, Trojan SQ, Trojan WM and L2 inv) for each dataset. In most cases, we observe that although the DNN classifies both clean and Trojaned samples to the same class (\emph{Class 6}), it generates different values for features in its penultimate layer. 
Embedding a trigger into samples from CIFAR10 and GTSRB using the L2 inv trigger is relatively less easy (compare separability of blue and red dots in \underline{Col 6, Row 2} and \underline{Col 6, Row 3}). 
  %  Any detection mechanism, therefore, can be expected to have a high false positive rate, where clean samples are (mistakenly) identified as Trojaned.
\label{fig:tsne}}
%\vspace{0.2 in}
\end{figure*}

\section{User Capability}\label{UserAbility}
%\noindent{\bf User Capability:}  
The user has limited computational resources, and uses ML platforms to train a model or uses publicly available models which have  high accuracy on clean samples. 
However, the user has sufficient resources to use an adversarial learning method~\cite{goodfellow2014FGS} to obtain the magnitude of adversarial noise that will result in misclassification of an input sample. 
We further assume that the user is aware of the possibility that the input sample or the model can be Trojaned, and aims to detect and remove malicious inputs. 
However, the user has no knowledge of the attacker strategy, and identities of the Trojan trigger and target class. 
The user is assumed to have access to a set of clean samples whose labels are known. 
%Further, these samples are assumed to be `balanced' across all possible output labels. 
The user has either \emph{white-box access} (access to weights and hyperparameters of the DNN model) or \emph{black-box access} (access to only outputs of the DNN model). 
We note that developing solutions for black-box model access is more difficult than for the white-box setting, where information about the DNN model is available. 
For both settings, our objective is to develop a method that would aid a user in identifying and discarding input samples that contain a Trojan trigger.% \emph{without retraining or pruning the DNN model}. 

\section{MDTD: Motivation and Design}\label{sec:motivation}

\begin{table*}[]
\caption{This table reports average (standard deviation) of certified radii for $200$ clean and $200$ Trojan samples for the CIFAR10, CIFAR100, GTSRB, and Flower102 datasets with \textbf{six different Trojan triggers} %(Badnets, Blend, Nature, Trojan SQ, Trojan WM and L2 inv) 
(noted in \textbf{bold} column titles). 
%\textcolor{blue}{
We observe that Trojan samples have a higher average certified radius than clean samples. Thus, the minimum magnitude of noise required to make a DNN model misclassify an input sample containing a Trojan trigger is larger.
}
 \label{tab:certifiedR}
    \centering
    \begin{tabular}{|c|c|c|c|c|c|c|c|}
    \hline
       Dataset & Clean &\textbf{Badnets} & \textbf{Blend} &  \textbf{Nature} & \textbf{Trojan SQ} & \textbf{Trojan WM} & \textbf{L2 inv}  \\ \hline
    %   CIFAR10 & 0.227 (0.258) & 0.557 (0.286) & 0.663 (0.208) & 0.873 (0.081) & 0.884 (0.044) & 0.863 (0.089) & 0.338 (0.172) & 0.475 (0.236)   \\  \hline
     CIFAR100 &   0.005 (0.045) & 0.657 (0.132) &0.103 (0.131) &  0.880 (0.017) & 0.759 (0.047) & 0.893 (0.0001) & 0.596 (0.036)  \\  \hline %& 0.335 (0.006)
     
       CIFAR10 & 0.015 (0.046) &  0.875 (0.035) &0.308 (0.216) & 0.893 (0.000) & 0.893 (0.0001) & 0.893 (0.0001) & 0.648 (0.059)   \\  \hline %& 0.619 (0.005)
      
       GTSRB & 0.308 (0.326) & 0.663 (0.234) & 0.749 (0.227) & 0.882 (0.068) & 0.857 (0.106) & 0.499 (0.235) & 0.239 (0.148)   \\  \hline %0.531 (0.170)
       SVHN &0.255 (0.268) &  0.727 (0.178) &0.651 (0.271) & 0.887 (0.035) & 0.879 (0.083) & 0.882 (0.062) & 0.893 (0.0001)  \\ \hline %& 0.893 (0.001)
    %   EuroSAT & 0.152 (0.323) & 0.159 (0.159) & 0.123 (0.160) & 0.485 (0.177) & 0.667 (0.166) & 0.001 (0.009) & 0.061 (0.101) & 0.000 (0.000)   \\  \hline
    Flower102 & $0.038 (0.173)$ &$ 0.189 (0.163)$ & $0.634 (0.302)$ & $0.875 (0.07)$ & $0.893 (0.001)$ & $0.893 (0.001)$ &$ 0.269 (0.148)$ \\ \hline
    \end{tabular}
    % \vspace{0.2 in}

\end{table*}

In this section, we describe the mechanism of \emph{MDTD}. 
We %first 
motivate the development of \emph{MDTD} by showing that feature values at intermediate layers of the DNN corresponding to clean and Trojaned samples are different. 
We hypothesize that as a consequence, clean and Trojaned samples will behave differently when perturbed with noise, and use the notion of a certified radius~\cite{cohen2019certified} to verify this. %our hypothesis. 
We then explain how \emph{MDTD} computes an estimate of the certified radius to effectively distinguish between Trojan and clean samples. %when a user has access to only a small set of clean samples. 

\subsection{Key Intuition behind MDTD}

\noindent{\bf Feature Value Visualization using t-SNE:}  
For a DNN model with $K$ layers, the first $(K-1)$ layers map an input $x$ to the \emph{feature space}, which typically has a lower dimension than the high-dimensional input space. 
The last layer of the DNN then uses the values in the feature space to make a decision about the input. 
For example, in an image classification task, the decision will be the identity of the class that the input is presumed to belong to. 
Thus, the output of the penultimate layer (layer $K-1$) of the DNN can then be interpreted as an indicator of the \emph{perspective} of the DNN about the given input sample.

We use a t-distributed stochastic neighbor embedding (t-SNE) \cite{van2008visualizing} to demonstrate that for the same output of the DNN model, values in the feature space for clean and Trojan samples are different. 
The t-SNE is a technique to visualize high-dimensional data through a two or three dimensional representation. 
For the CIFAR10, CIFAR100, GTSRB, SVHN, and Flowers102 datasets, we collect $200$ samples corresponding to each of six different Trojan trigger types that are classified by the DNN model as belonging to the class $y^d =6$ due to the presence of the trigger. 
We additionally collect $200$ clean input samples that do not contain a Trojan trigger for whom the output of the DNN model is $y = y^d$. 

Fig.~\ref{fig:tsne} shows t-SNE representations of the feature values of these samples, i.e., the outputs at the penultimate layer of the DNN. 
We observe that the clean samples (blue dots) can be easily distinguished from Trojan samples (red dots) for each of the Trojan trigger type in all five datasets. 
While the t-SNE visualization provides qualitative indicators of clean and Trojaned sample behavior, we are also interested in quantitative metrics. 
We use the %notion of 
certified radius to distinguish between clean and Trojaned input samples.
% This result motivates an analysis of other properties of these samples in order to effectively distinguish between clean and Trojan samples.\\ 
 
% The first K-1 layers of deep neural network map the input to feature space (a compressed variant of input) and then the last layer  makes a decision using feature values generated by K-1 first layers for the input. More precisely, the output of the penultimate layer (the last fully connected layer before the last layer) indicates the viewpoint of a deep neural network about the input. Therefore, we obtain the feature values generated by the penultimate layer for 200 clean that truly belong to the class $y^d$ and Trojan samples assigned to $y^d$ due to the presence of the trigger,  then we visualize these feature values in 2D using t-SNE.

% Figure~\ref{fig:tsne} indicates the t-SNE presentation of the output of the penultimate layer for both clean and  Trojan samples. This figure demonstrates that the penultimate layer generates different and distinguishable feature values for clean and Trojan samples even though both clean and Trojan samples are classified to the same class ($y^d =6$).  These results inspire the evaluation of the difference of the properties of these samples. Here, we demonstrate that the certified radius as one of properties that clean and Trojan samples differ in.\\

\noindent{\bf Certified Radius:} 
The certified radius is defined~\cite{cohen2019certified} as the radius of the largest ball %(according to the $l_2$ norm) 
centered at each input sample within which the DNN model returns the same output label. 
The certified radius is computed by estimating the distance to a decision boundary by perturbing samples with Gaussian noise with a predetermined/ known mean and variance. 
%\textcolor{blue}{
However, exactly computing the certified robustness has a high computational cost~\cite{li2020CertifiedComputationalCost}. Instead, we evaluate the robustness of Trojan-embedded samples by approximately computing the certified radius using a heuristic that perturbs a small number of clean and Trojaned samples with Gaussian noise centered at the input sample of interest. 
%}

Table~\ref{tab:certifiedR} presents average certified radii for $200$ clean and $200$ Trojan samples with six different Trojan trigger types (Badnets, Blend, Nature, Trojan SQ, Trojan WM and L2 inv) applied on CIFAR10, CIFAR100, GTSRB, SVHN and Flowers102 datasets. 
%\textcolor{blue}{
%From the table, we 
We observe that the certified radius is significantly higher for Trojaned samples %compared to that 
than for clean samples, clearly indicating a relatively larger distance to the decision boundary. 
Consequently, the minimum magnitude of noise required to make the DNN misclassify a Trojan sample will be larger than the noise required to misclassify a clean sample. \\
%We leverage this insight to inform the algorithmic design of \emph{MDTD}.\\ %, a mechanism to distinguish clean and Trojan samples without making any assumptions on the attacker's strategy.\\

%As we discussed, we investigate the certified radius of a sample as the property for a given sample for both Trojan and clean samples. The certified radius measures the distance of a sample to the decision boundary using Gaussian random noise~\cite{cohen2019certified}. Certified radius method finds the largest $l_2$  ball with radius of $r$ around each sample in which DNN classifier $f$ classifies any sample in that ball to the same class of the sample  most likely. Table~\ref{tab:certifiedR} indicates the average of $l_2$ certified radius obtained for 200 samples for both clean and Trojan samples. This table demonstrates that Trojan samples have larger certified radius, and consequently they are located further away from the decision boundary. For example for CIFA100 dataset, the certified radius of clean samples on average is $0.005$, while for a Trojan sample with badnets trigger is $0.65$.  Samples with higher certified radius are more robust to noise and consequently the minimum amount of noise required to make models misclassify them is larger. Using this insight, we propose a Trojan sample detection method, which distinguishes clean samples from Trojans by using the minimum adversarial noise for misclassification without having any assumptions about the attacker's strategy.\\

Although t-SNE visualizations and certified radius computation provide preliminary insights into differences between clean and Trojan samples, there are two significant challenges which limit their direct use in Trojan trigger detection. %in realistic settings. 
First, while t-SNE provides visual insights, it relies on access to (i) adequate number of clean and Trojan trigger-embedded samples, and (ii) outputs at intermediate layers of the DNN. 
However, t-SNE visualizations are not useful as a direct computational mechanism for Trojan detection since \emph{the user of a DNN model will typically not have access to adequate number Trojan trigger-embedded samples}. 
%The user of a DNN model will not have access to the required number of trigger-embedded samples to create meaningful t-SNE visualizations. 
%Obtaining outputs at intermediate layers of the DNN requires access to model weights. 
Further, access to intermediate layers is not feasible for users who only have black-box model access (i.e., access to only outputs of the model). 
Second, computing the certified radius is computationally expensive, and is known to be NP-complete~\cite{li2020CertifiedComputationalCost}, which limits their use in practice. 
We next present a two-stage algorithmic design for MDTD that leverages insights from t-SNE visualizations and certified radii.

\subsection{Two-Stage Design of MDTD}\label{sec:ourapproach}
%\subsection{Trojan Sample Detection} 
\emph{MDTD} %on the other hand, 
makes use of adversarial learning techniques%introduced in
~\cite{goodfellow2014FGS} 
to estimate distance of a sample from any decision boundary. It then computes the smallest magnitude of adversarial noise required to misclassify the sample 
%an input sample 
to infer whether the sample is Trojaned or not. 
\emph{MDTD} consists of two stages. In the first stage, MDTD estimates the distance of a given input sample to the decision boundary. 
In the second stage, distances estimated in the previous step and the distance of a small number of clean samples to the decision boundary are used to identify whether the sample contains a Trojan trigger or not. We describe each stage below.  

%then discards the inputs that their distances are not similar to the distances obtained from clean set samples of defender.

\noindent{\bf Stage 1: Estimating distance to decision boundary:}
%In order to overcome the computational expense associated with determining the certified radius~\cite{cohen2019certified}, 
\emph{MDTD} uses adversarial learning techniques~\cite{goodfellow2014FGS, moosavi2017universal} to estimate the minimum magnitude of noise perturbation, denoted $\delta$, that will cause the DNN model to misclassify a given sample. 
For an input sample $x$, the output of the DNN model is denoted $f(x)$. The value of the smallest perturbation $\delta$ is obtained by maximizing a loss function $\mathcal{L}(\cdot, \cdot)$ that quantifies the difference between the output label of the DNN for a perturbed variant of the input, denoted $f(x+\delta)$, and the true label of the input $y$. 
%We want the magnitude of $\delta$ to be small. 
This objective is well-defined and can be expressed as a regularized optimization problem with regularization constant $\lambda$ as~\cite{goodfellow2014FGS, moosavi2017universal}: 
%
%Adversarial learning methods~\cite{goodfellow2014FGS, moosavi2017universal} solve the following optimization problem to determine the value of $\delta$:
\begin{equation}
    \min_{\delta} -\mathcal{L}(f(x+\delta), y)+\lambda \|\delta\|, \label{eq:OptProb}
\end{equation}
%where $\lambda$ is a regularization constant. 
%$y$ is the output label corresponding to input sample $x$, and $\mathcal{L}$ is a loss function that quantifies the difference between the output label of the DNN model and the true label of samples, and $\lambda$ is a constant. 
In computing the value of $\delta$, we need to consider two types of user access to the DNN model, namely \emph{white-box} access and \emph{black-box access}. 
For a user with white-box access, we can use the Fast Gradient Sign Method (FGSM) and Iterative Fast Gradient Sign Method (IFGSM) to solve Eqn. (\ref{eq:OptProb}) due to their low computational cost~\cite{goodfellow2014FGS, kurakin2018}. 
When a user has only black-box access to the DNN model, we apply a SOTA adversarial learning method called HopSkipJump~\cite{chen2020hopskipjumpattack} which allows us to estimate the minimum magnitude of noise $\delta$ required for misclassification of the input.

\noindent{\bf Stage 2: Outlier detection:} 
Following our assumptions on user capability described in Sec.~\ref{sec:threatmodel}, information about the identity of the Trojan trigger and the target output class for Trojan samples is not available. 
Hence, the user will not be able to generate Trojan samples and estimate the distance of these samples to the decision boundary. 
However, due to the widespread availability of datasets, we assume that the user has access to a limited number of clean samples. 
%Instead, we assume that the user has access to a limited number of clean samples. 
We denote this set as $D_{user}=\{(x_1,y_1), (x_2,y_2),\cdots, (x_N,y_N)\}$. 
The user is assumed to have the ability to use the methods in \emph{Stage 1} above to estimate the minimum magnitudes of noise $\{\delta_1,\delta_2,\cdots,\delta_N\}$ required to misclassify these samples. 

In order to determine whether a given input sample contains a Trojan trigger, we use an outlier detection technique first proposed in~\cite{shewhart1931economic}. 
This method assumes that the distances of a clean sample (non-outliers) to the decision boundary follows a Gaussian distribution $\delta \sim \mathcal{N}(\mu, \sigma^2)$. 
\emph{MDTD} estimates the values of $\mu$ and $\sigma$ using the values of $\{\delta_1,\delta_2,\cdots,\delta_N\}$ determined from the set $D_{user}$. 
Then, for a threshold $\alpha$ on the maximum tolerable false positive rate, any sample whose distance to the decision boundary satisfies $| \delta - \mu| > \alpha \sigma$ will be identified as containing a Trojan trigger (outlier). 
%Here, $\alpha$ is a user-defined parameter chosen based on the maximum tolerable false positive rate- a 
A small value of $\alpha$ results in a lower rate of detection of Trojan samples; a large value results in more clean samples being incorrectly identified as Trojan. 

\noindent{\bf Choice of $\alpha$:} 
%Let $\gamma$ be the maximum tolerable false positive rate for a user of the DNN model. %be denoted as $\epsilon$. 
%\textcolor{blue}{
We show how to choose $\alpha$, given %the maximum tolerable false positive rate, 
an upper bound 
$\gamma$ for a user of the DNN model, depending on the size of the set $D_{user}$. 
When the size of $D_{user}$ is sufficiently large, $\alpha$ can be expressed using the tail distribution of a standard Gaussian $Q(\cdot)$ and the complementary error function $\text{erfc}(\cdot)$~\cite{abramowitz1964handbook}. 
With $\mu$ and $\sigma^2$ denoting sample mean and sample variance of entries in $D_{user}$, and $Q(\alpha)=\frac{1}{2}\text{erfc}(\frac{\alpha}{\sqrt{2}})$ and $\text{erfc}(z) = \frac{2}{\sqrt{\pi}}\int_{z}^\infty e^{-t^2}\text{d} t$, we choose $\alpha$ such that $2Q(\alpha) \leq \gamma$, which gives the minimum value of $\alpha$ as: 
%}
\begin{align}
\alpha &= \sqrt{2}~\text{erfc}^{-1}(\gamma). \label{DLarge}
\end{align}
%where $Q(\alpha)=\frac{1}{2}\text{erfc}(\frac{\alpha}{\sqrt{2}})$ and $\text{erfc}(z) = \frac{2}{\sqrt{\pi}}\int_{z}^\infty e^{-t^2}\text{d} t$. 

%If we have a sufficiently large data set $D_{user}$ to fit the Gaussian distribution $\mathcal{N}(\mu,\sigma^2)$, then parameter $\alpha$ should be chosen such that $2Q(\alpha)\leq \gamma$, where $Q(\alpha)$ is the Q-function defined as $Q(\alpha)=\frac{1}{2}\text{erfc}(\frac{\alpha}{\sqrt{2}})$ with $\text{erfc}(z) = \frac{2}{\sqrt{\pi}}\int_{z}^\infty e^{-t^2}\text{d} t$ being the complementary error function \cite{abramowitz1964handbook}.Hence, parameter $\alpha$ can be computed as $\alpha = \sqrt{2}\text{erfc}^{-1}(\gamma)$ for any user chosen tolerable false positive rate $\gamma$. 

%\textcolor{blue}{
If the size of data set $D_{user}$ (denoted $N$) is quite small, then the sample mean $\mu$ can be estimated using a $t$-distribution with $\nu=N-1$ degrees of freedom~\cite{lange1989robust}. 
For a user-defined $\gamma$, we denote the \emph{critical $t$-value} as $T_{(1-\gamma/2),\nu}$. 
This represents the $(1-\frac{\gamma}{2})$ quantile of the $t$-distribution. 
%Given the user chosen maximum tolerable false positive rate $\gamma$, we denote the critical $t$-value as $T_{1-\gamma/2,\nu}$, which represents the $1-\frac{\gamma}{2}$ cutoff of the $t$-distribution \cite{lange1989robust}. 
%The critical $t$-value can be calculated by using the cutoff function of the $t$-distribution.
In order to satisfy the maximum tolerable false positive rate, the parameter $\alpha$ will need to satisfy~\cite{lange1989robust}: 
%}
\begin{align}
    \alpha\sigma &\geq T_{(1-\gamma/2),\nu}\frac{\mu}{\sqrt{N}} \Rightarrow \alpha =T_{(1-\gamma/2),\nu}\frac{\mu}{\sigma\sqrt{N}}. \label{Dsmall}
\end{align}
% which gives us 
% \begin{align}
%     \alpha &=T_{(1-\gamma/2),\nu}\frac{\mu}{\sigma\sqrt{N}}. \label{Dsmall}
% \end{align}
%\textcolor{blue}{
The $t$-distribution approximates a Gaussian as $N$ becomes large~\cite{lange1989robust}. 
In our experiments, we use the Gaussian in Eqn.~(\ref{DLarge}) when $N>30$, and the $t$-distribution in Eqn.~(\ref{Dsmall}) otherwise, as suggested in \cite{romano2005testing}. 
%}

%\textcolor{blue}{
While Eqns.~(\ref{DLarge}) and (\ref{Dsmall}) provide a mathematical characterization of the threshold on the maximum false positive rate, this threshold can also be empirically determined using ROC curves, which provide a graphical %representation of the 
relationship between true and false positive rates for varying values of $\alpha$ (Figs.~\ref{fig:rocimage} and \ref{fig:ROCGraph} of our evaluations in Sec. \ref{sec:evauations}). 
We also provide an upper bound %to characterize 
on the worst-case false positive rate (false alarm) of MDTD in Appendix A. 

\section{MDTD: Evaluation Results}
\label{sec:evauations}

\begin{table*}[htb]
\caption{This table shows the classification accuracy (Acc.) for clean samples and attack success rate (ASR) of Trojan samples for \textbf{six different Trojan triggers} (Badnets, Blend, Nature, Trojan SQ, Trojan WM and L2 inv- noted by \textbf{bold} column titles) on five image-based datasets- CIFAR100, CIFAR10, GTSRB, SVHN, and Flower102 in the absence of any Trojan detection mechanism. %In the natural backdoor attack, the trigger is trained on the clean model, while for other backdoor attacks, a pre-defined trigger is embedded in Trojan model.
    We observe that the classification accuracy of Trojaned models and clean models is comparable; however, Trojaned models have high values of attack success rate. Note that the ASR value for the clean model is not defined (NA).}
    \label{tab:accImageTask}
    \centering
   
    \scalebox{0.85}{
    \begin{tabular}{|c|c|c|c|c|c|c|c|c|c|c|c|c|c|c|}
    %'natural','blend', 'nature', 'l2_inv', 'trojan_sq', 'trojan_wm','BadNets'
    \hline
         Datasets & \multicolumn{2}{c|}{Clean} & \multicolumn{2}{c|}{\textbf{BadNets} } &\multicolumn{2}{c|}{ \textbf{Blend}} &  \multicolumn{2}{c|}{\textbf{Nature}}  & \multicolumn{2}{c|}{\textbf{Trojan SQ}} & \multicolumn{2}{c|}{\textbf{Trojan WM}} &\multicolumn{2}{c|}{\textbf{L2 inv}} \\ \cline{2-15} 
         & Acc. & ASR & Acc. & ASR & Acc. & ASR & Acc. & ASR & Acc. & ASR & Acc. & ASR & Acc. & ASR\\ \hline
          CIFAR100& $55.69\%$ & NA & $53.01\%$ &  $95.51\%$ & $52.30\%$ &  $99.99\%$ & $53.88\%$ &  $100\%$ & $53.71\%$ &  $100\%$ & $53.8\%$ &  $100\%$ & $51.96\%$ &  $99.98\%$ \\ \hline %& $55.69$ &  $100$ 
CIFAR10& $82.57\%$ & NA  & $81.18\%$ &  $97.4\%$ & $81.11\%$ &  $99.95\%$ & $81.52\%$ &  $99.99\%$ & $81.69\%$ &  $100\%$ & $81.63\%$ &  $100\%$ & $81.46\%$ &  $99.95\%$ \\ \hline %  & $82.57$ &  $100$
GTSRB& $88.57\%$ & NA  & $84.19\%$ &  $91.91\%$ & $88.5\%$ &  $95.45\%$ & $87.41\%$ &  $99.98\%$ & $88.31\%$ &  $99.03\%$ & $85.24\%$ &  $99.76\%$ & $87.89\%$ &  $90.06\%$ \\ \hline %88.55 &  99.28
SVHN& $89.63\%$& NA  & $89.46\%$ &  $95.59\%$ & $89.79\%$ &  $99.28\%$ & $90.44\%$ &  $99.92\%$ & $90.26\%$ &  $99.72\%$ & $90.48\%$ &  $99.84\%$ & $91.36\%$ &  $99.69\%$ \\ \hline %$89.64\%$ &  $93.86\%$
Flower102& $50.59\%$ & NA  & $47.25\%$&  $89.61\%$ & $46.18\%$ &  $99.12\%$ & $46.37\%$ &  $100\% $& $46.67\%$ &  $100\%$ & $48.14\%$&  $100\%$ & $44.71\%$&  $97.16\%$ \\ \hline
         
    \end{tabular}
    
    }
    
\end{table*}

%\textcolor{blue}{
In this section we evaluate and compare \emph{MDTD} against four SOTA Trojan detection methods on image, graph, and audio-based input datasets. 
Our evaluation of MDTD uses exact network structures, parameters, training algorithms reported in the literature~\cite{gao2019strip, hamilton2017, tran2018samplefilter1, xi2021, MNTD}. 
%}
%datasets with varied input domains. 
%We carry out classification tasks on image and graph datasets, and a summarization task for a text dataset. 
For each case, we provide a brief overview of the datasets, describe our experimental setups, and present our results. 
We use metrics introduced in Sec. \ref{subsec:metrics} to evaluate the performance of MDTD. 
Our code is available at https://github.com/rajabia/MDTD. 
%We will make our code available when submitting the final version.% on this paper. 
%We test our approach on image and graph datasets for classification task and text dataset for summarization task with different types of Trojan. In each section, we describes our dataset and experimental setups first and then present our results.

\subsection{Image Inputs}
%In this section, We first introduce the image datasets. Then we describe the DNN classifiers used for evaluations and later we validate our method. 

\noindent{\bf Datasets:} 
We consider the following five datasets: CIFAR10~\cite{krizhevsky2009CIFARs}, CIFAR100~\cite{krizhevsky2009CIFARs}, SVHN~\cite{svhn}, GTSRB~\cite{GTSRB013}, and Flower102~\cite{Flower102}. 
The CIFAR10 and CIFAR100 datasets each consist of $60000$ color images that belong to one of $10$ or $100$ classes respectively. 
SVHN contains $600000$ images of house numbers obtained from Google Street View. 
GTSRB is a dataset containing $52000$ images of traffic signs, and Flower102 contains images of $102$ common flowers in the UK. 
In all our experiments, we use an image resolution of $32\times32$, and partition the dataset into $80\%$ for training and $20\%$ for test. 
For each dataset, we train one clean and six Trojan models with different trigger types (see Fig.~\ref{fig:TrojanSamples}). 
%We assume that the desired output class of the DNN model for an adversary carrying out a backdoor attack is $y^d = 6$. 
We experimentally verified that classification accuracy and attack success rate was not affected by the adversary's choice of target class. 
Consequently, we set the adversary-desired target class when carrying out a backdoor attack as $y^d = 6$. 

\noindent{\bf DNN Structure:} 
%\textcolor{blue}{
For CIFAR100, CIFAR10, and Flowers102, we used WideResnet DNNs~\cite{Zagoruyko2016WRN}. For GTSRB and SVHN, we used a DNN with 4 convolutional layers with kernel sizes $32,32,64$ and $64$ respectively followed by a maxpooling layer and a fully connected layer of size $4096$. 
We train models for 100 epochs with batch size of $64$ using a stochastic gradient descent (SGD) optimizer. 
We tuned the model and set the learning rate to $0.001$ and momentum to $0.9$.
%}
% For training robust DNNs, we use the robust learning technique proposed in~\cite{madry2017towards}. We set the step size to $ 0.00784$ and examined two different noise levels ($\epsilon=0.01, 0.1$). We used identical model structures, parameters, and setup reported in~\cite{hamilton2020} for our experiments on GNNs for graph-based inputs. 
%We use resolution of $32\times32$ for all datasets. For each dataset, we train a clean and 6 Trojan models using different trigger types with a randomly selected target class ($y_t=6$) as the desired output of an adversary. Our experiments demonstrate that the choice of target class does not affect classification accuracy or attack success rate.  The classification accuracy on clean sample (in test set) and  attack success rate for Trojan of Trojan models used for our evaluations are presented in Table~\ref{tab:accImageTask}.   

% clean CIFAR10 classifier has the classification accuracy of $???\%$ on clean samples.
% For example, Trojan model with embedded nature trigger trained on CIFAR10 has achieved $81.52\%$ classification accuracy on clean samples and $99.99\%$ attack success rate for Trojan samples.
% We describe the detail of our datasets and the   architecture DNN models in Appendix~\ref{sec:AppImage}. The codes of MDTD will be publicly available, after receiving the reviewers' decision.

\noindent{\bf Trojan triggers:} 
We consider six different Trojan triggers that an adversary can embed into image inputs provided to the DNN: %model: 
%For our embedded backdoor attacks, we consider spectrum of the following triggers: 
white colored square (BadNets)~\cite{badnet}, 
image of a coffee mug (Nature)~\cite{chen2017targetedbackdoor}, 
`Hello Kitty' image blended into the background (Blend)~\cite{chen2017targetedbackdoor}, 
multicolored square (Trojan SQ)~\cite{liu2017trojaning}, 
colored circular watermark (Trojan WM)~\cite{liu2017trojaning}, and an 
`invisible' trigger based on an $L2$-regularization of the input image (L2 inv)~\cite{li2019invisible}. 
%\textcolor{blue}{
Our choice of triggers and training methods for Trojan-embedded models follow the SOTA~\cite{gao2019strip, tran2018samplefilter1, MNTD}. 
%}

Table~\ref{tab:accImageTask} compares classification accuracy (Acc.) at test time of clean samples %(input samples without a trigger) 
and attack success rate (ASR) without any defense for samples embedded with six different Trojan triggers. 
We observe that Acc. values of Trojaned models is comparable to a clean model, while simultaneously achieving high ASR.% value. 

\noindent {{\bf Setup:}} 
%\textcolor{magenta}{
Defense against Trojans can be broadly categorized into solutions that (a) modify the supervised training pipeline of DNNs with secure training algorithms, (b) detecting backdoors in DNN models, or (c) detecting and eliminating input samples containing any Trojan trigger. 
The works in \cite{huang2022backdoor, chen2022effective, tran2018samplefilter1, zeng2021rethinking} belong to (a), \cite{wang2019neuralcleans, MNTD} belong to (b), and \cite{zeng2021rethinking, gao2019strip, tran2018samplefilter1, chen2018detecting} belong to (c). 
Our MDTD also belongs to category (c). 
Hence, we evaluate MDTD against four similar SOTA Trojan detection methods that also aim to detect and eliminate input samples containing a trigger: (i) DCT-based~\cite{zeng2021rethinking}, ii) STRIP~\cite{gao2019strip}, (iii) spectral signature~\cite{tran2018samplefilter1}, and (iv) activation clustering~\cite{chen2018detecting}.
%
%Hence, we chose to compare our results of evaluating MDTD with the papers in \cite{zeng2021rethinking, gao2019strip, tran2018samplefilter1, chen2018detecting} since they belong to cateogry (c). 
%
%Defense against Trojans can be broadly categorized into solutions that (a) modify the supervised training pipeline of DNNs with secure training algorithms \cite{huang2022backdoor, chen2022effective}, (b) detecting backdoors in DNN models \cite{wang2019neuralcleans, MNTD}, or (c) detecting and eliminating input samples containing a trigger at inference-time without having to retrain the DNN \cite{zeng2021rethinking, gao2019strip, tran2018samplefilter1, chen2018detecting}. 
% Our MDTD belongs to solution category (c). Hence, we evaluate MDTD against four similar SOTA Trojan detection methods that also aim to detect and eliminate input samples containing a trigger at inference time:  (i) DCT-based~\cite{zeng2021rethinking}, ii) STRIP~\cite{gao2019strip}, (iii) spectral signature~\cite{tran2018samplefilter1}, and (iv) activation clustering~\cite{chen2018detecting}.
%}
%
% We evaluate \emph{MDTD} against four SOTA Trojan input detection methods 
% (i) DCT-based detector~\cite{zeng2021rethinking}, (ii) STRIP~\cite{gao2019strip}, (iii) spectral signature~\cite{tran2018samplefilter1}, and (iv) activation clustering~\cite{chen2018detecting}. 
% \textcolor{blue}{
% We choose these four methods since, similar to our MDTD, they all aim to detect triggers in inputs given to a DNN model (in contrast to Trojan detection by analyzing DNN models, such as in \cite{wang2019neuralcleans, MNTD}). 
% }
We %briefly 
describe each detection method below:
%Both these methods were developed to detect Trojan input samples without any information on attacker strategy. 
%Moreover, we use STRIP in the first step of MDTD to measure the robustness of an input to noise, which has led to significant improvement in $F_1$-score in comparison to the original variants of STRIP.
%In the following we describe our settings for each method: 

% For STRIP, we blend each input with $20$ randomly selected clean samples, and consider it as Trojan input if the classifier returns the same outputs of the original input for at least for $10$ blended variants of inputs. 

\noindent{\underline{\it DCT-based detector:}} 
This method uses the discrete cosine transform (DCT) to analyze different frequencies present in an image. 
The authors of~\cite{zeng2021rethinking} showed that clean and Trojan samples consist of signals of different frequencies, which could be used to effectively distinguish between them. 
%\textcolor{blue}{
We follow experiment settings suggested in~\cite{zeng2021rethinking}- we use the complete set of training samples for each dataset and perturb clean samples by adding the %BadNets 
Trojan trigger and Gaussian random noise. 
However, DCT-based detection %is known to require 
requires 
using the entire training set~\cite{zeng2021rethinking}, which is computationally expensive. 
%}

\noindent{\underline{\it STRIP:}} The authors of~\cite{gao2019strip} demonstrated that inputs containing a Trojan trigger were more robust to noise than clean inputs. Therefore, DNN classifiers will be less likely to change their decisions when these inputs are `mixed' with other clean samples. 
We follow the setup from~\cite{gao2019strip} in our experiments. 
We select $20$ clean images at random, and `mix' these with each input sample before providing it to the DNN classifier. 
An input is considered Trojan if the classifier returns the same output for at least $10$ of the `mixed' variants of the input, and is considered clean otherwise. 

\noindent{\underline{\it Spectral Signature~\cite{tran2018samplefilter1}, Activation Clustering~\cite{chen2018detecting}:}} 
Spectral signature methods~\cite{tran2018samplefilter1} use an insight that clean and Trojan samples have different values of covariances of features learned by the DNN model. Activation clustering leverages differences between clean and Trojaned samples can be characterized in terms of the values of DNN `activations', which represent how the model made a classification decision for a given input sample.  
Both methods take a set of samples (Trojan and clean) as input and partition the set into clean and suspicious clusters using clustering and feature reduction techniques (e.g., PCA, FastICA). 
The goal of these methods is to detect and eliminate Trojan samples from the training set in order to prevent embedding of a trigger during model training. %These methods might mistakenly eliminate some clean samples in training set which might not affect significantly on the accuracy of final model, however it makes these methods unsuitable for Trojan detection during inference time. 

%This method found that the Trojan inputs have different frequencies in their DCT domains. Therefore, a DNN detector can be trained on the DCT domain of  a set of  Trojan and clean samples to detect unseen Trojan samples. For training a DCT-based detector, we use the whole training set and create a set of perturbed samples by adding the BadNets Trojan trigger and Gaussian random noise to clean samples as suggested in~\cite{zeng2021rethinking}.  This method found that it can detect Trojan samples containing unseen triggers. A DCT-based detector requires being trained on the whole training set, which is computationally as expensive as training the main task classifier.

\noindent{\underline{\it MDTD: (\textbf{OURS})}} 
MDTD uses estimates of distances to the decision boundary in order to distinguish between clean and Trojan samples. 
We consider cases when the user has \emph{white-box} and \emph{black-box} access to the DNN model. 
In the white-box setting, MDTD uses FGSM and IFGSM~\cite{goodfellow2014FGS} adversarial learning methods to compute the minimum magnitude of noise $\delta$ required to misclassify a sample. 
In the black-box setting, MDTD uses the HopSkipJump adversarial learning method~\cite{chen2020hopskipjumpattack} to estimate distances ($\delta$) to the decision boundary only based on outputs of the DNN model. 
Since the user does not have information about the Trojan trigger or target output class, MDTD uses a set of $500$ clean samples randomly selected from the training set to determine a threshold distance to the decision boundary. 
A sample is identified as Trojan if %the estimated distance 
$\delta$ is beyond this threshold. 

\begin{table}[]
    \caption{This table reports $F_1$-scores for spectral signature~\cite{tran2018samplefilter1}, activation clustering~\cite{chen2018detecting} DCT-based detectors~\cite{zeng2021rethinking}, STRIP~\cite{gao2019strip}, and MDTD (\textbf{ours}) for \textbf{six different Trojan triggers} on five image-based datasets. 
MDTD obtains the highest $F_1$-score in most cases (\textbf{bold} numbers), indicating high true positive rates and low false positive rates. 
%$F_1$-scores are higher for variants of MDTD that have white-box model access (MDTD (FGSM), MDTD (IFGSM)) compared to when only black-box access is available (MDTD (HSJ)). 
%The lower $F_1$-scores of MDTD for the SVHN and GTSRB datasets in some cases is because the certified radii of clean and Trojaned samples are comparable in these cases (see Table~\ref{tab:certifiedR}). 
Spectral signature and activation clustering methods have low $F_1$-scores due to high false positive rates. 
%\textcolor{blue}{
DCT-based detectors show high variation in $F_1$-score values depending on the number of frequency components contained in input image samples, which restricts its applicability. 
%}
% Large $F_1$-scores for DCT-based detectors for the SVHN dataset is because images in this dataset contain only a few colors, which makes their frequency components easily separable. This results in large true positive and small false positive rates. 
% The lower $F_1$-scores of MDTD for the SVHN and GTSRB datasets in some cases is because the certified radii of clean and Trojaned samples (see Table~\ref{tab:certifiedR}) are comparable in these cases. Lower $F_1$-score values for STRIP and DCT (in some cases) is due to large false positive rates. 
\emph{We show %values of 
true/ false positives in \textbf{Appendix B}}.
  %  
%     For datasets that have high certified radius for their clean models such as SVHN and GTSRB, DCT-based detector works well. In general, 
%     STRIP and DCT-based usually have lower $F
% _1$-score compared to MDTD, since they discard many clean samples (high false positive rates), even though  their Trojan sample detection rate is  high. MDTD returns better $F_1$-score since it obtains similar detection rate with lower false positive rates. See Figures~\ref{fig:tprImages} and~\ref{fig:fprImages} for more details. 
}
    \label{tab:imagesresults} 
    \scalebox{0.75}{
    \begin{tabular}{|c|c|c|c|c|c|c|c|c|}
    \hline
          & Trojan &  Spec & AC & DCT & STRIP & \shortstack{MDTD\\ (FGSM)}&  \shortstack{MDTD \\(IFGSM)}&  \shortstack{MDTD \\(HSJ)} \\ \hline
         \multirow{6}{*}{\rotatebox{90}{CIFAR100}} & BadNets & $0.303$ & $0.039$& $0.88$   & $0.94$& $\mathbf{0.95}$&  $0.94$ &  $0.86$ \\ \cline{2-9}
                                   & Blend   &  $0.302$ & $0.0001$&  $0.93$    & $0.95$ &  $0.97$ &  $\mathbf{0.98}$ &  $0.96$ \\ \cline{2-9}
                                    & Nature & $0.301$ & $0.20$ &  $0.93$   & $0.95$ &  $\mathbf{0.98}$ &  $\mathbf{0.98}$ &  $\mathbf{0.98}$  \\ \cline{2-9}
                                    & Trojan SQ & $0.302$ & $0.19$&  $0.93$ & $0.96$ &  $\mathbf{0.98}$ &  ${0.97}$ &  ${0.97}$  \\ \cline{2-9}
                                    & Trojan WM & $0.303$ & $0.26$&  $0.94$  & $0.95$  & $0.96$ &  $\mathbf{0.98} $ &  $0.8$ \\ \cline{2-9}
                                    & L2 inv & $0.302$ & $0.0001$& $0.94 $ & $0.95$ & $\mathbf{0.98}$ &  $0.97$ &  $0.97$ \\ \hline
                                    
                                    \hline
                                    
            \multirow{6}{*}{\rotatebox{90}{CIFAR10}} & BadNets   & $0.316$ & $0.348$&  $0.69 $   & $0.81$ &  $0.78$ &  $0.77$ &  $\mathbf{0.82}$ \\ \cline{2-9}
                                    & Blend   & $0.316$ & $0.354$ &  $0.70$   & $0.84$  &  $\mathbf{0.92}$ &  $0.91$ &  $0.84$ \\ \cline{2-9}
                                    & Nature   & $0.314$ & $0.346$ &  $0.7$    & $0.81$ & $\mathbf{0.94}$ &  $0.92$ &  $0.91$ \\ \cline{2-9}
                                    & Trojan SQ   & $0.315$ & $0.353$ &  $0.7$ & $0.84$ & $0.91$ &  $\mathbf{0.93} $&  $0.9$ \\ \cline{2-9}
                                    & Trojan WM   & $0.315$ & $0.0002$ & $0.7$  & $0.83$  & $0.91$ &  $0.91$ &  $\mathbf{0.93}$ \\ \cline{2-9}
                                    & L2 inv   & $0.316$ & $0.361$ &  $0.7$  & $0.84$  &  $\mathbf{0.91}$ &  $\mathbf{0.91}$ & $ 0.85$ \\ \hline
                                    
                                    \hline

              \multirow{6}{*}{\rotatebox{90}{GTSRB}} & BadNets   &  $0.304$& $0.345$& $0.83$   & $0.79$  &  $0.88$ & $ \mathbf{0.9}$ &  $0.82$  \\ \cline{2-9}
                                    & Blend   & $0.304$ & $0.335$ &  $0.71$  & $0.74$ &  $0.75$ &  $\mathbf{0.83} $&  $0.76$ \\ \cline{2-9}
                                    & Nature   & $0.304$ & $0.334$ & $0.84$  & $0.76$ &  $0.76$ &  $\mathbf{0.9}$ &  $0.76$ \\ \cline{2-9}
                                    & Trojan SQ   & $0.299$ & $0.339$ & $\mathbf{0.94}$ & $0.79$  &  $0.8$ &  $0.89$ &  $0.79$ \\ \cline{2-9}
                                    & Trojan WM   & $0.303$ & $0.334$ &  $\mathbf{0.99}$ & $0.76$  &  $0.87$&  $0.92$ &  $0.91$   \\ \cline{2-9}
                                    & L2 inv   & $0.303$ & $0.342$ &  $\mathbf{0.93}$ & $0.74$  &  $0.7$ &  $0.74$ &  $0.12$ \\ \hline
                                    
             \hline                       
             \multirow{6}{*}{\rotatebox{90}{SVHN}} & BadNets & $0.322$ & $0.364$ & $\mathbf{0.99} $  &   $ 0.87$   & $0.8$ &  $0.87$ &  $0.82 $\\ \cline{2-9}
                                    & Blend   & $0.323$ & $0.111$ &  $\mathbf{0.99}$  & $0.76$  &  $0.76$ &  $0.89$ &  $0.83$ \\ \cline{2-9}
                                    & Nature   & $0.323$  & $0.043$& $\mathbf{0.99}$ & $0.76$ & $0.78$ &  $0.94$ &  $0.82$ \\ \cline{2-9}
                                    & Trojan SQ   & $0.323$ & $0.06$ & $\mathbf{0.99}$  & $0.74$  & $0.71$ &  $0.91$ &  $0.82$ \\ \cline{2-9}
                                    & Trojan WM   & $0.323$  & $0.04$ & $\mathbf{0.99}$  & $0.76$  &  $0.83$&  $0.93$&  $0.89$  \\ \cline{2-9}
                                    & L2 inv& $0.323$ & $0.137$& $\mathbf{0.99}$ & $0.63$ & $ 0.87$ &  $0.86$ &  $0.91$ \\ \hline

            \hline
            \multirow{6}{*}{\rotatebox{90}{Flower102}} & BadNets   &  $0.323$& $0.347$ &  $0$  & $\mathbf{0.48}$&  $0.09$ &  $0.08$ &  $0.5$ \\ \cline{2-9}
                                    & Blend   & $0.305$ & $0.351$ &  $0$  &  $0.45$ &  $0.78$ &  $\mathbf{0.84}$ &  $0.65$ \\ \cline{2-9}
                                    & Nature   & $0.305$ & $0.456$& $0$  &  $0.45$  &   $0.89$ &  $\mathbf{0.93}$ &  $0.9$  \\ \cline{2-9}
                                    & Trojan SQ    & $0.305$ & $0.35$ &  $0$  & $0.88$ & $\mathbf{0.93}$ &  $0.92$ &  $0.92$ \\ \cline{2-9}
                                    & Trojan WM   & $0.305$ & $0.352$ & $0$ & $0.75$  & $\mathbf{0.92}$ &  $\mathbf{0.92}$ &  $\mathbf{0.92}$ \\ \cline{2-9}
                                    & L2 inv    & $0.305$ & $0.347$ &  $0$ &  $\mathbf{0.79}$  & $0.4$ &  $0.7$ & $0.03$ \\ \hline       
         
    \end{tabular}}
    % \vspace{0.1 inch}

\end{table}

\begin{figure*}
    \centering
    \begin{tabular}{c c c c }
         & MDTD (FGSM) & MDTD (IFGSM) &MDTD (HopSkipJump)  \\
         \rotatebox{90}{$\:\:\:\:\:\:\:\:\:\:\:\:\:$ CIFAR100}& 
         \includegraphics[scale=0.35, trim={1cm 0cm 1cm 0cm}]{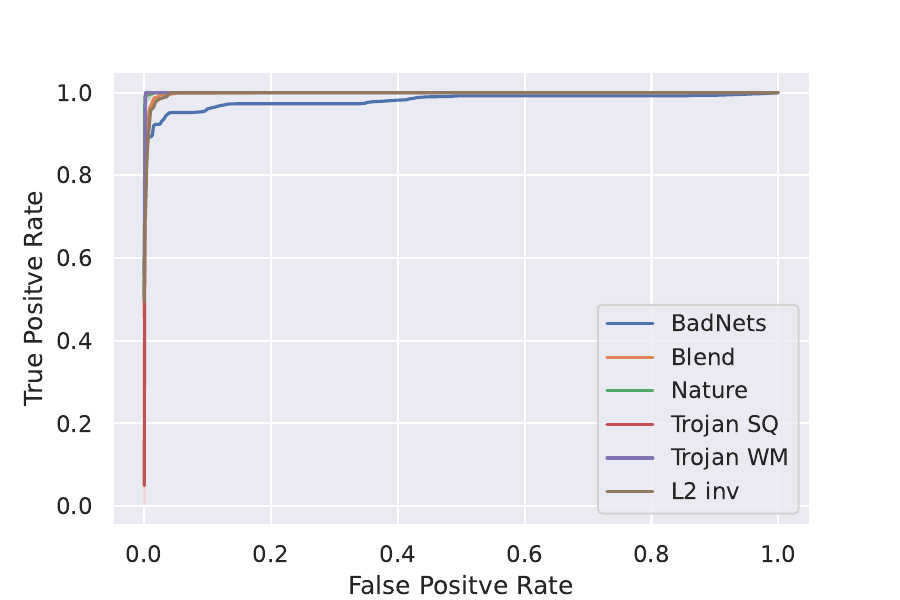}&
         \includegraphics[scale=0.35, trim={1cm 0cm 1cm 0cm}]{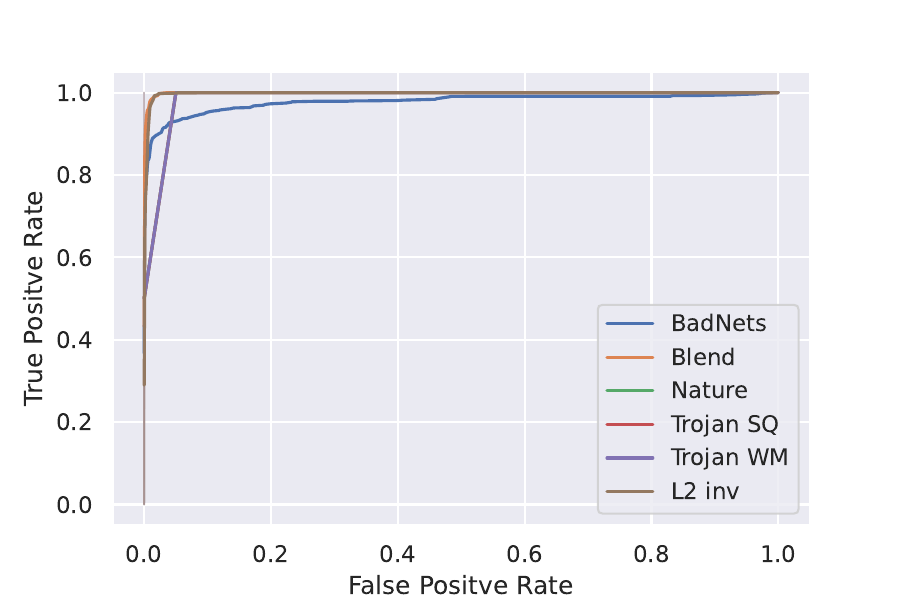}&
         \includegraphics[scale=0.35, trim={1cm 0cm 1cm 0cm}]{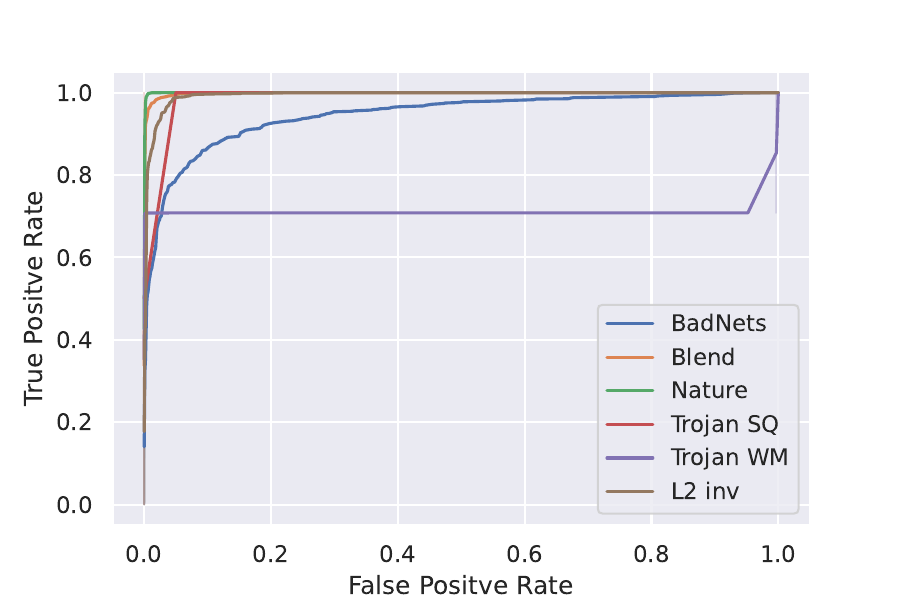}\\

         \rotatebox{90}{$\:\:\:\:\:\:\:\:\:\:\:\:\:$ CIFAR10}& 
         \includegraphics[scale=0.35, trim={1cm 0cm 1cm 0cm}]{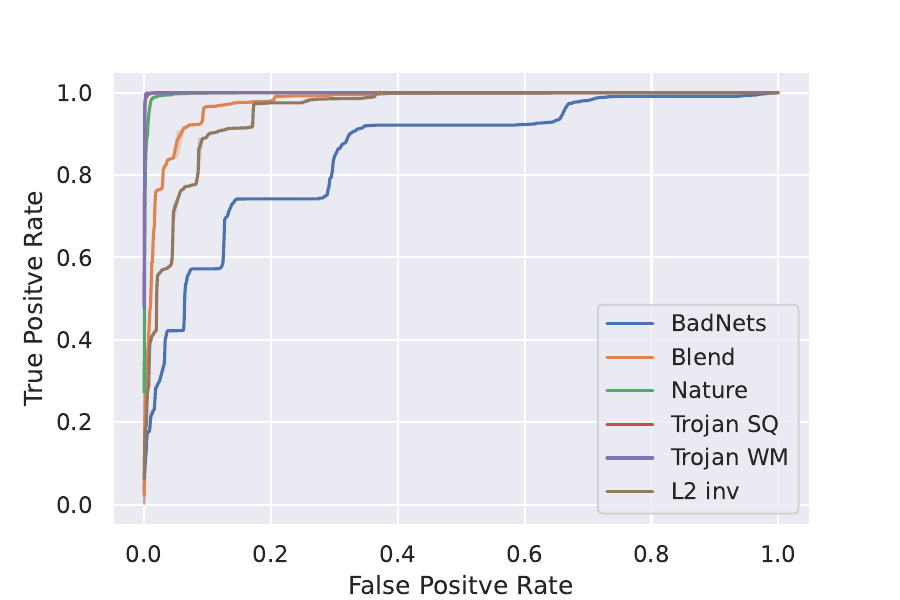}&
         \includegraphics[scale=0.35, trim={1cm 0cm 1cm 0cm}]{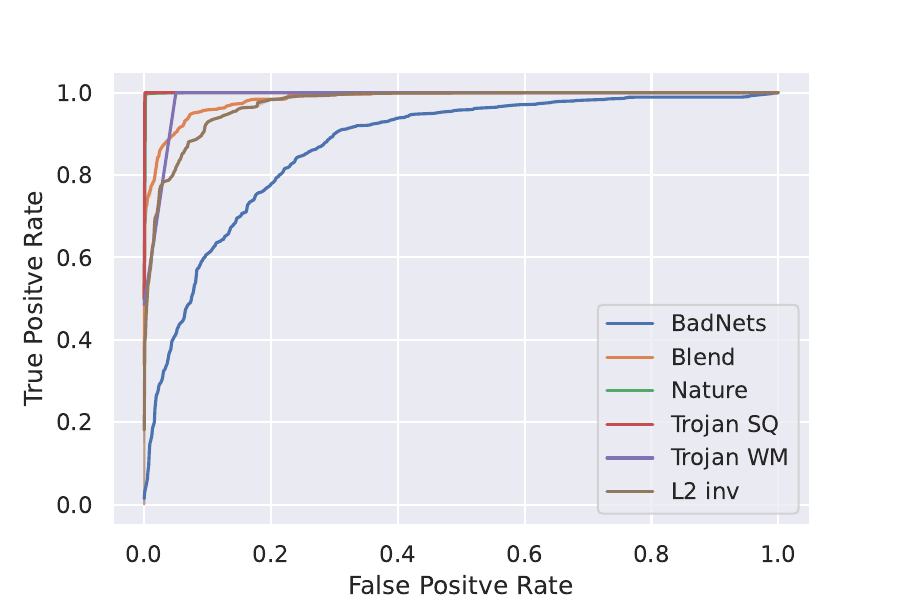}&
         \includegraphics[scale=0.35, trim={1cm 0cm 1cm 0cm}]{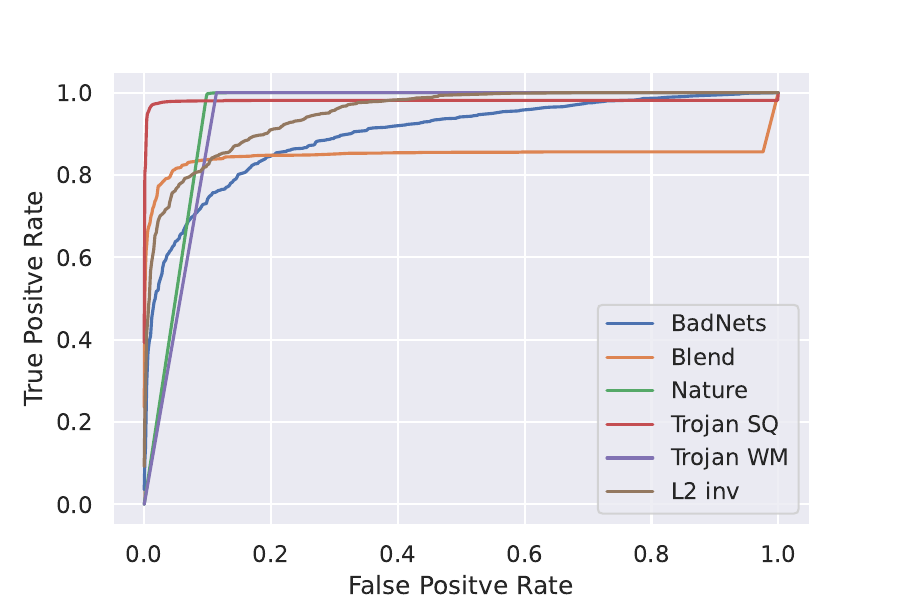}\\

         \rotatebox{90}{$\:\:\:\:\:\:\:\:\:\:\:\:\:$ GTSRB}& 
         \includegraphics[scale=0.35, trim={1cm 0cm 1cm 0cm}]{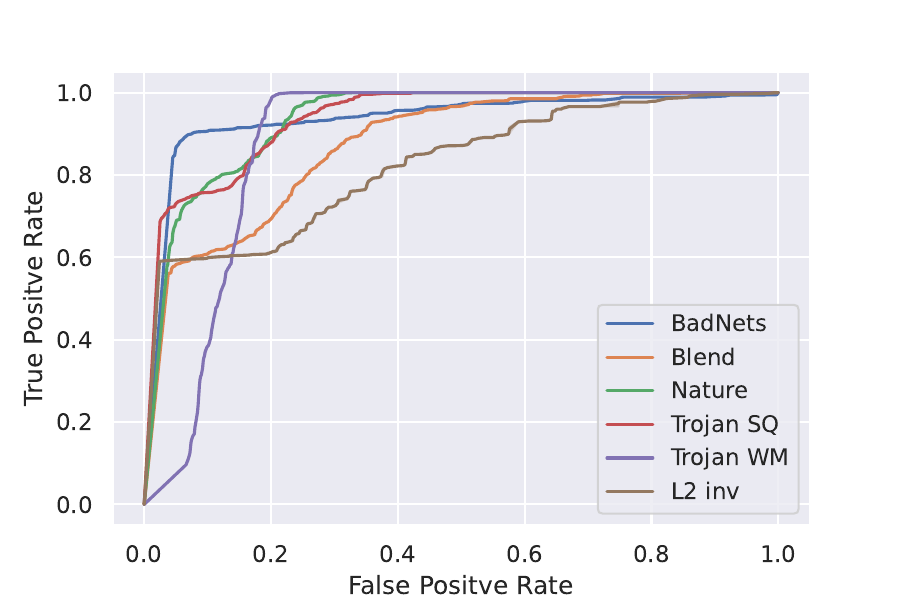}&
         \includegraphics[scale=0.35, trim={1cm 0cm 1cm 0cm}]{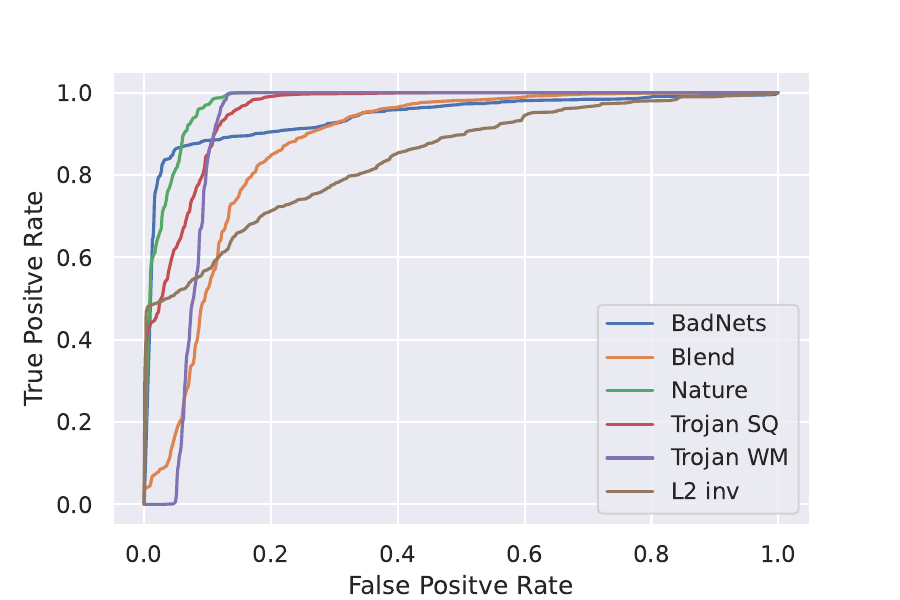}&
         \includegraphics[scale=0.35, trim={1cm 0cm 1cm 0cm}]{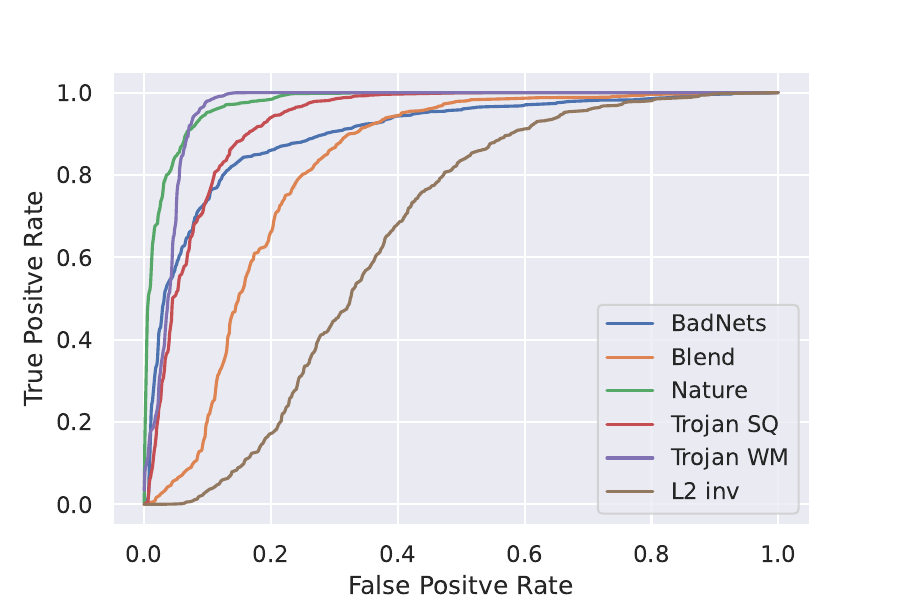}\\

         \rotatebox{90}{$\:\:\:\:\:\:\:\:\:\:\:\:\:\:\:\:$ SVHN}& 
         \includegraphics[scale=0.35, trim={1cm 0cm 1cm 0cm}]{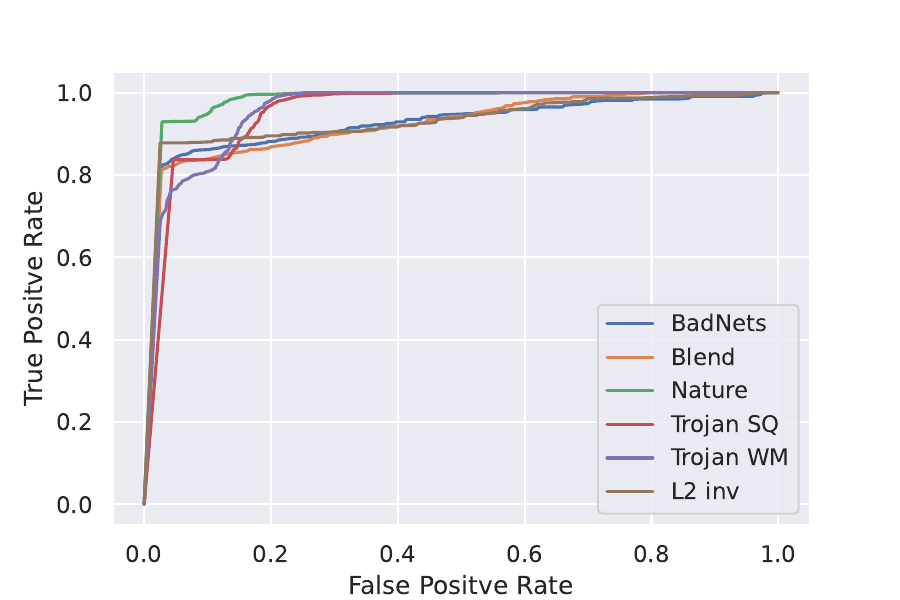}&
         \includegraphics[scale=0.35, trim={1cm 0cm 1cm 0cm}]{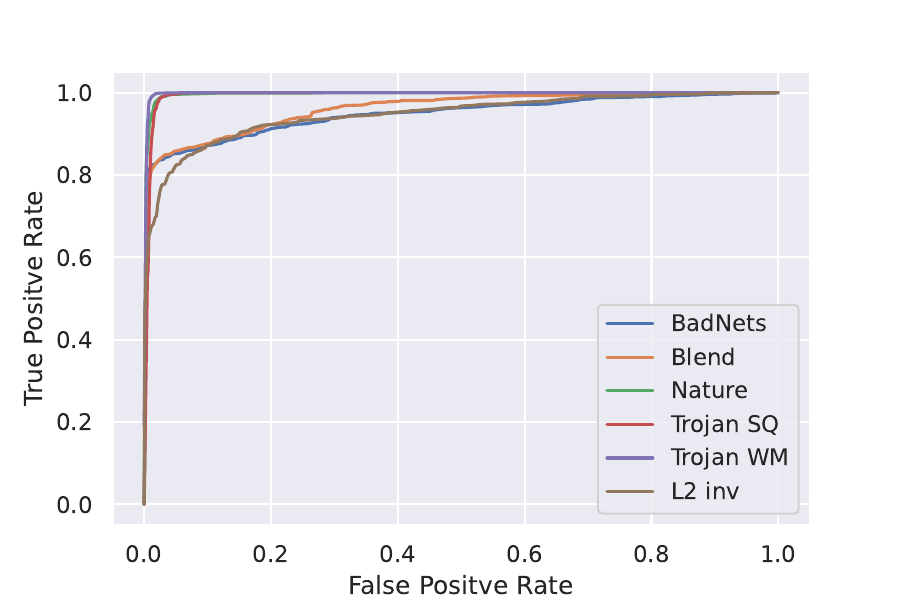}&
         \includegraphics[scale=0.35, trim={1cm 0cm 1cm 0cm}]{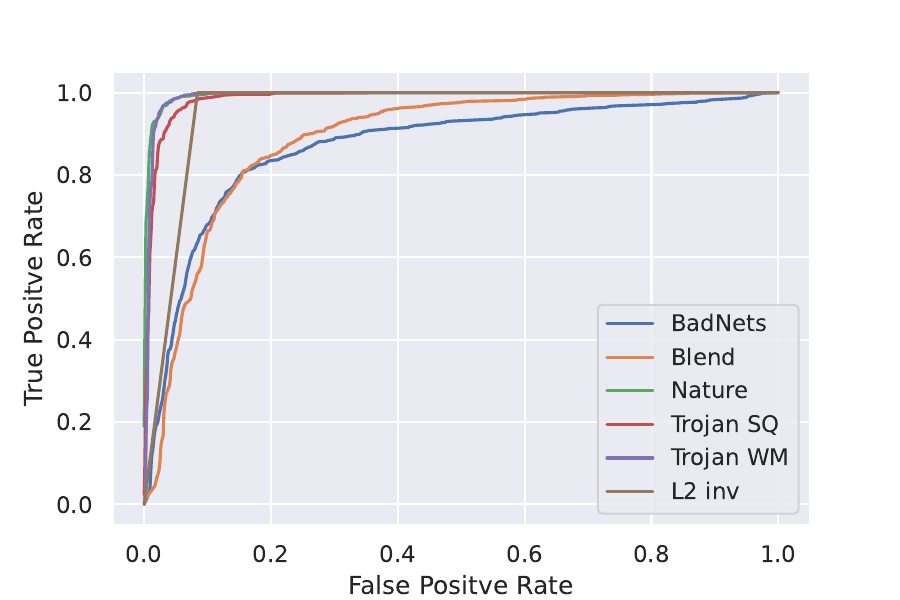}\\

         \rotatebox{90}{$\:\:\:\:\:\:\:\:$ Flower102}& 
         \includegraphics[scale=0.35, trim={1cm 0cm 1cm 0cm}]{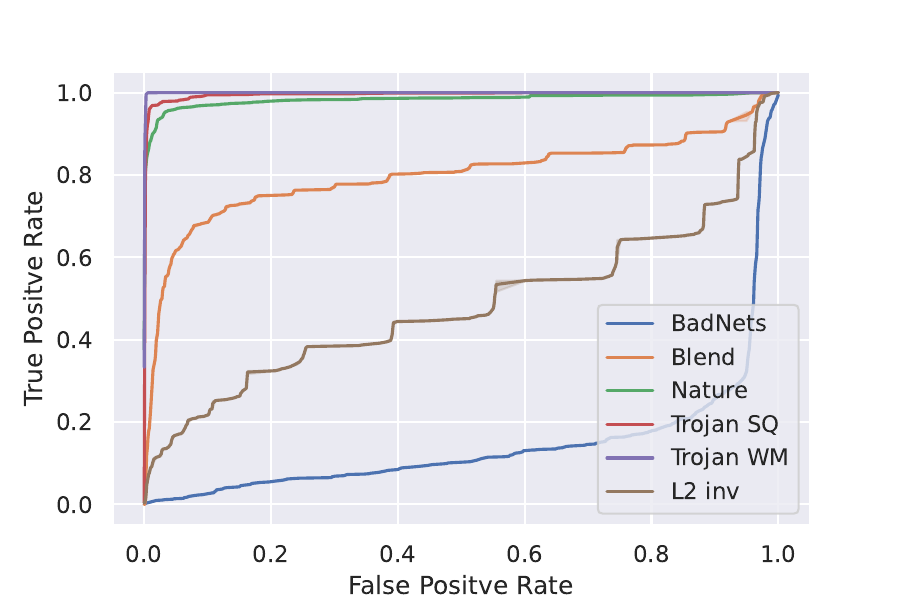}&
         \includegraphics[scale=0.35, trim={1cm 0cm 1cm 0cm}]{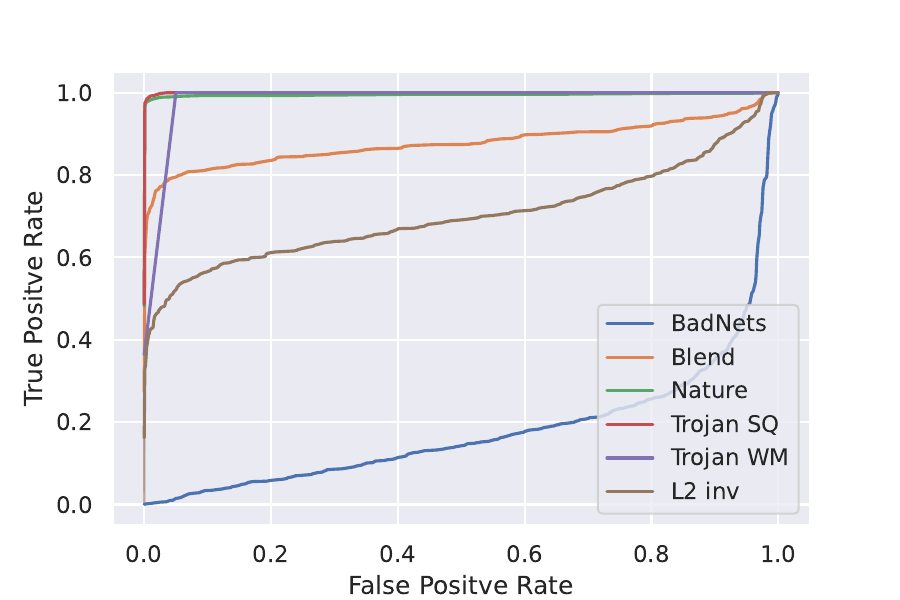}&
         \includegraphics[scale=0.35, trim={1cm 0cm 1cm 0cm}]{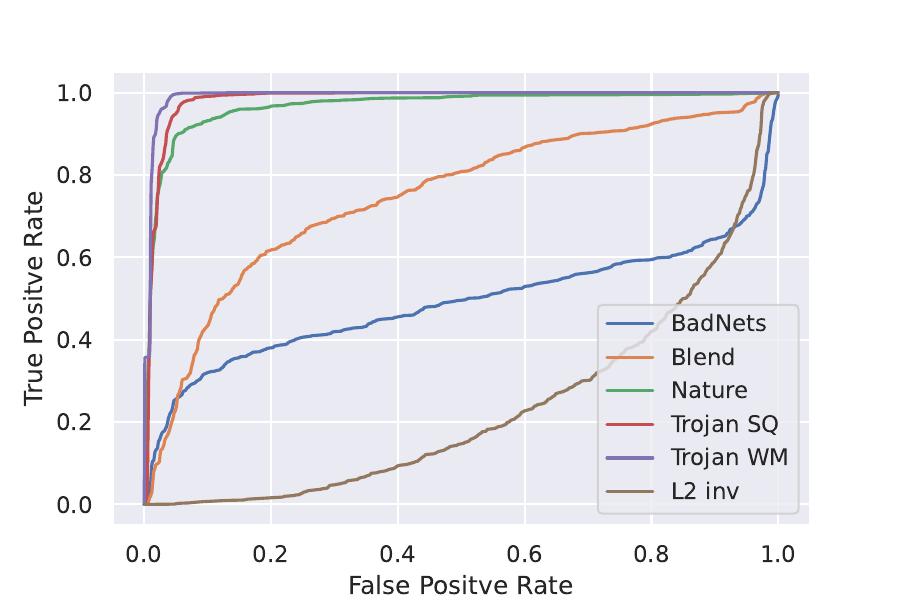}\\

    \end{tabular}
    \caption{This figure plots ROC curves showing change in accuracy of Trojan sample detection (True positive) with the change in the maximum tolerable false positive rate $
    \alpha$ for \emph{MDTD} using FGSM, IFGSM, and HopSkipJump adversarial learning methods for \textbf{six different types of Trojan triggers}- Badnets, Blend, Nature, Trojan SQ, Trojan WM, and L2 inv- for \textbf{five image-based datasets}- CIFAR100, CIFAR10, GTSRB, SVHN, and Flower102. 
    In each case, we observe that the threshold $\alpha$ for the false positive rate plays a critical role in determining values of the true positive rate. 
    Low values of true positive rates despite higher thresholds $\alpha$ when samples in the Flowers102 dataset are embedded with a BadNets (white square) or L2 inv Trojan triggers could be because (uniformly) selected clean input samples in this dataset includes white-colored flowers. 
    %This observation is also exemplified in the last row of Table 1 where values of the average certified radius for clean samples in the Flowers102 dataset and Trojan samples with the BadNets and L2 inv triggers are comparable and their variance is high. 
    %
    %In each case, we see that the true positive rate increases as the threshold $\alpha$ for the false positive rate is increased. 
    % This figure indicates that among all adversarial learning methods, IFGSM provides better distinguishability between Trojan and clean samples and natural backdoor is the most difficult backdoor attack to detect.
    % Note that MDTD chooses a threshold for minimum required adversarial noise for misclassification which affects both true positive and false positive rates. Higher thresholds increase the true positive rate (Trojan detection accuracy) at the cost of increasing the false positive rate. For example for MDTD (IFGSM), using a small threshold can detect Trojan samples perturbed by an embedded trigger with low positive rate, but  it has  low detection rate for  natural Trojan. 
    }
    \label{fig:rocimage}
    
\end{figure*}

\noindent{\bf Evaluating MDTD:}
Table~\ref{tab:imagesresults} shows the $F_1$-scores obtained for images embedded with different types of Trojan triggers when using spectral signature~\cite{tran2018samplefilter1}, activation clustering~\cite{chen2018detecting}, a DCT-based detector~\cite{zeng2021rethinking}, STRIP~\cite{gao2019strip}, and \emph{MDTD}. 
Recall from Eqn. (\ref{eqn:F1score}) that the $F_1$-score is defined as $F_1= \frac{2*TPR* TNR}{TPR + TNR}$, with $TNR=1-FPR$, where and $TPR$ and $FPR$ denote true and false positive rates. 

From Table~\ref{tab:imagesresults}, we observe that MDTD achieves the highest $F_1$-scores in almost all cases.  
This indicates that MDTD is simultaneously able to achieve high true positive rates (detection) and small false positive rates (false alarm). 
MDTD obtains a lower $F_1$-score for only $2$ pairs of cases (Badnets and L2 inv Trojan triggers for Flower102 dataset); we elaborate on these cases in Sec. \ref{sec:discussion}. 

%A lower $F_1$-score of MDTD in a few cases (e.g., Badnets Trojan trigger in the FLower102 dataset) is due to the fact that the difference in magnitudes of the certified radii for clean and Trojaned samples is smaller for these datasets (see Table~\ref{tab:certifiedR}). 
%$F_1$-scores are typically higher for variants of MDTD that have white-box model access (FGSM, IFGSM) compared to the case when only black-box access (HopSkipJump) to the model is available. 
%We further observe that the $F_1$-score is higher for the variant of MDTD that uses the FGSM or IFGSM adversarial learning methods compared to HopSkipJump. This is because when using HopSkipJump, MDTD has access only to outputs of the model (black-box access). 

%\textcolor{blue}{
Compared to MDTD, SOTA methods that analyze input samples- spectral signature~\cite{tran2018samplefilter1} and activation clustering~\cite{chen2018detecting}- have low $F_1$-scores due to high false positive rates. 
DCT-based detectors, on the other hand, show large variations in $F_1$-scores across different image-based datasets. 
Unsurprisingly, DCT-based detectors have very low false positive rates when the number of frequency components in input images is limited (e.g., input samples from SVHN), but they also have very high false positive rates ($\sim 100\%$) when input image samples contain a large number of frequency components (e.g., samples from Flower102). 
In Appendix B, we report true and false positive rates for each case. 
\noindent{\bf Effect of $\alpha$:} 
Fig.~\ref{fig:rocimage} plots ROC curves showing change in the true positive rate %(Trojan samples being correctly identified as Trojan) 
with varying values of the maximum tolerable false positive rate threshold $\alpha$ for \emph{MDTD} using FGSM, IFGSM, and HopSkipJump adversarial learning methods for six different Trojan triggers for all five image-based input datasets. 
For most datasets, and across Trojan trigger types, MDTD consistently accomplishes high true positive rates for smaller values of $\alpha$. 

\subsection{Graph Inputs}

\begin{figure*}
    \centering
    \begin{tabular}{c c}
    MDTD (FGSM) & MDTD (IFGSM)\\
        \includegraphics[scale=0.23, trim={1cm 0cm 1cm 0cm}]{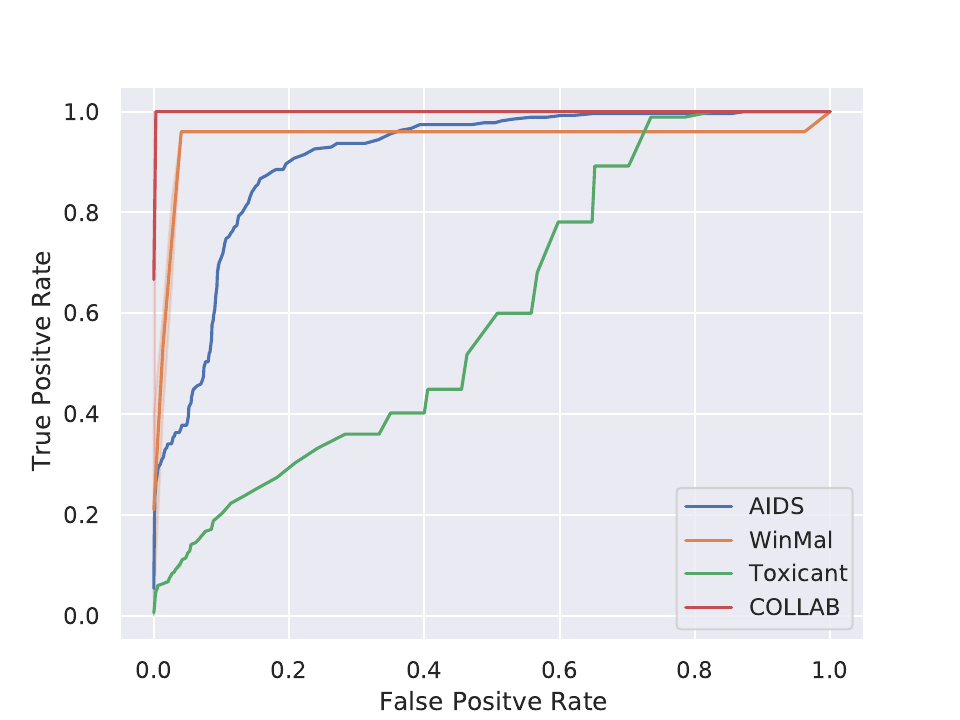} &  
         \includegraphics[scale=0.23, trim={1cm 0cm 1cm 0cm}]{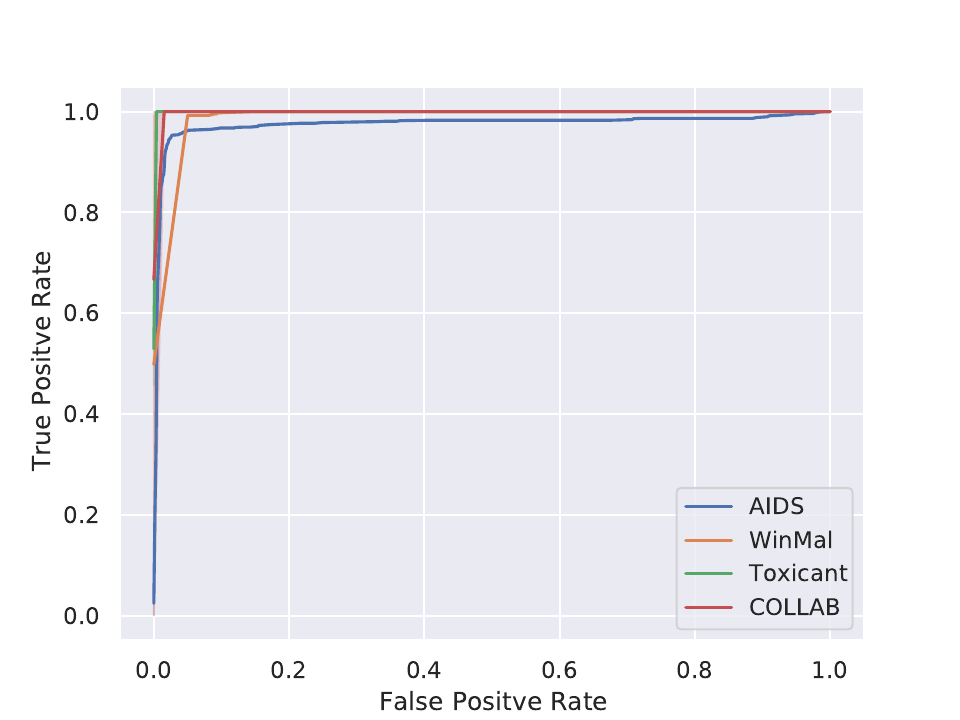}
    \end{tabular}
    \caption{This figure plots ROC curves showing change in accuracy of Trojan sample detection (true positive) with the change in maximum tolerable false positive rate $\alpha$ for MDTD using FGSM and IFGSM adversarial learning methods for \textbf{four graph-based datasets}- AIDS, WinMal, Toxicant, and COLLAB. 
    The threshold $\alpha$ for the false positive rate plays a critical role in determining values of the true positive rate. \\
    %Due to the iterative nature of the IFGSM algorithm, MDTD (IFGSM) is more effective in distinguishing between clean and Trojan samples than MDTD (FGSM).\\  
    %Our ROC curves demonstrate that IFGSM could provide better $\delta$ values for distinguishing clean and Trojan samples.  
    }
    \label{fig:ROCGraph}
\end{figure*}

\begin{figure*}
    \centering
    \begin{tabular}{c c c c }
     \includegraphics[scale=0.08, trim={4cm 2.5cm 4cm 4.5cm}]{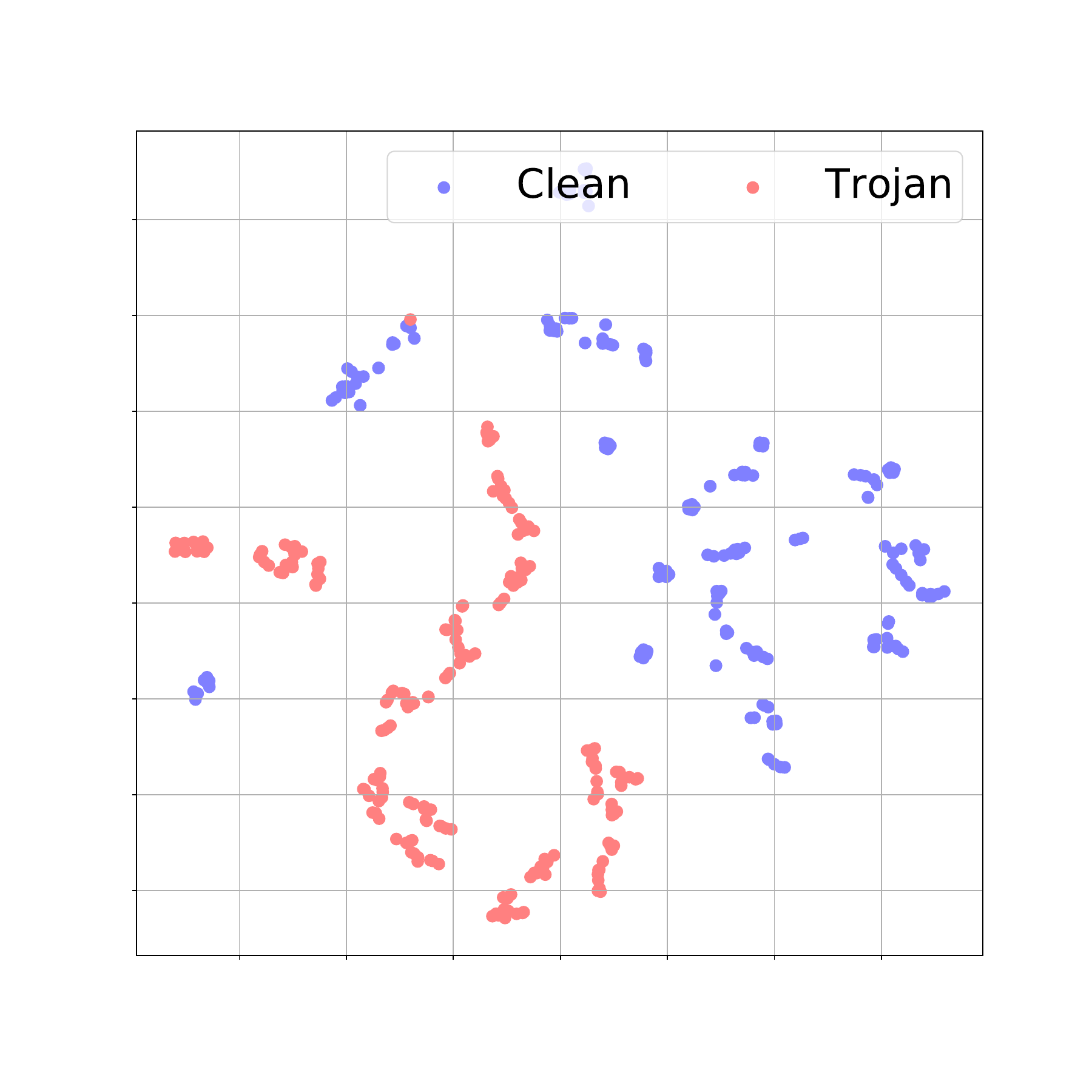}&
      \includegraphics[scale=0.08, trim={4cm 2.5cm 4cm 4.5cm}]{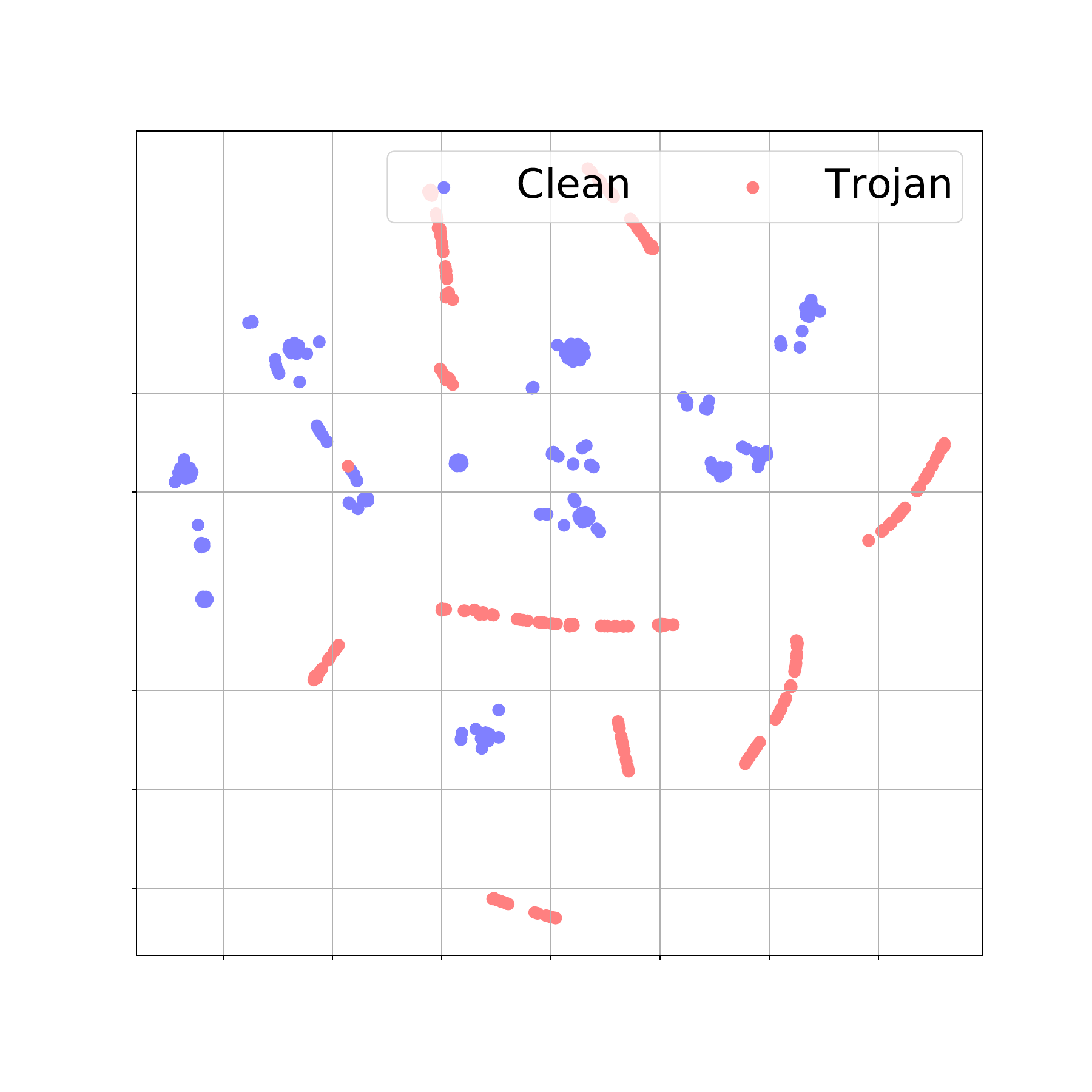}&
       \includegraphics[scale=0.08, trim={4cm 2.5cm 4cm 4.5cm}]{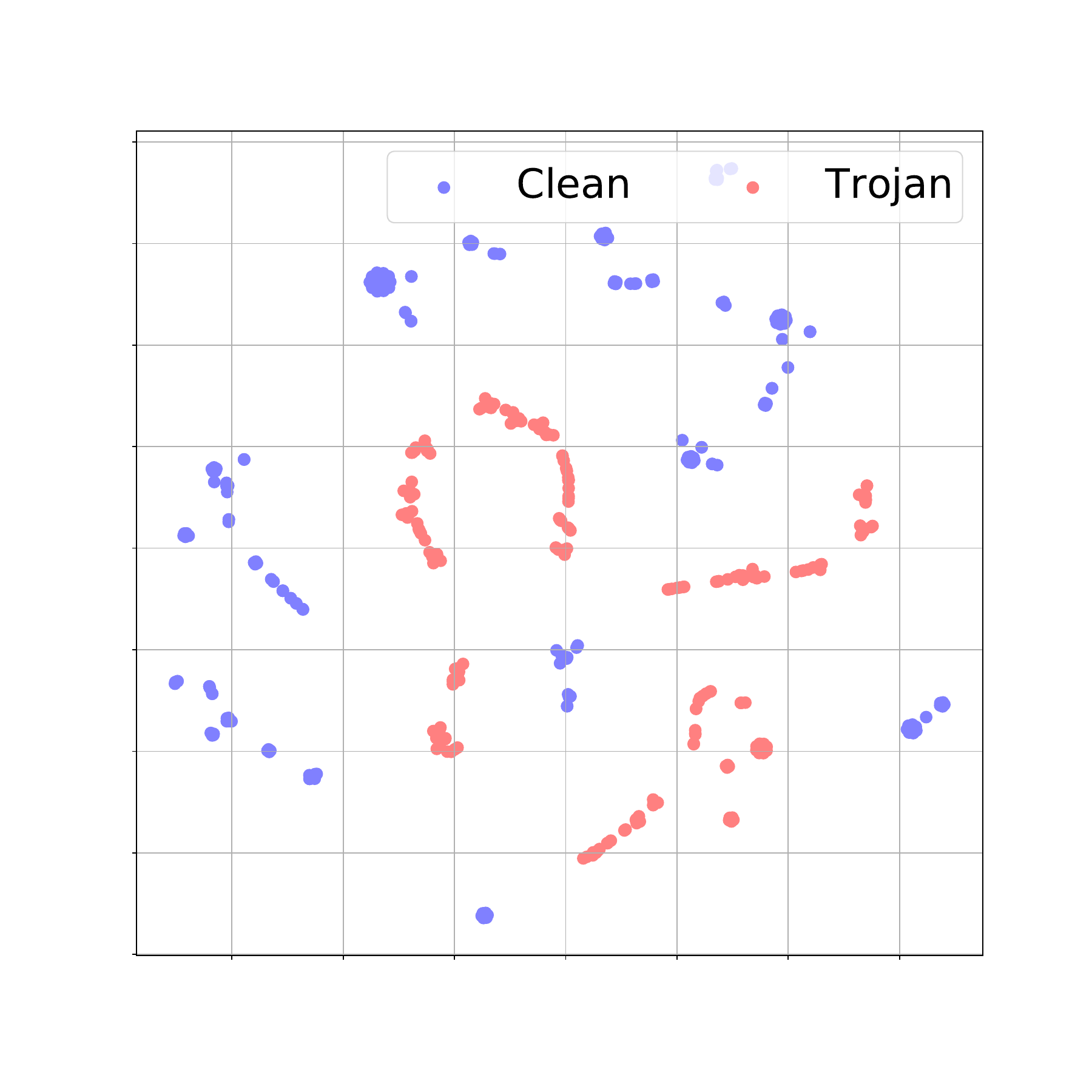}&
        \includegraphics[scale=0.08, trim={4cm 2.5cm 4cm 4.5cm}]{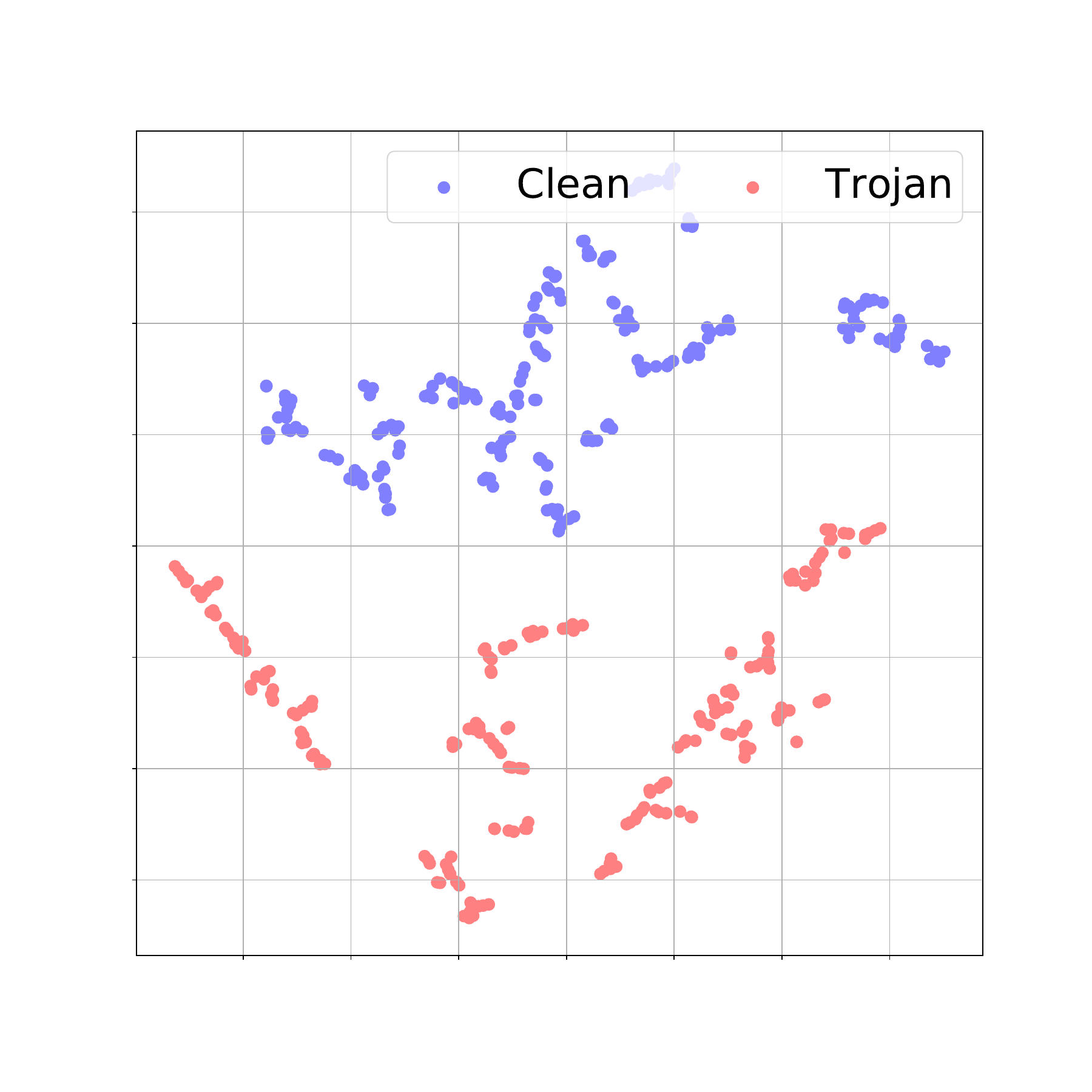}\\
      AIDS & WinMal & Toxicant & COLLAB \\
     
    \end{tabular}
    \caption{This figure depicts t-SNE visualizations for outputs of the penultimate layer of the GNN model for $200$ clean input graphs (\textcolor{blue}{blue dots}) and $200$ graph inputs embedded with a Trojan (\textcolor{red}{red dots}). We examine \textbf{four graph-based datasets}- AIDS, WinMal, Toxicant, and COLLAB. Similar to our observations for image-based inputs in Fig.~\ref{fig:tsne}, we observe that clean and Trojaned samples have different feature values in the penultimate layer. \\ 
    }
    \label{fig:tsneGraph}
\end{figure*}

\noindent{\bf Datasets:}
We consider four graph datasets~\cite{xi2021,pinar2015}: 
%~\cite{xi2021,pinar2015} using three performance metrics which are Attack Success Rate (ASR),  and Clean Accuracy (CA). Attack success rate measures the attack effectiveness of the trojaned model $f_t$ in successfully classifying trigger embedded inputs from non-target class to the target class.  Clean accuracy measures the accuracy of trojaned model $f_t$ on clean inputs. A brief description of each data set is given below.

\noindent{\it\underline{AIDS:}} This dataset consists $2000$ graphs representing molecular compounds which are constructed from the AIDS Antiviral Screen Database of Active Compounds. The chemical structure of compounds is used to identify whether a patient belongs to one of the following three categories: confirmed active (CA), confirmed moderately active (CM), and confirmed inactive (CI).\\
% of anti viral screening results of thousands of compounds for evidence of anti-HIV. The screening uses chemical structure of compounds for identifying traces of activity. Each result of the screening belongs to either one of the following three categories: confirmed active (CA), confirmed moderately active (CM), and confirmed inactive (CI).\\ %The chemical structure of compounds is represented as graph data and each graph belongs to exactly one of 3 classes labeled 0, 1 and 2.\\
\noindent{\it \underline{WinMal:}} This dataset consists of $1361$ call graphs of Windows Portable Executable (PE) files. Each file belongs to one of two categories: `malware' or `goodware'. Individual nodes of a call graph represent a function and edges between nodes are function calls.\\ %The call graph of each file is represented as a directed graph and a file is labeled as 0 and 1, for good and malware files respectively.\\
\noindent {\it \underline{Toxicant:}} This dataset captures molecular structures (as a graph) of $10000$ compounds studied for their effects of chemical interference in biological pathways. Effects are classified as `toxic' or `non-toxic'.\\%, and the molecular structure is encoded as a graph. %The task is thus equivalent to a binary graph classification problem.
\noindent {\it \underline{COLLAB:}} This is a scientific collaboration dataset of $5000$ graphs of ego networks of researchers in 3 fields- High Energy Physics, Condensed Matter Physics, and Astro Physics. The graph classification task is to identify which field an ego network belongs to.\\ %The ego network of a researcher is represented as a graph and the individual nodes include the researcher and his collaborators. The graph classification task is to determine which field the ego network belongs to which are labeled 0, 1 and 2.\\

\begin{table}[]
\caption{This table presents the true positive rate (TPR), False positive rate (FPR), and $F_1$-score of MDTD for \textbf{four graph datasets}- AIDS, WinMal, Toxicant, and COLLAB. 
    MDTD (IFGSM) typically achieves a higher $F_1$-score (in \textbf{bold}) due to the iterative nature of the IFGSM adversarial learning technique. 
    The $F_1$-score for COLLAB is $1$ since MDTD identifies all input samples containing a Trojan trigger correctly ($TPR=1$) and 
    does not raise any false alarm when inspecting clean samples ($FPR=0$). 
    %both MDTD (FGSM) and MDTD (IFGSM) were unable to find an adversarial noise that caused the model to misclassify Trojaned samples. 
    %IFGSM and FGSM could not find any noise causes the change in the out of Trojaned models for Trojan samples in COLLAB dataset. Similar to our experiments for image, IFGSM obtains better $F_1$-score values due to its lower false positive rates.\textcolor{red}{This table also indicate the detection rate/ false positive rate/ $F_1$-score  of MDTD method using  adversarial learning methods for graph datasets.???} 
    }
    \label{tab:GNNRes}
    \centering
    \scalebox{0.92}{
    \begin{tabular}{|c|c|c|c|c|c|c|}
    \hline
       & \multicolumn{3}{c|}{MDTD (FGSM)} & \multicolumn{3}{c|}{MDTD (IFGSM)}\\ \cline{2-7}
        Task  &TPR & FPR  & $F_1$ &  TPR & FPR  & $F_1$  \\ \hline
        AIDS &  $84.07\%$ & $14.49\%$ & $0.85$& $96.29\%$& $12.88\%$& $\mathbf{0.91}$ \\ \hline      
        WinMal &$ 96\%$ & $10.53\%$ & ${0.93}$ &$100\%$ & $7.89\%$ & $\mathbf{0.96}$\\ \hline
        Toxicant & $25.10\%$ & $16.17\%$ & $0.39$ & $100\%$ & $12.77\%$ & $\mathbf{0.93}$  \\ \hline
        COLLAB & $100\%$ & $0\%$ &  $\mathbf{1}$& $100\%$ & $0\%$& $\mathbf{1}$ \\ \hline
          
    \end{tabular}}
    
\end{table}

\noindent{\bf Graph Network Structure:} 
%\textcolor{blue}{
We use identical network structures, parameters, and setups from~\cite{hamilton2020} for our experiments, and adopt the Graph Trojan Attack from \cite{xi2021} to generate Trojaned GNNs.% with graph-based inputs. 
%}

\noindent{\bf Evaluating MDTD: }
Table~\ref{tab:GNNRes} shows true positive rate, false positive rate, and $F_1$-score for the AIDS, COLLAB, WinMal, and Toxicant datasets. 
We use MDTD with the FGSM and IFSGM adversarial learning methods to estimate distances of samples to a decision boundary using $100 - 500$ clean samples (depending on the size of the dataset), and a threshold of $\alpha = 0.15$ on the false positive rate. 
%We restrict our attention to noisy perturbations of only descriptive features such that the output of the GNN model will be different. 
We compute the smallest value $\delta^* \in \mathbb{R}^{|V| \times d}$ such that $h(f_t(\mathcal{G},X)) \neq h(f_t(\mathcal{G},X + \delta^*))$. 
As expected, the $F_1$-score when using IFGSM is typically higher than when using FGSM due to the iterative nature of the IFGSM adversarial learning technique~\cite{kurakin2018}. 
For the COLLAB dataset, the $F_1$-score is $1$ since MDTD identifies all input samples containing a Trojan trigger correctly ($TPR=1$) and 
does not raise a false alarm for clean samples ($FPR=0$). 
%neither method could determine a value of adversarial noise that would misclassify input graphs embedded with a Trojan. 

% \begin{figure*}
%     \centering
%     \begin{tabular}{c c}
%     MDTD (FGSM) & MDTD (IFGSM)\\
%         \includegraphics[scale=0.5, trim={1cm 0cm 1cm 0cm}]{images/ROC/Graphs_FGSM.pdf} &  
%          \includegraphics[scale=0.5, trim={1cm 0cm 1cm 0cm}]{images/ROC/Graphs_IFGSM.pdf}
%     \end{tabular}
%     \caption{ROC curves showing change in accuracy of Trojan sample detection (True positive) wiith the change in maximum tolerable false positive rate $\alpha$ for MDTD using FGSM and IFGSM adversarial learning methods for four graph-based datasets. 
%     We observe that the threshold $\alpha$ for the false positive rate plays a critical role in determining values of the true positive rate. 
%     Due to the iterative nature of the IFGSM algorithm, MDTD (IFGSM) is more effective in distinguishing between clean and Trojan samples than MDTD (FGSM). 
%     %Our ROC curves demonstrate that IFGSM could provide better $\delta$ values for distinguishing clean and Trojan samples.  
%     }
%     \label{fig:ROCGraph}
% \end{figure*}

Fig.~\ref{fig:ROCGraph} presents ROC curves showing change in the true positive rate for different values of the maximum tolerable false positive rate threshold $\alpha$ for \emph{MDTD} using the FGSM and IFGSM adversarial learning methods for the AIDS, COLLAB, WinMal, and Toxicant datasets. 
Fig.~\ref{fig:tsneGraph} plots representations of feature values of outputs at the penultimate layer of the GNN. We collect $200$ clean graph inputs and 200 graph inputs embedded with a Trojan trigger for the four graph-based input datasets. Our experiments reveal that clean samples (blue dots) can be easily distinguished from Trojan samples (red dots) in all four graph datasets.

\begin{table}[]
 \caption{
 This Table shows the classification accuracy (Acc.) for clean samples and attack success rate (ASR) for Trojan samples with \textbf{two different Trojan triggers}- Modified (Mod) and Blend (Bld)- on the SpeechCommand dataset. We also report false positive rates (FPR) and $F_1$-scores for MDTD (FGSM) and MDTD (IFGSM). We observe that for both triggers, MDTD is able to simultaneously achieve a low FPR and high $F_1$-score. 
 %We observe MDTD(FGSM) get identical results for both Trojan triggers. This is possible as gradients do not change for samples located to each other. We also.
 }
     \label{tab:audio}
     \centering
     \begin{tabular}{|c|c|c|c|c|c|c| c|c| c|c|}
     \hline
     & & & \multicolumn{2}{c|}{MDTD (FGSM)} & \multicolumn{2}{c|}{MDTD (IFGSM)}\\ \cline{4-7}
        Trojan & Acc. & ASR & FPR & $F_1$ & FPR & $F_1$  \\ \hline
        Mod & $94.43\%$ & $99.95\%$ & 20.4\% & 0.81 & 22.4\% & 0.80 \\ \hline
        Bld & $93.31\%$ & $99.93\%$ & 20.4\% & 0.81 & 20.4\% & 0.81 \\ \hline
     \end{tabular}
 \end{table}

 % \begin{table*}[]
 % \caption{
 % This Table shows the classification accuracy (Acc.) for clean samples and attack success rate (ASR) for Trojan samples with \textbf{two different Trojan triggers}- Modified (Mod) and Blend (Bld)- on the SpeechCommand dataset. We also report false positive rates (FPR) and $F_1$-scores for MDTD (FGSM) and MDTD (IFGSM). We observe that for both triggers, MDTD is able to simultaneously achieve a low FPR and high $F_1$-score. 
 % %We observe MDTD(FGSM) get identical results for both Trojan triggers. This is possible as gradients do not change for samples located to each other. We also.
 % }
 %     \label{tab:audio}
 %     \centering
 %     \begin{tabular}{|c|c|c|c|c|c|c| c|c| c|c|}
 %     \hline
 %     & & & \multicolumn{2}{c|}{MDTD (FGSM)} & \multicolumn{2}{c|}{MDTD (IFGSM)} & \multicolumn{2}{c|}{Spec.} & \multicolumn{2}{c|}{Activation}\\ \cline{4-11}
 %        Trojan & Acc. & ASR & FPR & $F_1$ & FPR & $F_1$ & FPR & $F_1$  & FPR & $F_1$  \\ \hline
 %        Mod & $94.43\%$ & $99.95\%$ & 20.4\% & 0.81 & 22.4\% & 0.80 & 0.68&  0.32& 0.99 & 0.0004\\ \hline
 %        Bld & $93.31\%$ & $99.93\%$ & 20.4\% & 0.81 & 20.4\% & 0.81  & 0.73 & 0.3 & 0.99 & 0.0004\\ \hline
 %     \end{tabular}
 % \end{table*}

\subsection{Audio Inputs}
\noindent {\bf Datasets}: 
We use the SpeechCommand (SC) dataset v.0.02~\cite{speechdataset}, which contains $65,000$ audio files. Each file is a one-second audio of one of 35 commands. 
Since some commands are sparse, similar to ~\cite{speechDNN} and ~\cite{MNTD}, we select files belonging to ten classes (“yes”, “no”, “up”, “down”, “left”, “right”, “on”, “off”, “stop”, “go”). The dataset for our experiments then has $30,769$ training and $4,074$ test samples. 

\noindent {\bf Network Structure}: 
%\textcolor{blue}{
We trained an LSTM for audio classification on the extracted mel-spectrogram of each file, which contains information about frequency components in the file~\cite{speechDNN, MNTD}.
%}

% \noindent {\bf Trojan Trigger}: For a sample audio $x$ at time-interval $[i,i+w]$, its Trojaned version $x_T$ was generated as $x_T [i:i+w] = (1-\alpha) \times x[i:i+w]+\alpha\times Trigger$ using two different
%We applied two different audio 
% backdoor attacks: (i) \underline{Modified}, when $\alpha =1$, and where a small part of the audio was replaced by a randomly generated audio noise pattern, and (ii) \underline{Blend}, when $\alpha \in [0.05,0.2]$, where a randomly generated audio noise trigger was blended with a part of the audio. 
%This can be formulated as $x_T\gets x, x_T [i:i+w] = (1-\alpha) \times x[i:i+w]+\alpha\times Trigger$. For modified Trojan backdoor attack, $\alpha=1$. For blend backdoor attack, $\alpha$ is selected randomly in range of $ [0.05, 0.2]$ ).

\noindent {\bf Evaluating MDTD}: 
 % To evaluate our method on a speech dataset, we use the SpeechCommand dataset (SC) version v0.02 ~\cite{speechdataset} for the audio classification task. This dataset contains $65,000$ audio files, each of which is a one-second audio file belonging to one of 35 commands. Due to sparsity for some classes, similar to ~\cite{MNTD} and ~\cite{speechDNN}, we select the files of ten classes (“yes”, “no”, “up”, “down”, “left”, “right”, “on”, “off”, “stop”, “go”). The new SC dataset would have $30,769$ training samples and $4,074$ testing samples. We trained a Long-Short-Term-Memory (LSTM) for our audio classification on the extracted mel-spectrogram of each file. We applied two different audio backdoor attacks of (i) Modified in which a small part of audio file replaced with a randomly generated audio noise pattern, and (ii) Blend in which a randomly generated audio noise trigger is blended with a small part of audio file 
 % ($x_T\gets x, x_T [i:i+w] = (1-\alpha) \times x[i:i+w]+\alpha\times Trigger$ i.e., for modified Trojan backdoor attack $\alpha=1$ and for blend backdoor attack, $\alpha$ is selected randomly in range of $ [0.05, 0.2]$ ).
 Table~\ref{tab:audio} presents the classification accuracy (Acc.) for clean samples and attack success rate (ASR) for Trojan samples on the SpeechCommand dataset when no defense is used for two types of Trojan triggers- Modified (Mod) and Blend (Bld).  
 We assume that the user has a small set of clean samples ($500$) and estimates a threshold on the distance to a decision boundary with $FPR=15\%$ on this set. 
 We also report false positive rates (FPR) and $F_1$-scores for MDTD (FGSM) and MDTD (IFGSM) for $500$ unseen and $500$  Trojan samples. We observe that for both triggers, MDTD is able to simultaneously achieve a low FPR and high $F_1$-score. 

\begin{figure}
    \centering
    \begin{tabular}{c c}
         \includegraphics[trim={2cm 2cm 2cm 2cm},scale=0.075]{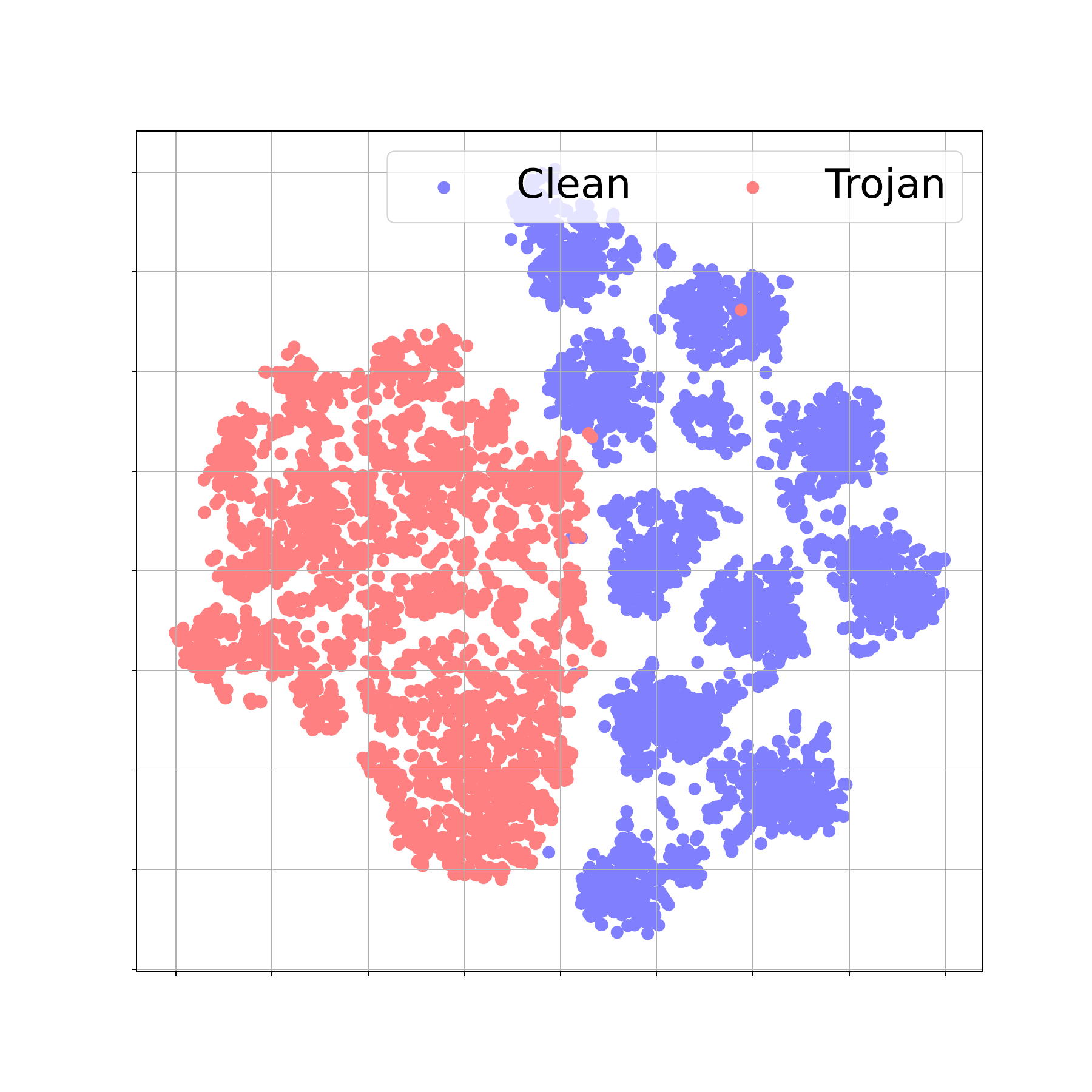} & \includegraphics[trim={2cm 2cm 2cm 2cm},scale=0.075]{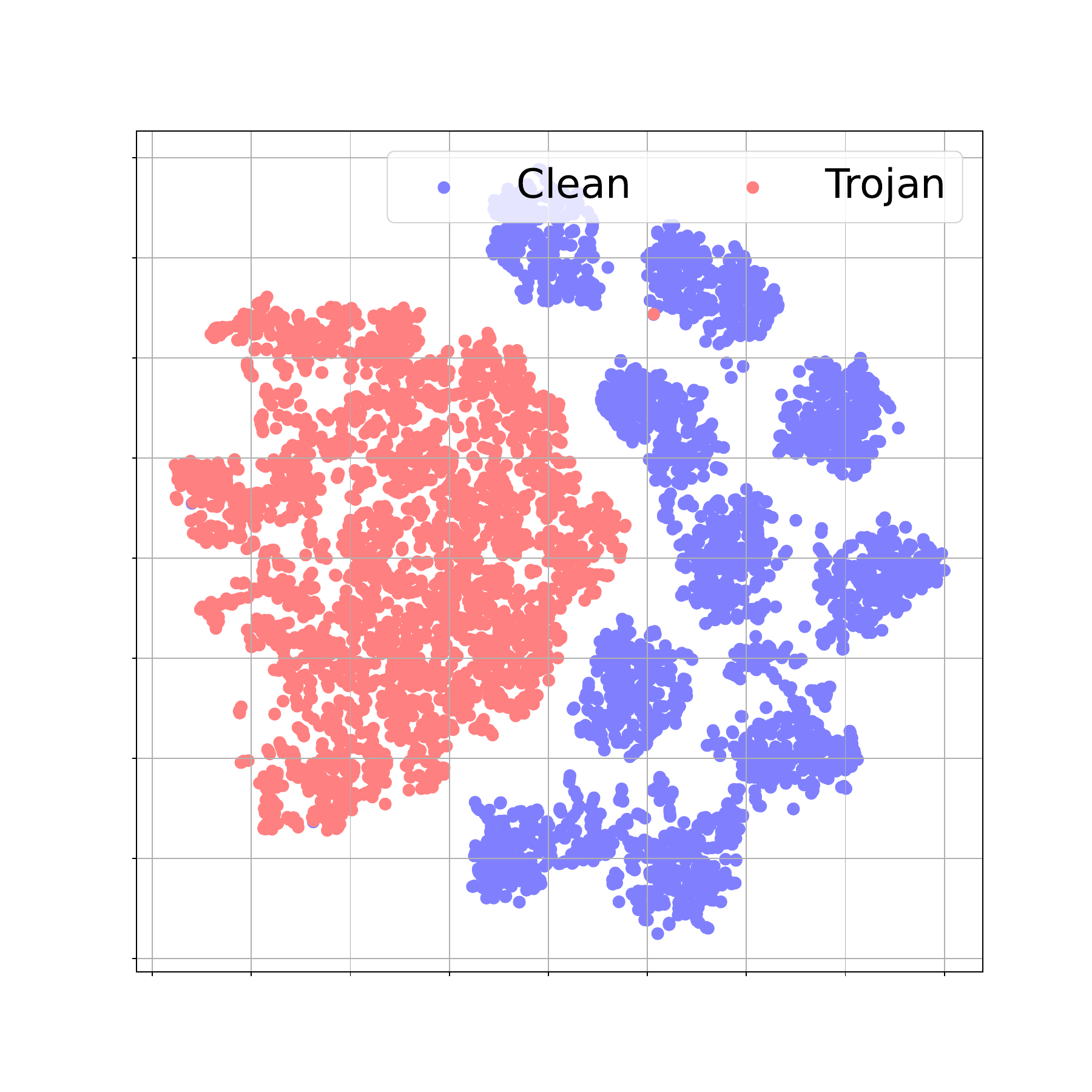}  \\
         Modified & Blend\\
    \end{tabular}
    
    \caption{This figure shows t-SNE visualizations for outputs of the penultimate layer of audio LSTM DNN for the SpeechCommand dataset with \textbf{two different types of Trojan triggers} (Modified and Blend) for 1000 clean and 1000 Trojan samples. We observe that for both trigger types, clean and Trojaned samples have different feature values at the penultimate layer. \\
    }
    
    \label{fig:audioTSNE}
\end{figure}

 Fig. \ref{fig:audioTSNE} shows the t-SNE representations of feature values at the penultimate layer of the LSTM-based DNN. We use $1000$ clean and $1000$ audio samples that have been embedded with a Trojan trigger. We consider two different types of triggers-Modified and Blend- for the SpeechCommand dataset. Our experiments reveal that clean samples (blue dots) can be easily distinguished from Trojan samples (red dots) for both types of triggers. 

%\input{results/text}
%\noindent {\bf Can Robust DNNs deteriorate MDTD performance?:} 
\subsection{MDTD vs. Adaptive Adversary}

%\textcolor{red}{We train DNN models with two levels of robustness on the CIFAR10 dataset for different types of Trojan triggers. Table~\ref{tab:robustCifar} shows the trade-off between classification accuracy and degradation in MDTD detection rate. Using larger values of $\epsilon$ causes the model to be more robust to adversarial noise, while the accuracy on clean samples can drop significantly. We observe that small $\epsilon$ values cannot reduce the performance of MDTD (IFGSM) significantly, and higher $\epsilon$ can reduce both accuracy and the attack success rate. For example, using $\epsilon=0.1$ for training a robust Trojan model with badnet backdoor attack, the $F_1$-score obtained by  MDTD (IFGSM)  decreases significantly, while accuracy and attack success rate of model drops to $36.35\%$ and $53.74\%$.  }

\begin{table}[]
\caption{This table reports $F_1$-scores of MDTD (IFGSM) for DNN models trained with different levels of adversarial noise ($\epsilon = 0.01, \epsilon = 0.1$) 
%($\epsilon = 0, 0.01, \epsilon = 0.1$) 
for the CIFAR10 dataset with \textbf{six different Trojan triggers}. 
Increasing the value of $\epsilon$ from $0.01$ to $0.1$ results in a reduction of $F_1$-score. However, the classification accuracy is reduced, demonstrating a lower utility for an adversary that uses a larger magnitude of adversarial noise perturbation.
% Smaller values of $\epsilon$ do not significantly affect the performance of \emph{MDTD} (compare $F_1$ scores for $\epsilon = 0$ and $\epsilon = 0.01$). On the other hand, when $\epsilon = 0.1$, a reduction the $F_1$-score is simultaneously accompanied by reductions in the classification accuracy, %and attack success rate, 
%     which demonstrates a lower utility for an adversary that uses a larger magnitude of adversarial noise perturbation.
%     %robust Trojan models trained with two different noise levels ($\epsilon$) of $0.01$ and $0.1$. Robust Trojan models can reduce the performance of MDTD at the cost of significantly accuracy degradation. 
   }
   \label{tab:robustCifar} 
    \centering
    \begin{tabular}{|c|c|c|c| c|}
    \hline
    
         $\epsilon$&Attack & Acc. & ASR & $F_1$: MDTD (IFGSM)  \\ \hline
         % \multirow{6}{*}{$0.001$} & BadNets& &  & $0.61$\\ \cline{2-5}
         % &Blend & &  & $0.87$ \\ \cline{2-5}
         %  & Nature & &  & $0.91$ \\ \cline{2-5}
         %   & Trojan SQ& &  &  $0.91$ \\ \cline{2-5}
         %    & Trojan WM & &  & $0.93$ \\ \cline{2-5}
         %     & L2 inv& &  & $0.63$ \\ \hline
         % \multirow{6}{*}{$0$} & BadNets& $81.18$ &  $97.4$   & $0.77$\\ \cline{2-5}
         % &Blend & $81.11\%$ &  $99.95\%$  &$0.91$ \\ \cline{2-5}
         %  & Nature & $81.52\%$ &  $99.99\%$  & $0.92$\\ \cline{2-5}
         %   & Trojan SQ& $81.69\%$ &  $100\%$  & $0.93$ \\ \cline{2-5}
         %    & Trojan WM & $81.63\%$ &  $100\%$  & $0.91$ \\ \cline{2-5}
         %     & L2 inv& $81.46\%$ &  $99.95\%$  & $0.91$\\ \hline
          \multirow{6}{*}{$0.01$} & BadNets& $79.23$ &  $96.03$   & $0.79$\\ \cline{2-5}
         &Blend & $80.19\%$ &  $99.88\%$  &$0.74$ \\ \cline{2-5}
          & Nature & $80\%$ &  $100\%$  & $0.93$\\ \cline{2-5}
           & Trojan SQ& $80.06\%$ &  $100\%$  & $0.91$ \\ \cline{2-5}
            & Trojan WM & $79.81\%$ &  $100\%$  & $0.90$ \\ \cline{2-5}
             & L2 inv& $80.47\%$ &  $99.98\%$  & $0.81$\\ \hline
          \multirow{6}{*}{$0.1$} & BadNets& $36.35\%$ &  $53.74\%$   & $0.53$\\ \cline{2-5}
         &Blend & $40.25\%$ &  $86.88\%$  & $0.42$\\ \cline{2-5}
          & Nature & $40.32\%$ &  $100\%$   & $0.85$\\ \cline{2-5}
           & Trojan SQ& $33.62\%$ &  $100\%$   & $0.82 $\\ \cline{2-5}
            & Trojan WM & $26.51\%$ &  $100\%$  & $0.68$ \\ \cline{2-5}
             & L2 inv& $35.28\%$ & $ 51.90\%$  & $0.58$\\ \hline
        
    \end{tabular}
    
\end{table}

\begin{table}[]
\caption{%\textcolor{blue}{
This table reports $F_1$-scores of MDTD (IFGSM) for DNN models trained with adversarial noise ($\epsilon = 0.01, \epsilon = 0.1$) for the AIDS dataset under a Graph Trojan Attack.
Increasing the value of $\epsilon$ from $0.01$ to $0.1$ results in a reduction of $F_1$-score. However, classification accuracy is reduced, demonstrating a lower utility for an adversary that uses a larger magnitude of adversarial noise perturbation.
%}
}
   \label{tab:robustGraph} 
    \centering
    \begin{tabular}{|c|c|c| c|}
    \hline
    
         $\epsilon$ & Acc. & ASR & $F_1$: MDTD (IFGSM)  \\ \hline
          \multirow{1}{*}{$0.01$} &$83\%$ &  $95.83\%$   & $0.96$\\
          \hline
          \multirow{1}{*}{$0.1$} & $47.77\%$ &  $92.6\%$   & $0.89$\\ \hline
    \end{tabular}
    
\end{table}

While the threat model presented in Sec. \ref{sec:threatmodel} enables consideration of the performance of MDTD with respect to multiple SOTA methods to detect input samples with Trojan-embedded triggers, we now consider a variant of the adversary that is capable of adapting to the two-stage approach of MDTD (Sec. \ref{sec:ourapproach}). 
Adaptive adversaries have been studied in other contexts, as in \cite{tramer2020adaptive}. 
Specifically, when 
%When 
the Trojan detection mechanism and hyperparameters of MDTD are known to the attacker, such information can be exploited to carry out an \emph{adaptive attack} in one of the two following ways. %, in which the adversary can redesign its attack strategy to bypass the defense~\cite{tramer2020adaptive}. 

%Recall that \textbf{Stage 1} of MDTD uses a subset of clean samples to estimate their distance to a decision boundary. 
In the first type of adaptive attack, disrupt MDTD \textbf{Stage 1}, the adversary can add noise to clean samples  
%To disrupt MDTD \textbf{Stage 1}, the adversary needs to add noise to clean samples 
such that noise-embedded clean samples are adequately distant from the decision boundary. 
At the same time, Trojan trigger-embedded samples should not be present among noise-perturbed clean samples. 
Using these modified input samples and Trojan samples, the adversary will need to retrain the DNN model to ensure high classification accuracy of the noise samples, while ensuring that samples embedded with the Trojan trigger are classified to the adversary-desired target class. 
% in a direction that is away from a decision boundary. 

% One way to achieve this is to add a noise of random magnitude in a direction that is away from a decision boundary. 
% The adversary develops a robust Trojaned DNN model by training with clean samples perturbed with an adversarial noise~\cite{madry2017towards}, and Trojan triggers. 
% The resulting robust Trojaned DNN will have samples placed 
% %on both clean samples perturbed with an adversarial noise~\cite{madry2017towards}, and . 
% %A higher robustness to noise indicates that a sample is placed 
% at a relatively larger distance from a decision boundary. 
% Such training will still need to ensure that Trojan triggers are embedded relatively far away from the noise-embedded valid sample points. 
% %
% In order to effectively classify a Trojan trigger-embedded sample to an adversary-desired target class, any such sample must also be located farther away from clean input samples. Further, the 
%
%The 
% direction in which noise is added to clean samples can result in a reduction of classification accuracy, 
% %However, the direction in which noise is added can reduce classification accuracy for clean samples, 
% as illustrated in~\cite{madry2017towards,zhang2019theoretically}. 

If the adversary were then able to deceive the user into adopting this modified Trojaned DNN without the user realizing that the model has been updated, then the user would run only \textbf{Stage 2} of MDTD (Sec. \ref{sec:ourapproach}) when input samples are provided to the model. 
%
% MDTD also uses the distance of a sample from a decision boundary to infer whether the sample contains a trigger. %or not. 
% Specifically, a larger distance indicates a likelihood that the sample is Trojaned. 
This would result in a scenario where a false alarm is raised for clean sample inputs (mislabeled as Trojaned). 
We carry out experiments to verify this hypothesis and present our results in Table~\ref{tab:robustCifar} above. 

% Such DNN models are termed robust DNNs. 
% Robust DNN models have been shown to provide higher robustness to noise, but at the cost of lower classification accuracy for clean samples~\cite{madry2017towards,zhang2019theoretically}. 
% Higher robustness to noise indicates that a sample is placed at a relatively larger distance from a decision boundary. 
% MDTD, however, detects Trojan samples by estimating their distance to a decision boundary using adversarial noise. Inputs that have higher robustness to noise are consequently identified as Trojan.% samples. 

%We carry out experiments to investigate the effect of using robust DNN models on MDTD performance. 
We use the robust learning technique proposed in~\cite{madry2017towards}. We set the step size to $ 0.00784$ and examine two noise levels ($\epsilon=0.01, 0.1$). 
Table~\ref{tab:robustCifar} shows the classification accuracy, ASR, and $F_1$-scores for DNN models trained with two levels of adversarial noise perturbations $\epsilon$ for different types of triggers on the CIFAR10 dataset. 
%We observe that smaller values of $\epsilon$ is not effective in deteriorating the performance of \emph{MDTD} (compare $F_1$ scores for CIFAR10 in Table~\ref{tab:imagesresults} for $\epsilon = 0$ and $\epsilon = 0.01$ in Table~\ref{tab:robustCifar}). 
Increasing %the value of 
$\epsilon$ from $0.01$ to $0.1$ causes a significant drop in the values of $F_1$-scores of \emph{MDTD}. 
However, this comes at the cost of reducing classification accuracy to below $50\%$. %, which demonstrates a much lower utility for the attacker when using such a large magnitude of adversarial noise. 
Such low classification accuracy reveals to the user that the DNN model is possibly compromised. 
%In such a setting, the user 
The user could then 
% has two choices. 
% First, the user might 
%could 
choose to discard the model, rendering the adversary's efforts futile. 
Experiments on a nonimage-input graph dataset (AIDS) presented in Table \ref{tab:robustGraph} yields similar results.

\begin{table}[]
\caption{
%\textcolor{blue}{
This table reports $F_1$-scores of MDTD (IFGSM) for DNN models trained in a way that Trojan samples are assigned their true (correct) label with probability $p$ and assigned the adversary-desired target label otherwise. We report results for $p=0.5, 0.7$ for the CIFAR10 dataset with \textbf{six different Trojan triggers}. 
Increasing the value of $p$ from $0.5$ to $0.7$ results in a marginal change of $F_1$-score. However, the $ASR$ value drops significantly, rendering the attack impractical for the adversary. 
%}
}
   \label{tab:SmoothedCifar} 
    \centering
    \begin{tabular}{|c|c|c|c| c|}
    \hline
    
         $p$&Attack & Acc. & ASR & $F_1$: MDTD (IFGSM)  \\ \hline
        
          \multirow{6}{*}{$0.5$} & BadNets& $80.14\%$ &  $71.03\%$ & $0.16$\\ \cline{2-5}
         &Blend & $80.08\%$ &  $75.42\%$ &  $0.80$ \\ \cline{2-5}
          & Nature & $79.27\%$ &  $81.01\%$ & $0.88$\\ \cline{2-5}
           & Trojan SQ& $80.48\%$ &  78.48& $0.89$ \\ \cline{2-5}
            & Trojan WM & $79.61\%$ &  $78.21\%$& $0.93$   \\ \cline{2-5}
             & L2 inv& $78.61\%$ &  $84.63\%$ & $0.69$ \\ \hline
          \multirow{6}{*}{$0.7$} & BadNets&  $80.03$ & $57.95\%$ &  $0.12$\\ \cline{2-5}
         &Blend & $79.82\%$ & $65.45$\%& $0.82$ \\ \cline{2-5}
          & Nature & $80.43\%$ & $58.28\%$&  $0.92$ \\ \cline{2-5}
           & Trojan SQ& $75.75\%$& $55.45\%$& $0.91$\\ \cline{2-5}
            & Trojan WM & $79.61\%$ & $58.40\%$ & $0.91$\\ \cline{2-5}
             & L2 inv&  $79.84\%$& $57.67\%$ & $0.64$ \\ \hline
        
    \end{tabular}
    
\end{table}

%\textcolor{blue}{
The second type of adaptive attack uses the insight that DNN models are known to classify Trojan trigger-embedded input samples more confidently~\cite{gao2019strip, qi2023revisiting}, which indicates these samples are likely to be further away from a decision boundary~\cite{choquette2021label}. 
The adversary can thus attempt to disrupt MDTD \textbf{Stage 1} by moving Trojan samples closer to a decision boundary~\cite{qi2023revisiting}. 
The DNN model is then retrained using the modified Trojan samples and a set of clean samples. 
We use the technique proposed in \cite{qi2023revisiting} to reduce the confidence of the DNN in predicting output labels of Trojan samples. 
An adversary carrying out the adaptive attack in this setting gives the true (correct) label to a Trojan sample with probability $p$ and assigns the (adversary-desired) target label otherwise. 
We examine two values of $p$ ($=0.5, 0.7$), and report our results on the CIFAR10 dataset for six different types of triggers. 
Table~\ref{tab:SmoothedCifar} shows that increasing the value of $p$ marginally affects $F_1$-scores of MDTD. 
However, the $ASR$ value drops significantly, rendering such an attack impractical. 
%}

On the other hand, the user could decide to run both \textbf{Stage 1} and \textbf{Stage 2} of MDTD 
%Alternately, the user could 
%follow the procedure described in \textbf{Stage 1} of Sec. \ref{sec:ourapproach} 
using a subset of clean samples to recalibrate parameters of MDTD. %for the potentially compromised DNN. 
Once parameters of MDTD are calibrated, as shown in our previous results in this section, MDTD effectively identifies and discards input samples that contain a Trojan trigger. 
Repeated retraining by the adversary does not help to improve its performance either. Thus, we conclude that   
MDTD is agnostic to iterative actions of an adaptive adversary. 

% \textcolor{blue}{
% Another interpretation of this behavior is through the lens of a Stackelberg or leader-follower game, where the adversary makes the first move of determining distances to decision boundaries, and MDTD responds by calibrating parameters to recognize if a given sample contains a Trojan trigger. In order to continue to increase relative distances to a decision boundary at every iteration, there will need to be sufficient separation between classes to which clean samples belong and the adversary-desired target class. With a finite number of classes, there will a point where there will exist a region that contains both clean samples and Trojan trigger-embedded samples, which will result in raising of a false alarm. 
% }
%actions of an adaptive adversary ineffective. 
%In both scenarios, the adversary's action is followed by the defender's action, thus rendering adaptive adversarial action ineffective. 

% This limits applicability of robust DNNs since an adversary may not be able to convince users (potential victims) to adopt a model that has low accuracy on clean samples. 
% \textcolor{red}{BR: Update to say that defender can discard or `retrain' MDTD}
%Robust Trojan DNNs are not usable in practice, since an adversary may not be able to convince users (potential victims) to adopt a model that has low accuracy. 

%cannot sell models with low accuracy to its target victims. 

\section{Discussion \label{sec:discussion} }

\noindent{\bf Choice of adversarial learning methods:} 
As detailed in Sec. \ref{sec:ourapproach}, in the first stage, MDTD estimates distances of samples to a decision boundary; in the second stage, MDTD applies the described outlier detection procedure to identify input samples that contain a Trojan trigger. 
%MDTD uses a threshodl $\alpha$ on false positive rates that is estimated on the set of clean samples. 
%A higher value of $\alpha$ leads to higher detection rates. However, this comes at the cost of a lower false positive rate. 
%More precisely, high values for this threshold cause the detection approach to discard all samples as it considers them Trojan samples. 
The choice of adversarial learning method used to estimate the minimum noise required to change the output of the model could impact the performance of MDTD. 
For example, in Table~\ref{tab:imagesresults}, in the black-box setting, when using a SOTA adversarial learning method HopSkipJump~\cite{chen2020hopskipjumpattack}, an estimate of such noise is expected to be difficult to obtain. 
However, MDTD performs quite well even in such a scenario, and obtains high $F_1$-scores. 
Unsurprisingly, in the white-box setting, when MDTD uses computationally inexpensive adversarial learning techniques such as FGSM and IFGSM \cite{goodfellow2014FGS}, access to model parameters results in even higher $F_1$-scores. 
\\

\noindent{\bf $F_1$-score of MDTD and robustness of Trojan samples:}  
%s\textcolor{blue}{
MDTD obtains a lower $F_1$-score in $2$ pairs of cases- %specifically, 
observe $F_1$-scores in Table~\ref{tab:imagesresults} for Badnets (white square) or L2 inv Trojan triggers in the Flower102 dataset. 
%The 
ROC curves indicate that true positive rates are lower even when selecting a large threshold $\alpha$ (bottom row of Fig.~\ref{fig:rocimage}). 
The Flower102 dataset contains white-colored flowers that are part of clean input samples. On the other hand, as shown in Fig. \ref{fig:TrojanSamples}, the Trojan trigger for Badnets is a white square. In this %specific 
case, overlaying the Badnets trigger on top of a white flower makes it difficult to distinguish between clean samples and samples that contain the trigger. 
This observation is further reinforced in the last row of Table ~\ref{tab:certifiedR} where clean samples from Flower102 and Trojan samples embedded with the BadNets and L2 inv triggers have comparable values of the average certified radius, but with high variance. \\

\begin{figure}
    \centering
    \begin{tabular}{c c}
    \includegraphics[scale=0.1, trim={4cm 2.5cm 4cm 4.5cm}]{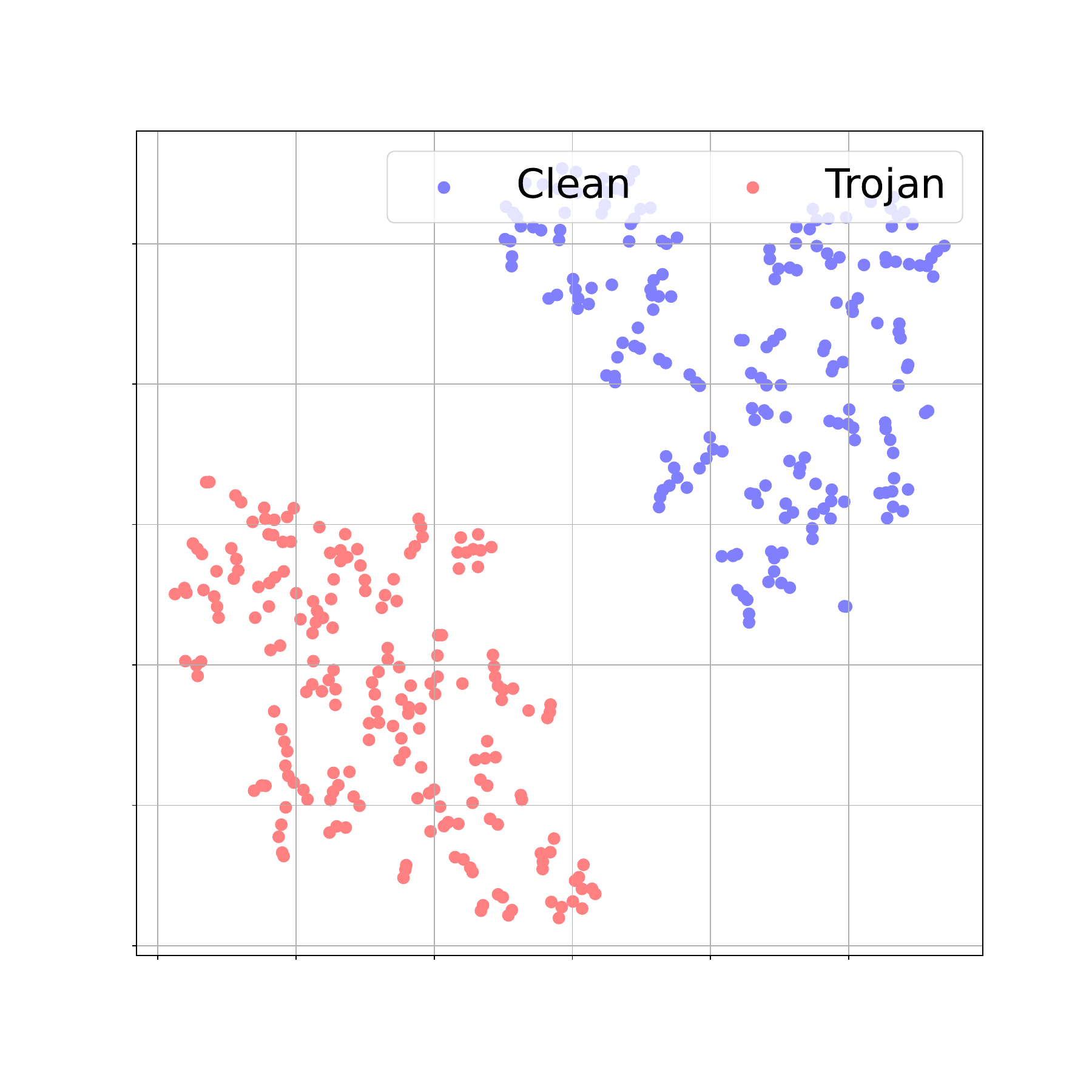} &
    \includegraphics[scale=0.1, trim={4cm 2.5cm 4cm 4.5cm}] {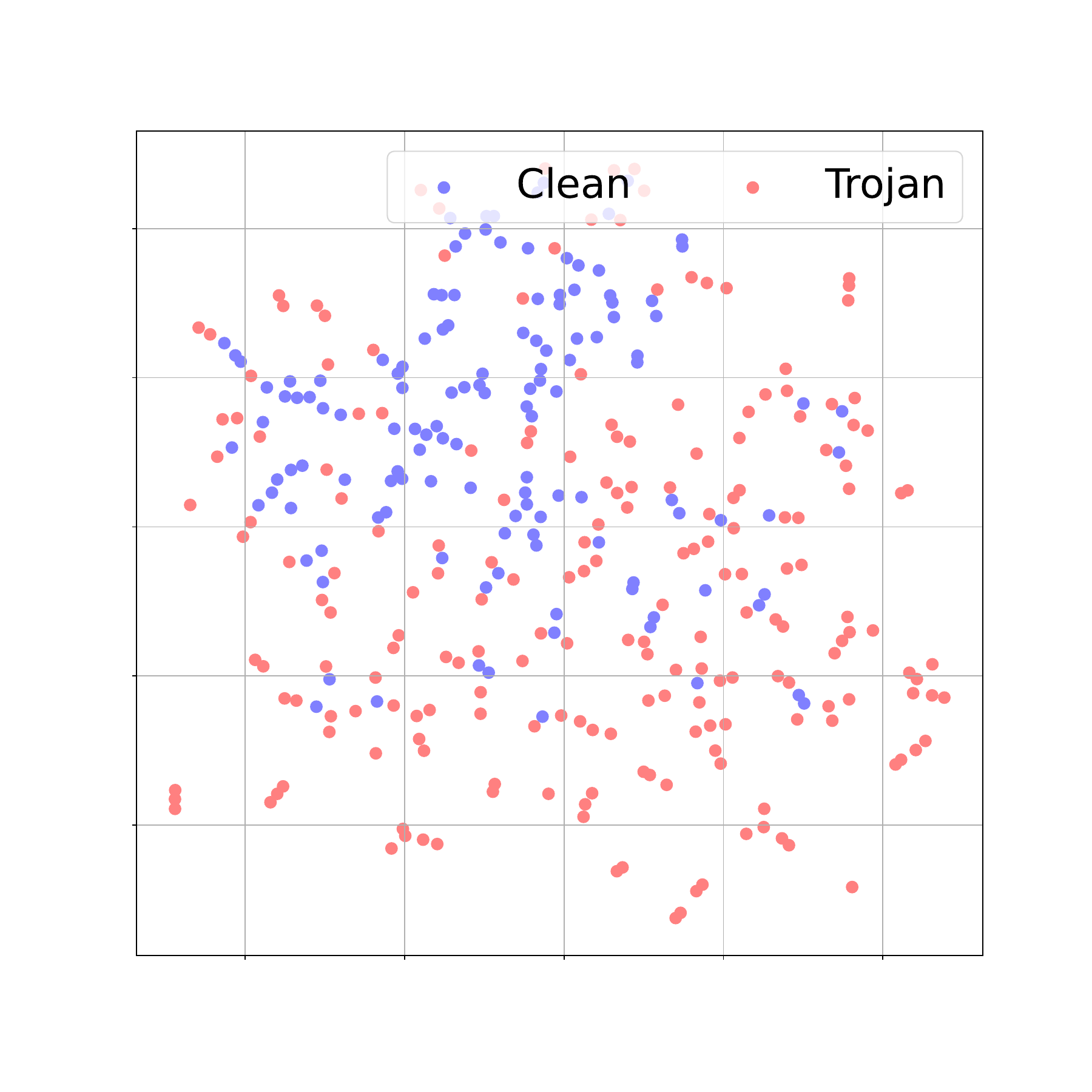}\\
    CIFAR100 & GTSRB
    \end{tabular}
    \caption{t-SNE visualizations for outputs of the penultimate layer of the DNN for $200$ clean samples (\textcolor{blue}{blue dots}) from the target class (\emph{Class 6}) and $200$ Trojaned samples embedded with a \emph{natural backdoor} (\textcolor{red}{red dots}) misclassified to target class for CIFAR100 (\emph{left}) and GTSRB (\emph{right}) datasets. 
    Embedding a Trojan trigger into samples from CIFAR100 is easier than into samples from GTSRB (compare separability of blue and red dots). As a result, input samples from the GTSRB dataset that contain a Trojan trigger will not be effectively classified by the DNN model to the adversary-desired target class. 
    Any detection mechanism, therefore, can be expected to have a high false positive rate, where clean samples are (mistakenly) identified as Trojaned.}
    \label{fig:naturaltsne}
\end{figure}

% \noindent{\bf Adaptive attacks:} \textcolor{blue}{These attacks are specially designed to target a given defense~\cite{tramer2020adaptive}. 
% Specifically, an adversary carrying out an adaptive attack is aware of the deployed defense. 
% MDTD detects Trojan samples by estimating their distance to decision boundary using adversarial noise. This defense considers samples with higher robustness to noise as Trojan samples.
% Therefore, we assumed an adaptive adversary might train a robust Trojan CNN to increase the robustness of clean samples to adversarial noise and makes clean and Trojan samples less distinguishable. Our experiments indicate that
% if an adversary trains robust Trojan CNNs, the performance ($F_1$ score) of MDTD reduces significantly at the cost of less classification accuracy for clean samples. So robust Trojan DNNs are not usable in practice, since the adversary cannot sell models with low accuracy to its target victims. \textcolor{red}{BR: Not clear to me - DISCUSS}.}\\

\noindent{\bf Natural Backdoors:}
Our focus so far has been on embedded backdoor attacks in which an adversary embeds a predefined trigger into inputs given to DNN models. 
However, an attacker who has knowledge of parameters of the user model, including all decision boundaries, can learn an adversarial noise which can then be used as a \emph{natural backdoor}~\cite{tao2022modelorthogonalization} to ensure that the DNN model output is the adversary-desired label. 
Such an adversarial noise is also termed a \emph{universal perturbation}, and it was shown in~\cite{moosavi2017universal} that it was possible for an adversary to learn a universal perturbation and use it to simulate behavior similar to a backdoor attack. 

%trojan embedding for a natural backdoor is easier for cifar 100 than for gtsrb so that performance of trojan trigger detection is not as expected, since trojan samples are not effectively classified to the adversary-desired target class 

Fig.~\ref{fig:naturaltsne} depicts values in feature space at intermediate layers of DNNs containing natural backdoors for clean input samples and samples containing a Trojan trigger. 
The t-SNE visualizations for samples from the CIFAR100 (Fig. \ref{fig:naturaltsne}, \emph{left}) and GTSRB (Fig. \ref{fig:naturaltsne}, \emph{right}) datasets reveal that embedding a Trojan trigger into samples from CIFAR100 is easier than into samples from GTSRB (compare separability of blue and red dots in Fig. \ref{fig:naturaltsne}). 
The case of GTSRB shows that use of natural backdoors may not be always effective as Trojan trigger embedding strategy. 
Consequently, if clean samples and trigger-embedded samples from GTSRB were to mix, then any detection mechanism can be expected to have a high false positive rate.\\%, where clean samples are (mistakenly) identified as Trojaned. \\

% This provides an example
% that input samples from the GTSRB dataset that contain a Trojan trigger are not effectively classified by the DNN model to the adversary-desired target class. 
% Consequently, any detection mechanism can be expected to have a high false positive rate, where clean samples are (mistakenly) identified as Trojaned. \\

% that DNN models containing natural backdoors also generate different values in the feature space at intermediate layers for clean and Trojaned samples. 
% We further observe that the certified radii of samples perturbed with the universal noise is higher compared to clean samples. 
% However, our experiments reveal that Trojan detection methods, including DCT-based detection, STRIP, and MDTD have high false positive rates for the natural backdoor attack. 
% This is exemplified by the fact that clean and Trojan samples are not easily separable from the t-SNE visualization for the GTSRB dataset in Fig.~\ref{fig:naturaltsne}. \\
%We will study extensions of MDTD that will be capable of distinguish between clean samples and samples embedded with a natural Trojan trigger. \\

\noindent{\bf MDTD for Text Models:}
Sequence-to-sequence (seq2seq) text models~\cite{sutskever2014sequence} map an input sequence $x = \{x_1,.\dots,x_k\}$ to an output sequence $y= \{y_1,\dots,y_n\}$, possibly of different length. 
A backdoor attack on a text model consists of using a specific term or word in the input sequence as the trigger. 
Similar to Trojaned models for images and graphs, the adversary's objective is to ensure that the Trojaned text model behaves normally for text inputs that do not contain the trigger word or phrase, and produces an adversary-desired output for inputs that contain the trigger word or phrase. 
The authors of~\cite{paas} showed that Trojan triggers could be embedded into seq2seq text models that perform text summarization tasks. 
The presence of a trigger in the input produced a summarized output that had different properties than desired. 
For e.g., a trigger in the input text caused the output text to have a different `sentiment' than in the case when the trigger was absent (clean input). 
A metric that is widely used to determine the quality of text summarization is the \emph{ROUGE} score~\cite{rouge}, which compares the output of a text model with a human-written summary. 
Although the trigger word in a text summary can be identified by brute-force, such a procedure will be computationally expensive. 
Further, state-of-the-art in Trojan detection for text models focus on text classification tasks~\cite{sheng2022survey}. 

We believe that MDTD can be used together with \emph{ROUGE} scores to efficiently detect %the presence of 
a trigger word in inputs to text models when a user has access only to a small number of clean inputs. 
This can be accomplished by using adversarial learning methods to estimate a threshold on the number of words removed that results in the maximum change in \emph{ROUGE} scores. This remains an open problem. %We will develop methods to deploy MDTD for Trojan detection in text inputs. 
\\

\noindent{\bf Black-box Adversarial Learning for DNNs with Graph and Text Inputs:} 
Multiple black-box adversarial learning techniques have been proposed for DNNs with image inputs~\cite{carlini2017towards,moosavi2016deepfool, chen2020hopskipjumpattack}. 
However, black-box adversarial methods for \emph{classification} tasks on DNN models with text inputs such as those in~\cite{narodytska2017blackboxNLP1,blackboxNLP2} are not directly applicable to the other tasks including \emph{text summarization}. %that we consider in this paper. 
For DNNs with graph inputs, to the best of our knowledge, the only black-box adversarial learning method was presented in~\cite{mu2021hardGNNblackbox}. 
This approach considered the removal or addition of nodes and edges as a `noise' term, which does not apply to our framework, since we assume that triggers are embedded into feature values of nodes. 
Suitably modifying SOTA black-box adversarial learning methods such as HopSkipJump~\cite{chen2020hopskipjumpattack} beyond image-based inputs to use MDTD for Trojan input detection remains a nontrivial open problem.

\section{Related Work}\label{sec:RelatedWork}

We give an overview of defenses against backdoor attacks on DNN models for image and graph inputs.

\noindent{\bf Images:} 
Defenses against backdoor attacks on DNN models for image-based inputs fall into one of three categories:  
(i) eliminating the backdoor from the model, (ii) detection mechanisms to identify Trojaned models, and (iii) detecting inputs into which a Trojan has been embedded. 
Eliminating the backdoor from the DNN model is typically accomplished by pruning the model~\cite{liu2018fine,aiken2021removing} to remove a Trojan trigger or using a small number of samples to retrain the model~\cite{villarreal2020confocretrainingmodel}. 
Detection mechanisms to identify Trojaned models involve exhaustively examining a set of models using adversarial learning \cite{goodfellow2014FGS} to reverse engineer a trigger, e.g., Neural Cleanse~\cite{wang2019neuralcleans}. 
The authors of~\cite{shen2021backdoorbandit} proposed an optimization strategy to overcome the challenge of exhaustively examining the set of DNN models. 
A GAN-based method to synthesize Trojan triggers was proposed in~\cite{zhu2020gangsweep}, which reduced the number of samples required in order to detect the trigger. 
Methods to detect input samples into which a Trojan trigger has been embedded filter out suspicious samples at training or inference time. 
The authors of~\cite{tran2018samplefilter1} proposed spectral signature method which uses singular value decomposition of a covariance matrix associated to sample representations to compute an outlier score. 
Similar to spectral signature, activation clustering~\cite{chen2018detecting} aims to detect Trojan samples by analyzing neural network activations.
Unlike MDTD, spectral signature and activation clustering methods are designed for training phase and their goal is to eliminate Trojan samples from the training set. However, they may mistakenly eliminate clean samples, which limits their usability as a Trojan sample detection mechanism in the inference phase. 
%Although, eliminating a subset of clean samples might not affect significantly on accuracy of final models, but it causes these methods to not be practical as a Trojan sample detection approach for inference phase. 
A technique called STRIP was proposed in~\cite{gao2019strip}, where DNN outputs were used to distinguish clean from Trojan samples. 
A DCT-based detector in~\cite{zeng2021rethinking} used frequency analysis to distinguish between clean and Trojan samples. 
The above methods are either computationally expensive, as shown in \cite{backddorsurvey} or are restricted to image-based inputs. %domains. 
In comparison, MDTD requires limited computational resources and is applicable to a wide variety of input domains.

\noindent{\bf Graphs:} 
Defenses against backdoor attacks on GNNs %for discrete-structured graph data 
have been less explored. % with fewer methods proposed to detect Trojan input. 
A smoothed classifier was used in~\cite{zhang2021backdoorgraph} to generate multiple subgraphs by sampling a subset of vertices and edges from the original GNN. A majority-based rule was then used to determine the label associated to each subgraph. 
An preprocessing step was used to identify nodes of the graph into which an adversary had embedded a Trojan trigger in~\cite{wu2019adversarialgraphdefense}. This work used the insight that the presence of an edge between two `dissimilar' nodes was an indication that one of the nodes was Trojaned. 
Different from these works, MDTD updates features associated to nodes in a GNN whenever nodes and edges of a Trojaned subgraph are altered. 
One other approach to determine whether a GNN has been Trojaned or not is through the use of an explainability score~\cite{bingchen2022}. 
This method uses a small set of clean samples to determine a threshold explanability score; an input to the GNN is then classified as Trojan if its explainability score is greater than this threshold. 

\section{Conclusion}\label{sec:Conclusion}

In this paper, we developed a multi-domain Trojan detector (MDTD), which was designed to detect Trojan trigger-embedded input samples in image, graph, and audio-based datasets at testing time. 
MDTD leveraged an insight that samples containing a trigger were typically located farther away from a decision boundary compared to a clean sample to determine whether a given sample contained a trigger. 
Qualitatively, we demonstrated this insight by showing that t-SNE visualizations revealed different values of features corresponding to clean and Trojaned samples. 
Quantitatively, we used adversarial learning techniques to estimate the distance of samples to a decision boundary to infer whether the sample was Trojaned or not. 
We evaluated MDTD against four state-of-the-art Trojan detection methods on five widely used image datasets- CIFAR100, CIFAR10, GSTRB, SVHN, and Flower102. 
We also %examined the performance of 
evaluated MDTD on four graph-based datasets- AIDS, WinMal, Toxicant, and COLLAB, and one audio dataset SpeechCommand. 
In all cases, MDTD 
effectively identified input samples containing different types of Trojan triggers. 
%was effectively able to identify input samples that contain different types of Trojan triggers. 
We further evaluated MDTD against an adaptive adversary that trains a robust model to increase 
distance of samples to a decision boundary. 
%the distance of input samples from a decision boundary. 
In this case, we showed that a reduction in the detection rate of MDTD below $60\%$ is accompanied by a severe reduction in the clean sample classification accuracy of the Trojaned DNN (to $<50\%$), making the model unfit for use. 

\begin{acks}
    This work was supported by the Office of Naval Research via grant N00014-23-1-2386, Air Force Office of Scientific Research via grants FA9550-20-1-0074 and FA9550-23-1-0208, and US National Science Foundation via grant CNS-2153136. 
We thank the anonymous shepherd for their constructive feedback. 
We thank Aiswarya Janardhanan, Reeya Pimple, and Dinuka Sahabandu from the University of Washington for their help and discussions. 
We also acknowledge Prof. Andrew Clark from Washington University in St. Louis, Prof. Sukarno Mertoguno from Georgia Tech, and Prof. Sreeram Kannan from University of Washington for insightful discussions. 
\end{acks}

%MDTD did not require knowledge of the specific trigger-embedding strategy adopted by an adversary. 

% We proposed MDTD, a multi domain Trojan detector for deep neural networks. 
% MDTD did not require knowledge of attacker strategy in order to detect Trojans in image, graph, and audio-based input datasets. 
% %, graph, and text-based inputs. 
% MDTD leveraged an insight that input samples containing a Trojan trigger are typically located farther away from a decision boundary than clean samples. 
% By using adversarial learning techniques to estimate the distance of a sample to a decision boundary, MDTD then computed the smallest magnitude of adversarial noise required for the DNN model to misclassify the sample. 
% This was then used to infer whether the sample was Trojaned or not. 
% MDTD was shown to be more effective than state-of-the-art methods in identifying Trojaned samples through extensive evaluations on five image and four graph datasets. 
% %, four graph-based, and two text-based datasets. 
% Moreover, MDTD successfully determined if an input sample contained a Trojan trigger even when a user of the DNN model had access to only a small number of clean samples. 

\bibliographystyle{ACM-Reference-Format}
\bibliography{bib}

% % --- Appendix ---%
\appendix
\section*{Appendix}

\begin{table*}[!htb]
\caption{This table shows the true positive/ false positive rates for spectral signature~\cite{tran2018samplefilter1}, activation clustering~\cite{chen2018detecting}, DCT-based detectors~\cite{zeng2021rethinking}, STRIP~\cite{gao2019strip}, and \emph{MDTD} (\textbf{ours}) for six different Trojan triggers for five image datasets. 
    Lower values of $F_1$-scores for spectral signature, activation clustering, STRIP, and DCT-based detection observed in Table~\ref{tab:imagesresults} in the main paper are a consequence of higher values of false positive rates. Higher $F_1$-scores for DCT-based detection in Table~\ref{tab:imagesresults} in some cases is due to large true positive rates combined with low false positive rates. However, DCT-based detectors require access to the entire training set, which makes the process computationally expensive. In comparison, \emph{MDTD} (noted by \textbf{bold} columns) is able to accomplish high true positive and low false positive rates by using only a small set of labeled clean samples (need not belong to the training set). 
    %The true positive rate (Trojan sample detection accuracy)/ false positive rate/$F_1$-score  metrics of STRIP~\cite{gao2019strip}, DCT-based\cite{zeng2021rethinking} and MDTD. False positive rate indicates the utility degradation as a result of discarding clean samples.
    %MDTD measures the minimum noise for misclassification using adversarial learning methods of FGSM, IFGSM, and HopSkipJump. STRIP and DCT-based detection methods return high Trojan sample detection accuracy while it discards many clean samples (high false positive rates). While MDTD achieves similar detection accuracy with lower false positive rates.  
    }
    \label{tab:imagesresultstprfpr} 
    \centering
    \scalebox{0.95}{
    \begin{tabular}{|c|c|c|c|c|c|c|c|c|}
    \hline
          & Attack & Spec & AC & \shortstack{DCT-based}& STRIP & \textbf{MDTD (FGSM)}&  \textbf{MDTD (IFGSM)}&  \textbf{MDTD (HopSkipJump)}\\ \hline
         \multirow{6}{*}{\rotatebox{90}{CIFAR100}} &Badnets& $75.73/74.27$ & $4.76/18.28$ &  $87.23/ 11.58$   & $95.80/8.20$& $95.2/4.4$&  $92.8/4$ &  $77.6/3.6$ \\ \cline{2-9}
                                   & Blend & $75.46/74.54$ & $0.01/38.88$ &  $99.53/ 11.58$    & $99.4/9.1$ &  $99.6/4.6$ &  $100/3.60$ &  $99.8/6.6$ \\ \cline{2-9}
                                    & Nature & $75.25/74.75$ & $31.20/19.81$ & $99.62/ 11.58$   & $100/8.7$ &  $100/3.6$ &  $100/4.4$ &  $100/3$  \\ \cline{2-9}
                                    & Trojan SQ & $75.73/74.27$ & $27.47/16.48$ & $99.62/ 11.58$ & $100/8.30$ &  $100/4.8$ &  $100/6$ &  $100/6.6$  \\ \cline{2-9}
                                    & Trojan WM & $75.75/74.25 $ & $42.01/19.05$  & ${100}/ 11.58$  & $100/9.3$  & $100/6.8$ &  $100/3.2 $ &  $70.8/7$ \\ \cline{2-9}
                                    & L2 inv & $75.60/74.40$ & $0.01/36.50$& ${100}/ 11.58$ & $99.5/9.8$ & $100/3.8$ &  $99.9/5.8$ &  $97.8/3.8$ \\ \hline
                                    
                                    \hline
                                    
            \multirow{6}{*}{\rotatebox{90}{CIFAR10}} & Badnets & $79.06/70.94$ & $100.00/87.43$  &  $95.20/ 44.92$   & $93/28.3$ &  $69.9/11.4$ &  $70.90/16.60$ &  $80.6/17$ \\ \cline{2-9}
                                    & Blend & $79.04/70.96$  &  $100/82.40$  &  ${97.71}/ 44.92$   & $98/27.2$  &  $97.2/12.8$ &  $97.2/13.6$ &  $84.50/15.6$ \\ \cline{2-9}
                                    & Nature & $78.60/71.40$ & $100/88.60$  & $99.91/ 44.92$    & $96.3/29.8$ & $100/11$ &  $100/14$ &  $100/16.4$ \\ \cline{2-9}
                                    & Trojan SQ & $78.80/71.20$ & $100/82.90$  & $99.87/ 44.92$ & $99.1/27.8$ & $100/16.4$ &  $100/13 $&  $98.10/17$ \\ \cline{2-9}
                                    & Trojan WM & $78.88/71.12$ & $0.03/44.10$  & $99.87/ 44.92$  & $99.9/28.9$  & $100/16.2$ &  $100/17$ &  $100/13.60$ \\ \cline{2-9}
                                    & L2 inv &  $78.89/71.11$&  $100/76.70$ &  ${99.99}/ 44.92$  & $98.5/26.4$  &  $90.7/8.8$ &  $95.4/13.8$ & $ 88.4/17.4$ \\ \hline
                                    
                                    \hline

              \multirow{6}{*}{\rotatebox{90}{GTSRB}} & Badnets & $75.89/74.01$ & $100/89.97 $  & $77.31/{9.08}$   & $95.8/32.2$  &  $91.70/15.8$ & $ 88.5/8.8$ &  $84.9/2$  \\ \cline{2-9}
                                    & Blend & $75.89/74.01 $ & $100/98.69 $  & $58.95/9.08$  & $94.1/39$ &  $91.9/36.2$ &  $78.9/13.4 $&  $91.50/34.4$ \\ \cline{2-9}
                                    & Nature & $75.89/74.01$ &$100/99.05$ &   $78.24/ {9.08}$  & $99.8/38.6$ &  $100/38.6$ &  $100/18.2$ &  $100/38.6$ \\ \cline{2-9}
                                    & Trojan SQ & $74.81/75.09$ & $100/94.92$  & $98.58/ 9.08$ & $97/33.1$  &  $99.6/32.6$ &  $98.4/18.8$ &  $99.2/34.2$ \\ \cline{2-9}
                                    & Trojan WM & $75.71/74.20$ &  $100/99.05$ &  $99.21/ {9.08}$ & $96.4/37$  &  $100/22.6$&  $100/15.6$ &  $100/16.2$   \\ \cline{2-9}
                                    & L2 inv & $75.84/74.06$ &  $100/92.07$ &  $95.25/ 9.08$ & $84.9/34.2$  &  $65.3/24.6$ &  $66.1/15.4$ &  $6.7/12.40$ \\ \hline
                                    
             \hline                       
             \multirow{6}{*}{\rotatebox{90}{SVHN}} & Badnets & $80.56/69.44$ & $100/74.70$ & $99.96/ 0.21$  &   $ 86.20/12.60$   & $90.8/28.8$ &  $88.60/13.6$ &  $80.9/17.4$\\ \cline{2-9}
                                    & Blend &  $80.70/69.30$& $18.20/45.52$ &  $99.97/ 0.21$  & $98.8/38.3$  &  $90.70/34.60$ &  $88.8/10.2$ &  $88.80/22.2$ \\ \cline{2-9}
                                    & Nature & $80.70/69.30$ & $4.82/7.59 $  &  $99.97/{ 0.21}$ & $99.7/38.7$ & $99.90/36.60$ &  $99.9/10.6$ &  $100/30.4$ \\ \cline{2-9}
                                    & Trojan SQ &$80.70/69.30$  & $6.89/7.59$  & $99.97/{ 0.21}$  & $99.4/41.2$  & $100/45.2$ &  $100/16$ &  $99.9/30$ \\ \cline{2-9}
                                    & Trojan WM & $80.67/69.32$ & $5.21/7.47$  &$99.97/ {0.21}$  & $99.60/38.7$  &  $100/29.40$&  $100/13.80$&  $100/20.6$  \\ \cline{2-9}
                                    & L2 inv & $80.65/69.35$ & $23.29/46.93$ & $99.97/{ 0.21}$ & $67.5/41.6$ & $ 89/15.4$ &  $90.70/17.4$ &  $100/16.4$ \\ \hline

            \hline
            \multirow{6}{*}{\rotatebox{90}{Flower102}} & Badnets & $74.80/70.20$ & $100/88.04$   & $99.98/ 100$  & $31.8/1.2$&  $4.7/17$ &  $4.40/13.2$ &  $35.10/13.2$ \\ \cline{2-9}
                                    & Blend & $74.71/70.29$ & $100/84.51$ & ${100}/ 100$  &  $28.7/0.6$ &  $73/16.6$ &  $82.60/15$ &  $51.2/12.20$ \\ \cline{2-9}
                                    & Nature & $74.71/70.29$ & $100/19.22$  & ${100}/ 100$  &  $29/0.40$  &   $97.50/18$ &  $99.30/13.2$ &  $95.90/14.4$  \\ \cline{2-9}
                                    & Trojan SQ & $74.71/70.29$ & $100/85.69$   & ${100}/100$  & $78.8/0.4$ & $99.5/12.8$ &  $100/14.20$ &  $99.50/14.60$ \\ \cline{2-9}
                                    & Trojan WM & $74.71/70.29$ &  $100/84.12$ & $100/100$ & $60.1/0.5$  & $100/15.6$ &  $100/14.2$ &  $100/15$ \\ \cline{2-9}
                                    & L2 inv & $74.80/70.20$ &  $100/87.84$  & ${100}/ 100$ &  $66/1$  & $25.5/11.2$ &  $58.9/13$ & $ 1.3/18.2$ \\ \hline
         
    \end{tabular}}
    
\end{table*}

\subsection*{Appendix A: Worst-Case False Positive Rate}
This Appendix presents a theoretical characterization of the worst-case false positive rate of MDTD. 

\emph{Theorem: } 
Assume that the function $f$ in Eqn. (\ref{eq:OptProb}) is differentiable, and that its gradient $\nabla_x f$ is locally Lipschitz. 
For any clean sample $x$, the worst-case false positive rate of MDTD is given by
\begin{equation}
    \mathbb{P}(\text{MDTD identifies } x \text{ as Trojan})\leq e^{-(\zeta+\alpha\sigma)^2/2\sigma^2},
\end{equation}
where $\zeta = \frac{L\|\nabla_f\mathcal{L}\|\|x-\hat{x}\|}{2\lambda-L\|\nabla_f\mathcal{L}\|}$, and $\hat{x}=\argmin_{x'\in D_{user}}\|x'-x\|_1$ is the sample from $D_{user}$ that belongs to the same class as $x$.

\begin{proof}
%The proof consists of two steps. 
We first quantify the range of values of the distance of a clean sample $x$ to a decision boundary in terms of its certified radius $\delta$. %of clean sample $x$. 
We then use these range of values to determine the worst-case false positive rate of MDTD.

Let $\hat{x}=\argmin_{x'\in D_{user}}\|x'-x\|_1$ and $\hat{\delta}$ be the certified radius of $\hat{x}$. 
Since $f$ is differentiable, $\delta$ and $\hat{\delta}$ must be critical points of:% the following: 
\begin{align*}
    -\nabla_f  \mathcal{L}\nabla_x  f|_{x+\delta} + 2\lambda\delta=0; 
    -\nabla_f  \mathcal{L}\nabla_x  f|_{\hat{x}+\hat{\delta}} + 2\lambda\hat{\delta}=0,
\end{align*}
respectively. 
We thus have that
\begin{subequations}\label{eq:dist}
\begin{align}
    &~\|2\lambda\delta-2\lambda\hat{\delta}\|_1 
    =\|\nabla_f  \mathcal{L}\nabla_x  f|_{x+\delta}-\nabla_f  \mathcal{L}\nabla_x  f|_{\hat{x}+\hat{\delta}}\|_1\\
    \leq&~\|\nabla_f  \mathcal{L}\|_1L\|\hat{x}-x+\hat{\delta}-\delta\|_1\label{eq:dist-1}\\
    \leq &~ \|\nabla_f  \mathcal{L}\|_1L(\|\hat{x}-x\|_1+\|\hat{\delta}-\delta\|_1)\label{eq:dist-2}
\end{align}
\end{subequations}
where inequalities \eqref{eq:dist-1} and \eqref{eq:dist-2} hold by the assumption that $\nabla_xf$ is locally Lipschitz and the triangle inequality, respectively.
Rearranging \eqref{eq:dist} yields 
\begin{equation}\label{eq:delta dist}
    \|\delta-\hat{\delta}\|_1\leq \frac{L\|\nabla_f\mathcal{L}\|_1\|x-\hat{x}\|_1}{2\lambda-L\|\nabla_f\mathcal{L}\|_1}\triangleq \zeta.
\end{equation} 
Given parameters $\mu$ and $\sigma$ which are estimated using the clean data set $D_{user}$, we can quantify the worst-case false positive rate for clean sample $x$ as 
\begin{align}
    \mathbb{P}(\text{MDTD identifies } x \text{ as Trojan})
    =&~ \mathbb{P}(|\delta - \mu|\geq\alpha\sigma)\\
    \leq&~\mathbb{P}(|\hat{\delta}-\zeta-\mu|\geq \alpha\sigma). \label{eq:sufficient bound}
\end{align}
Eqn. (\ref{eq:sufficient bound}) follows from Eqn. (\ref{eq:delta dist}) and the fact that we focus on the upper tail of the Gaussian distribution.
Using the tail bound of Normal distribution \cite{mitzenmacher2017probability}, we have 
$\mathbb{P}(|\hat{\delta}-\zeta-\mu|\geq \alpha\sigma)\leq e^{-(\zeta+\alpha\sigma)^2/2\sigma^2}$.
\end{proof}

\subsection*{Appendix B: True and False Positive Rates for Trojan Triggers on Image Datasets}
% \onecolumn 
In Sec. 5,  Table~\ref{tab:imagesresults} presented $F_1$ scores for spectral signature~\cite{tran2018samplefilter1}, activation clustering~\cite{chen2018detecting}, DCT-based detectors~\cite{zeng2021rethinking}, STRIP~\cite{gao2019strip}, and \emph{MDTD} for six different Trojan triggers for five image datasets- CIFAR100, CIFAR10, GTSRB, SVHN, and Flowers102. 
Table~\ref{tab:imagesresultstprfpr} presents values of true and false positive rates for spectral signature~\cite{tran2018samplefilter1}, activation clustering~\cite{chen2018detecting}, DCT-based detectors~\cite{zeng2021rethinking}, STRIP~\cite{gao2019strip}, and \emph{MDTD} for six different Trojan triggers for the same five datasets. 

We examine variants of MDTD when a user has white-box access (MDTD (FGSM) and MDTD (IFGSM)) and black-box access (MDTD (HopSkipJump)) to the DNN model. 
We observe that DCT-based detectors are able to simultaneously achieve high TPR and low FPR for the SVHN and Flowers102 datasets. 
However, the DCT-based detector method requires access to the entire training dataset, which makes it computationally expensive. 
On the other hand, MDTD is able to simultaneously achieve high true positive and low false positive rates even when a user has access to only a small set of labeled clean samples that need not be a part of the training set.

\subsection*{Appendix C: Adaptive Adversary for Audio}
%\textcolor{blue}{
In this appendix, we evaluate MDTD against an adversary seeking to disrupt MDTD \textbf{Stage 1} by adding noise to clean samples for nonimage audio-based inputs such that noise-embedded clean samples are adequately distant from the decision boundary. 
%}

%\textcolor{blue}{
Table~\ref{tab:robustAudio} shows $F_1$-scores of MDTD (IFGSM) for DNNs trained without adversarial noise ($\epsilon = 0.000$) vs. with adversarial noise ($\epsilon = 0.001$) for the SpeechCommand audio dataset with two different Trojan triggers- Modified (Mod) and Blend (Bld). 
Increasing the value of $\epsilon$ results in a reduction of $F_1$-score for the Mod trigger. 
This indicates that an adaptive adversary is more effective in reducing effectiveness of MDTD for audio inputs with a Mod trigger than when using a Bld trigger. 
However, different from our results for image (Table \ref{tab:robustCifar}) and graph (Table \ref{tab:robustGraph}) inputs, classification accuracy is only marginally affected. 
We provide a possible explanation below. 
%}

%\textcolor{blue}{
Each audio sample in the SpeechCommand dataset can be of a different length or time-duration, and different samples can belong to widely different ranges of values. 
This is unlike for images, where each pixel takes a value in a known interval- e.g., $[0,1]$ for black-white and $[0, 255]$ for color images. 
Such large variations in length and value make training of robust DNNs challenging, since adding (small amounts of) adversarial noise can result in minimal impact on sample values, and consequently, on classification accuracy. 
%}

% Unlike images in which each pixel has a value between [0,1], for an audio, each sample can have different length and its elements can have very large values. Therefore training a robust DNN can be challenging. Here, the robust Trojan model for Audio classification is trained using FGSM learning methods with $\epsilon=0.0001$. The larger amount of $\epsilon$  caused the model obtains very low accuracy on adversarial examples. The small amount as shown in Table~\ref{tab:audioRobust} does not degrade the detection rate of MDTD significantly fro IFGSM. FGSM approach could not obtain high detection rate since the model was trained on the adversarial examples generated by this approach. However, the robustness against FGSM adversarial noise does not guarantee the robustness against  adversarial noise trained using other adversarial training methods.

\begin{table}[]
\caption{This table reports $F_1$-scores of MDTD (IFGSM) for DNN models trained without adversarial noise ($\epsilon = 0.000$) vs. with adversarial noise ($\epsilon = 0.001$) 
%($\epsilon = 0, 0.01, \epsilon = 0.1$) 
for the SpeechCommand audio dataset with \textbf{two different Trojan triggers}. 
Increasing the value of $\epsilon$ results in a reduction of $F_1$-score for the Mod trigger. However, classification accuracy is only marginally affected, indicating that an adaptive adversary is more effective in reducing effectiveness of MDTD for audio inputs than for image inputs without sacrificing classification accuracy (compare with results in Tables \ref{tab:robustCifar}. \ref{tab:robustGraph}). 
   }
   \label{tab:robustAudio} 
    \centering
    \begin{tabular}{|c|c|c|c| c|}
    \hline
    
         $\epsilon$&Attack & Acc. & ASR & $F_1$: MDTD (IFGSM)  \\ \hline
         % \multirow{6}{*}{$0.001$} & BadNets& &  & $0.61$\\ \cline{2-5}
         % &Blend & &  & $0.87$ \\ \cline{2-5}
         %  & Nature & &  & $0.91$ \\ \cline{2-5}
         %   & Trojan SQ& &  &  $0.91$ \\ \cline{2-5}
         %    & Trojan WM & &  & $0.93$ \\ \cline{2-5}
         %     & L2 inv& &  & $0.63$ \\ \hline
         % \multirow{6}{*}{$0$} & BadNets& $81.18$ &  $97.4$   & $0.77$\\ \cline{2-5}
         % &Blend & $81.11\%$ &  $99.95\%$  &$0.91$ \\ \cline{2-5}
         %  & Nature & $81.52\%$ &  $99.99\%$  & $0.92$\\ \cline{2-5}
         %   & Trojan SQ& $81.69\%$ &  $100\%$  & $0.93$ \\ \cline{2-5}
         %    & Trojan WM & $81.63\%$ &  $100\%$  & $0.91$ \\ \cline{2-5}
         %     & L2 inv& $81.46\%$ &  $99.95\%$  & $0.91$\\ \hline
          \multirow{2}{*}{$0.000$} & Mod& $94.43$ &  $99.95$   & $0.80$\\ \cline{2-5}
         &Bld & $93.31\%$ &  $99.93\%$  &$0.81$ \\ 
          \hline
          \multirow{2}{*}{$0.001$} & Mod& $94.65\%$ &  $99.75\%$   & $0.69$\\ \cline{2-5}
         &Bld & $94.25\%$ &  $94.97\%$  & $0.82$\\  \hline
        
    \end{tabular}
    \end{table}
\end{document}